\documentclass [11pt]{article}
\pdfoutput=1

\usepackage{jheppub}

\usepackage[utf8]{inputenc}

\usepackage{amsfonts}

\usepackage{amsmath,amssymb}

\usepackage{mathtools}

\usepackage{array}

\usepackage{float}

\usepackage{color}

\usepackage{mathrsfs}

\usepackage{ytableau}

\def\nn{\nonumber}

\newcommand{\n}{\mathbf{n}} 

\newcommand{\cO}{\mathcal{O}}

\newcommand{\<}{\langle}
\renewcommand{\>}{\rangle}

\DeclareMathOperator*{\SumInt}{%
	\mathchoice%
	{\ooalign{$\displaystyle\sum$\cr\hidewidth$\displaystyle\int$\hidewidth\cr}}
	{\ooalign{\raisebox{.14\height}{\scalebox{.7}{$\textstyle\sum$}}\cr\hidewidth$\textstyle\int$\hidewidth\cr}}
	{\ooalign{\raisebox{.2\height}{\scalebox{.6}{$\scriptstyle\sum$}}\cr$\scriptstyle\int$\cr}}
	{\ooalign{\raisebox{.2\height}{\scalebox{.6}{$\scriptstyle\sum$}}\cr$\scriptstyle\int$\cr}}
}

\title{\boldmath Two-point Functions and Bootstrap Applications in Quantum Field Theories}

\author[a,b]{Denis Karateev}

\affiliation[a]{Fields and Strings Laboratory, Institute of Physics\\ École Polytechnique Fédéral de Lausanne (EPFL)
	\\ Route de la Sorge, CH-1015 Lausanne, Switzerland}

\affiliation[b]{
	Philippe Meyer Institute, Physics Department\\
	\'Ecole Normale Sup\'erieure (ENS), Universit\'e PSL\\
	24 rue Lhomond, F-75231 Paris, France}

\abstract{
We study two-point functions of local operators and their spectral representation in UV complete quantum field theories in generic dimensions focusing on conserved currents and the stress-tensor.
We establish the connection with the central charges of the UV and IR fixed points. We re-derive ``c-theorems'' in 2d and show the absence of their direct analogs in higher dimensions. We conclude by focusing on quantum field theories with a mass gap. We study the stress tensor two-particle form factor, derive implications of unitarity and define concrete bootstrap problems in generic dimensions.
}

\begin{document}
\maketitle

\section{Introduction}
The numerical S-matrix bootstrap program was recently revived in \cite{Paulos:2016fap, Paulos:2016but, Paulos:2017fhb} and received further attention in \cite{Doroud:2018szp, He:2018uxa, Cordova:2018uop, Guerrieri:2018uew, Homrich:2019cbt, EliasMiro:2019kyf, Cordova:2019lot, Bercini:2019vme, Correia:2020xtr, Bose:2020shm,Guerrieri:2020bto,Hebbar:2020ukp}. This program allows to numerically construct scattering amplitudes which obey crossing and unitarity at all energies.  In \cite{Karateev:2019ymz} the authors proposed to extend the S-matrix bootstrap program to accommodate form factors and spectral densities of local operators in a general number of dimensions.\footnote{In $d=2$ this direction is rather well explored in the case of integrable models, see for example \cite{Karowski:1978vz,Cardy:1990pc,Babujian:1998uw,Babujian:2013roa,Mussardo:1993ut,Delfino:1994ea,Delfino:1996jr,Delfino:2003yr}.} When preparing \cite{Karateev:2019ymz} it became clear that the systematic treatment of two-point functions, spectral densities and their relation to central charges in a generic number of dimensions is missing in the literature.

The first goal of this work is to fill this gap. The second goal is to define concrete bootstrap problems in higher dimensions. In sections \ref{sec:two-point_functions} and \ref{sec:spectral_densities} we provide the main definitions and setup the formalism. The main results are given in sections \ref{sec:conformal_spectral}, \ref{sec:spectral_densities_and_central_charges} and \ref{sec:discussion}.  More precisely, in section \ref{sec:conformal_spectral} we compute explicitly spectral densities of conserved currents and the stress-tensor in conformal field theories. In section \ref{sec:spectral_densities_and_central_charges} we show that in generic quantum field theories in $d\geq 3$ the asymptotic behavior of spectral densities of conserved currents and the stress-tensor is driven by the central charges, in $d=2$ instead we obtain the integral sum-rules for the central charges which lead to the ``c-theorems''.\footnote{These were first discussed in \cite{Cappelli:1990yc,VilasisCardona:1994ri}, see also \cite{Cappelli:1991ke}.} In section \ref{sec:discussion} focusing on quantum field theories with a mass gap we discuss the stress-tensor two-particle form factor and partial amplitudes. We then derive semi-positive definite constraints coming from unitarity and discuss applications to bootstrap.
Various computations and technical details supporting the main text are given in appendices \ref{app:2pt_CFTs} - \ref{app:FF_normalization}.

All our results and conclusions are clearly stated in sections \ref{sec:conformal_spectral}, \ref{sec:spectral_densities_and_central_charges} and \ref{sec:discussion}. As a consequence we do not dedicate a separate section to conclusions. In order to somewhat compensate for this and also to facilitate the reading of the paper we provide however an extended summary of the paper and its key points.

\section*{Summary of the paper}

In section \ref{sec:two-point_functions} we study Euclidean two-point functions. Their most general form compatible with rotational and translational invariance is given by \eqref{eq:2pt_J} and \eqref{eq:2pt_T} for conserved currents and the stress-tensor respectively. In the presence of conformal symmetry, the two-point functions of conserved currents and the stress-tensor are completely fixed up to the numerical  coefficients $C_J$ and $C_T$ called the central charges, see \eqref{eq:2pt_J_conformal} and \eqref{eq:2pt_T_cft_main}. In a generic quantum field theory (QFT) we assume that both its UV and IR fixed points are described by the UV and the IR conformal field theory (CFT).\footnote{Fixed points are scale invariant. In $d=2$ scale invariance implies conformal invariance \cite{Zamolodchikov:1986gt,Polchinski:1987dy}. In $d\geq 3$ there is no general proof, for some recent discussion see \cite{Bzowski:2014qja,Dymarsky:2014zja,Dymarsky:2013pqa}.} This requirement at the level of two-point functions translates into conditions \eqref{eq:limit_1_J}, \eqref{eq:limit_2_J} and \eqref{eq:limit_1_E}, \eqref{eq:limit_2_E}. The most important result of section \ref{sec:two-point_functions} are the integral expressions for the difference of the UV and IR central charges given in \eqref{eq:integral_J} and \eqref{eq:integral_expression_T_form2}. In section \ref{sec:two-point_functions} we also show that two-point functions might contain a parity odd part in $d=2$ and $d=3$ dimensions. In $d=2$ it is completely fixed by the global anomaly $C^\prime_J$ for conserved currents and by the gravitational anomaly $C^\prime_T$ for the stress-tensor.\footnote{After this paper came out we became aware of the works \cite{Bastianelli:1996gh,Bastianelli:1997hb} where the same conclusions were made.} In $d=3$ the parity odd part does not contain any information about the UV and IR fixed points.

In section \ref{sec:spectral_densities} we study Wightman and time-ordered two-point functions in the Lorentzian signature.\footnote{For the formal definition of Wightman two-point functions see appendix \ref{app:wightman}.} We define in section \ref{sec:wightman} the spectral densities as Fourier transformed Wightman two-point functions. We define components of the spectral densities as the coefficients in their decomposition into a basis of tensor structures. This basis is constructed from the projectors (objects mapping finite irreducible representations of the Lorentz group into the finite irreducible representations of the Little group), see \eqref{eq:projectors_vector_main_text} and \eqref{eq:projectors} for their explicit expressions. We then show the non-negativity of the components of spectral densities. The explicit spectral decomposition of the two-point function of conserved currents and the stress-tensor is given in \eqref{eq:spectral_representation_2} and \eqref{eq:spectral_representation_T_final1} respectively. We study time-ordered two-point functions in section \ref{sec:time-ordered}. Their spectral decomposition, known as the K\"{a}ll\'en-Lehmann representation, in the case of conserved currents and the stress-tensor is given in \eqref{eq:spectral_J} and
\eqref{eq:spectral_TT} respectively. Under the Wick rotation the time-ordered two-point functions get mapped precisely to the Euclidean two-point functions. This allows to define the K\"{a}ll\'en-Lehmann spectral decomposition of Euclidean two-point functions.

In conformal field theories the Wightman two-point functions are completely fixed by the conformal symmetry, hence the spectral densities are also completely fixed. In section \ref{sec:conformal_spectral} we explicitly compute the components of the spectral densities in the case of Lorentz spin one and Lorentz spin two operators, see \eqref{eq:J_cft_density} and \eqref{eq:T_density}.

In section \ref{sec:spectral_densities_and_central_charges} we show that the central charges $C_J$ and $C_T$ in $d\geq 3$ define the asymptotic behavior of certain components of the spectral densities, see \eqref{eq:asymptotics_elaborated} and \eqref{eq:asymptotics_T2hat}. In $d=2$ we recover the known integral expressions \eqref{eq:sum_rule_J_2d_main_text_kkbar} and \eqref{eq:sum_rule_T_2d_main_text_kkbar}. The latter prove immediately the ``c-theorems'' in $d=2$.
In order to show all these statements systematically we employ the sum-rules \eqref{eq:integral_J} and \eqref{eq:integral_expression_T_form2} and perform the K\"{a}ll\'en-Lehmann decomposition of its integrands. We provide the technical details of this strategy in appendix \ref{app:technical_dentails}.

In section \ref{sec:discussion} we discuss bootstrap applications. We start in section \ref{sec:form_factor} by studying the two-particle form factor of the stress-tensor. We discuss its generic form, the relation to the stress-tensor spectral density and its projections to definite Little group spin, see \eqref{eq:stress-tensor_form_factor}, \eqref{eq:components_spectral_ff} and \eqref{eq:projected_FF}. In section \ref{sec:unitarity_constraints} we derive the unitarity constraints as semi-positive conditions on the matrices involving partial amplitudes, the stress-tensor form factor and the stress-tensor spectral density, see \eqref{eq:positivity_S}, \eqref{eq:positivity_theta} and \eqref{eq:positivity_2_final}. In section \ref{sec:bootstrap_problems} we define concrete bootstrap problems which can be studied with modern numerical techniques.

\section*{Notation}
Let us comment on the notation of the paper. We will use Latin letters to indicate the Euclidean space
\begin{equation}
a,b = 0,1,2,\ldots,d-1.
\end{equation}
Instead we will use Greek letters to indicate the Lorentzian space
\begin{equation}
\mu,\nu=0,1,2,\ldots,d-1.
\end{equation}
We attribute the meaning of time to $x^0$ component. It should be clear from the context if $x^0$ is Euclidean or Lorentzian time. Sometimes we will also indicate it explicitly. In the Lorentzian signature we will use the mostly plus metric
\begin{equation}
\eta_{\mu\nu} = \eta^{\mu\nu} = \{-, +, +, \ldots \}.
\end{equation}
Throughout the text we will also use the following (manifestly translation invariant) objects
\begin{equation}
x_{ij}^a \equiv x_i^a-x_j^a,\qquad
x_{ij}^\mu \equiv x_i^\mu-x_j^\mu.
\end{equation}
We will also use sometime vector notation for spatial coordinates
\begin{equation}
\vec x = \{x^1,x^2,\ldots x^{d-1} \}.
\end{equation}

\section{Euclidean two-point functions}
\label{sec:two-point_functions}
We start by studying two-point functions in Euclidean signature. We refer to them as the Euclidean two-point functions. We attribute $a=0$ component of the Euclidean coordinate $x^a$ to Euclidean time. The Euclidean two-point functions are ``time-ordered'' with respect to this Euclidean time, see appendix \ref{app:wightman} for details.
In what follows we will study Euclidean two-point functions of conserved currents and the stress-tensor at non coincident points.\footnote{Treating coincident points correctly is very difficult due to presence of contact terms, see for example \cite{Bzowski:2015pba} and section 3.1 of \cite{Bzowski:2013sza} for a discussion of two-point functions in CFTs. Luckily in position space one can often avoid talking about them. The situation is different in momentum space where one has to integrate over the whole space including the coincident points. From this perspective working with momentum space correlators is much more difficult. For works on CFT correlators in momentum space see \cite{Bzowski:2015pba, Bzowski:2013sza,Bzowski:2017poo,Bzowski:2018fql, Gillioz:2018mto, Gillioz:2019lgs, Gillioz:2020mdd, Bautista:2019qxj, Gillioz:2020wgw, Gillioz:2021sce, Gillioz:2021kps}.} We will derive their most general form fixed by the rotational and translational invariance.\footnote{For a concrete perturbative computation of time-ordered two-point functions of the stress-tensor in gauge theories see \cite{Coriano:2020zap}.} We will define central charges and derive integral expressions (sum-rules) they satisfy. This section develops on the ideas presented in \cite{Cardy:1988tj,Cardy:1988cwa}.

\subsection{Conserved currents}
\label{sec:J}
Consider the local conserved current $J^a(x)$. Such an operator is generally present in systems with  a $U(1)$ symmetry. The generalization to the case of non-Abelian symmetries is trivial.\footnote{In case the system under consideration is invariant under a non-Abelian group, the corresponding conserved operator would be $J^a_A(x)$, where $A$ is the index in the adjoint representation of the non-Abelian group. The two-point function in \eqref{eq:2pt_J} gets an additional overall tensor structure which depends on the adjoint indices, namely $\<0|J_A^a(x_1) J_B^b(x_2)|0\>_E\sim \text{tr}(t_At_B)$, where $t^A$ are the generators of the symmetry in the adjoint representation. One can always choose a basis of these generators such that $\text{tr}(t_At_B)=\delta_{AB}$. } Due to rotational and translational invariance the Euclidean two-point function of conserved currents has the following generic form
\begin{equation}
\label{eq:2pt_J}
\begin{aligned}
\<0|J^a(x_1) J^b(x_2)|0\>_E &= 
\frac{1}{r^{2(d-1)}}\times
\left(
h_1(r) \delta^{ab} + h_2(r)\,\frac{x_{12}^a\, x_{12}^b}{r^2}
+\sum_n i g_n(r) \mathbf{T}_n^{ab}(x_1,x_2)
\right),\\
r &\equiv |x_{12}|,
\end{aligned}
\end{equation}
where $h_1(r)$, $h_2(r)$ and $g_n(r)$ are dimensionless functions which contain dynamical information of a particular theory and $\mathbf{T}_n^{ab}$ are the parity odd tensor structures (structures containing a single Levi-Civita symbol).\footnote{The imaginary unit $i$ in the parity odd part of \eqref{eq:2pt_J} is introduced for future convenience.} Since the form of the Levi-Civita symbol depends on the number of dimensions, the parity odd tensor structures should be discussed separately for each dimension. We postpone this discussion until the end of this section.  Notice, that since $r$ is a dimensionful quantity one needs at least one dimensionful parameter in the theory in order for the functions $h_1$, $h_2$ and $g_n$ not to be simply constants. Suppose we have a single dimensionful parameter $a$ in the theory with the mass dimension $[a]=1$. Then the functions in \eqref{eq:2pt_J} would have the following arguments
\begin{equation}
h_1(a r),\quad h_2(a r),\quad g_n(ar).
\end{equation}
Notice also that we exclude the $r=0$ point from the discussion in order to remove the contact terms which do not play any role in our further investigation.

Since the Euclidean two-point functions are time-ordered, the following symmetry condition must be obeyed
\begin{align}
\label{eq:symmetry}
\<0|J^a(x_1) J^b(x_2)|0\>_E =
\<0|J^b(x_2) J^a(x_1)|0\>_E. 
\end{align}
Clearly, the parity even structures in \eqref{eq:2pt_J} satisfy this condition automatically.
In the presence of conformal symmetry there are further constraints on the two-point function \eqref{eq:2pt_J}. We derive them in appendix \ref{app:2pt_CFTs}. Here we simply quote the final result
\begin{equation}
\label{eq:2pt_J_conformal}
\<0|J^a(x_1) J^b(x_2)|0\>_{E,\;CFT} = 
\frac{C_J}{r^{2(d-1)}}\times
\mathcal{I}^{ab}(x_1,x_2)+
\frac{i\,C^\prime_J}{r^{2(d-1)}}\times\delta_{d,2}\,\mathcal{E}^{ab}(x_1,x_2),
\end{equation}
where we have defined
\begin{align}
\label{eq:objects_IE}
\mathcal{I}^{ab}(x_i,x_j) \equiv \delta^{ab} - 2\,\frac{x_{ij}^a x_{ij}^b}{x_{ij}^2},\qquad
\mathcal{E}^{ab}(x_i,x_j) \equiv \epsilon^{ab} +
2 \frac{ x_{ij}^a\epsilon^{bc} x_{ij}^c}{x_{ij}^2}.
\end{align}
The constants $C_J$ and $C^\prime_J$ (partly) characterize the dynamics of the conformal field theory. They are called the central charges of two currents.\footnote{\label{foot:C_J}For a generic local operator in a CFT the constant appearing in its two-point function defines the normalization of this operators. Its value can be set to one by rescaling the normalization of this operator. For conserved operators this is no longer the case since they obey a particular symmetry algebra which fixes their normalizations. In the Abelian case we have $Q\equiv\int d^d x J^0(x)\delta(x^0)$ and $[Q,\cO]=q_\cO \cO$, where $q_\cO$ are the charges. In the non-Abelian case instead we have $Q^A\equiv\int d^d x J^{0}_A(x)\delta(x^0)$ and $[Q_{A},Q_{B}]=i f_{ABC} Q_{C}$, where $A$, $B$ and $C$ are the adjoint indices of some non-Abelian group and $f$ are its structure constants.} In unitary theories $C_J>0$ and $-C_J\leq C^\prime_J\leq +C_J$, see \eqref{eq:conditions_CJ} and appendix \ref{app:unitarity_appendix} for details. The central charge $C_J$ was introduced in \cite{Osborn:1993cr}, it corresponds to the parity even structure $\mathcal{I}^{ab}$ and is a universal quantity in any number of dimensions. The central charge $C^\prime_J$ corresponds to the parity odd structure $\mathcal{E}^{ab}$ and can only be present in $d=2$ dimensions. Notice that the parity odd structure $\mathcal{E}^{ab}$ automatically obeys the symmetry condition \eqref{eq:symmetry}. This is not obvious at first glance, but can be shown using the following identity
\begin{equation}
\label{eq:condition_2d}
x^a\epsilon^{bc}+x^b\epsilon^{ca}+x^c\epsilon^{ab}=0.
\end{equation}
The two-point function \eqref{eq:2pt_J_conformal} is a special case of \eqref{eq:2pt_J} where all the dimensionless functions $h_1(r)$, $h_2(r)$ and $g_n(r)$ are constants (since there are no dimensionful parameters in the CFTs) appropriately related to form conformally covariant tensor structures \eqref{eq:objects_IE}.

Given that we work with a UV complete QFT at high energies (UV) or equivalently at small distances we should recover conformal invariance, namely\footnote{Here we take the limit $r\rightarrow 0$ in such a way that $r$ is always positive. It can be infinitely close to zero but never becomes zero. In other words it does not probe contact terms.}
\begin{equation}
\label{eq:limit_1_J}
\lim_{x\rightarrow 0}\,
r^{2(d-1)}\times\left(
\<0|J^a(x) J^b(0)|0\>_E - \<0|J^a(x) J^b(0)|0\>_{E,\;UV\;CFT}
\right)=0.
\end{equation}
Analogously at low energies (IR) or equivalently at large distances we again recover conformal invariance
\begin{equation}
\label{eq:limit_2_J}
\lim_{x\rightarrow \infty}\,
r^{2(d-1)}\times\left(
\<0|J^a(x) J^b(0)|0\>_E -
\<0|J^a(x) J^b(0)|0\>_{E,\;IR\;CFT}
\right)=0.
\end{equation}
In quantum field theories with a mass gap such as QCD, the IR CFT is simply empty.

\subsubsection*{Parity even part}
\label{sec:integral_expressions_J}
Let us first focus on the parity even terms in \eqref{eq:2pt_J} and \eqref{eq:2pt_J_conformal}. We can rewrite the condition \eqref{eq:limit_1_J} and \eqref{eq:limit_2_J} as follows
\begin{equation}
\label{eq:boundary_J}
\begin{aligned}
\lim_{r\rightarrow 0} h_1(r) &= C_J^{UV},\qquad
\lim_{r\rightarrow 0} h_2(r) = -2\,C_J^{UV},\\
\lim_{r\rightarrow \infty} h_1(r) &= C_J^{IR},\qquad
\lim_{r\rightarrow \infty} h_2(r) = -2\,C_J^{IR}.
\end{aligned}
\end{equation}
In other words, the central charges $C_J^{UV}$ and $C_J^{IR}$ determine the asymptotic behavior of the functions $h_1(r)$ and $h_2(r)$.

We will now derive an integral expression for the UV and IR central charges in terms of the two-point function of conserved currents in a generic QFT. Conservation of the currents implies the following differential equation
\begin{equation}
\label{eq:conservation_J}
\partial_\mu J^\mu(x)=0
\quad\Rightarrow\quad
h'_1(r)+h'_2(r)=\frac{d-1}{r}\times\left(2h_1(r)+h_2(r)\right).
\end{equation}
Integrating both sides of \eqref{eq:conservation_J} and using the asymptotic conditions \eqref{eq:boundary_J}  we get
\begin{equation}
\label{eq:integral_J_raw}
C_J^{UV}-C_J^{IR}=(d-1)\times
\lim_{r_{min}\rightarrow 0}
\lim_{r_{max}\rightarrow \infty}\int_{r_{min}}^{r_{max}} \frac{dr}{r}\,\left(2h_1(r)+h_2(r)\right).
\end{equation}
This can be equivalently rewritten by using \eqref{eq:2pt_J} as\footnote{As we will see shortly, the parity odd tensor structures obey $\delta^{ab}\mathbf{T}^{ab}=0$ and $x^a x^b\mathbf{T}^{ab}=0$.}
\begin{equation}
\label{eq:integral_J}
C_J^{UV}-C_J^{IR}=
\lim_{r_{min}\rightarrow 0}
\lim_{r_{max}\rightarrow \infty}\int_{r_{min}}^{r_{max}}
dr\, r^{2d-3}\left(\delta^{ab}+(d-2) \frac{x^a x^b}{r^2}\right)
\<0|J^a(x) J^b(0)|0\>_E.
\end{equation}

As shown in appendix \ref{app:reflection_positivity} in unitary theories the following constraints hold
\begin{equation}
\label{eq:inequalities_J}
\forall r:\qquad
h_1(r)\geq 0,\quad
h_1(r)+h_2(r)\leq 0
\end{equation}
valid for $d\geq 2$.
As a result the integrand in \eqref{eq:integral_J_raw} does not have a definite sign and one cannot derive any inequality for the difference of the UV and IR central charges simply using \eqref{eq:inequalities_J}, in other words using \eqref{eq:inequalities_J} one cannot prove the ``c-theorem'' for conserved currents. The proof of the ``c-theorem'' for conserved currents exists however and will be given in section \ref{sec:spectral_densities_and_central_charges}.

\subsubsection*{Parity odd part}
Let us now focus on the parity odd terms in \eqref{eq:2pt_J} and \eqref{eq:2pt_J_conformal}. Since the number of indices in the Levi-Civita symbol depends on the number of dimensions, we will address the case of $d=2$, $d=3$ and $d\geq 4$ dimensions separately.

Let us start from $d=2$. Using rotational and translational invariance one can write two parity odd tensor structures in \eqref{eq:2pt_J}, namely 
\begin{align}
\sum_n ig_n(r) \mathbf{T}_n^{ab}(x_1,x_2)=
ig_1(r)\,\epsilon^{ab}+
ig_2(r)\,\frac{x_{12}^a \epsilon^{bc}x_{12}^c}{r^2}.
\end{align}
Requiring \eqref{eq:symmetry} and using \eqref{eq:condition_2d}
one obtains the following constraint on the unknown functions
\begin{equation}
g_2(r) = 2\,g_1(r).
\end{equation}
As a result we get the following most general form of the parity odd part of the two-point function of two currents
\begin{align}
\label{eq:current_odd_final}
\<0|J^a(x_1) J^b(x_2)|0\>_E^\text{odd} = 
\frac{ig_1(r)}{r^{2(d-1)}}\times
\left(
\epsilon^{ab}+
2\,\frac{x_{12}^a \epsilon^{bc}x_{12}^c}{r^2}
\right).
\end{align}
Conservation implies that
\begin{equation}
g'_1(r) =0
\quad\Rightarrow\quad
g_1(r)=const.
\end{equation}
In other words the expression for the parity odd part of the two-point function of conserved operators \eqref{eq:current_odd_final} is identical to the one of conformal field theories in \eqref{eq:2pt_J_conformal}.
The asymptotic conditions \eqref{eq:limit_1_J} and \eqref{eq:limit_2_J} imply that
\begin{equation}
\label{eq:anomaly}
g_1(r) = C_J^{\prime\,UV} = C_J^{\prime\,IR}.
\end{equation}
This requirement shows that the central charge $C_J^{\prime\,UV}$ is well defined along the flow and remains unchanged in the IR. It is nothing but the anomaly coefficient of the global $U(1)$ current.\footnote{For further reading on global anomalies in $d=2$ see for example section 19.1 in \cite{Peskin:1995ev}. See also section 6 in \cite{Bilal:2008qx}.} Using the standard anomaly matching argument of 't Hooft one can argue that the global anomaly must be an invariant quantity along the flow in accordance with \eqref{eq:anomaly}.

In $d=3$ one can write only a single parity odd tensor structure
\begin{equation}
\label{eq:current_odd_d=3}
\<0|J^a(x_1) J^b(x_2)|0\>_E^\text{odd} = 
\frac{i g_1(r)}{r^{2(d-1)}}\times
\frac{\epsilon^{abc}x_{12}^c}{r}.
\end{equation}
The expression \eqref{eq:current_odd_d=3} automatically complies with the condition  \eqref{eq:symmetry} and satisfies conservation. Since there are no allowed parity odd terms in the CFT two-point function in $d=3$ the asymptotic conditions \eqref{eq:limit_1_J} and \eqref{eq:limit_2_J} require
\begin{equation}
\lim_{r\rightarrow 0} g_1(r) = \lim_{r\rightarrow \infty} g_1(r) = 0.
\end{equation}
Moreover in unitary theories due to reflection-positivity the following condition holds
\begin{equation}
\forall r:\qquad
-h_1(r)\leq g_1(r) \leq +h_1(r).
\end{equation}
This can be shown by plugging \eqref{eq:current_odd_d=3} into \eqref{eq:reflection_positivity_current}.
The parity odd contribution to the two-point function \eqref{eq:current_odd_d=3} is the Chern-Simons like term.\footnote{See for example chapter 5 of David Tong's  lectures on the Quantum Hall Effect \cite{Tong}. } It does not contain any information about the UV or IR fixed points and thus it will not be studied further in this paper.

In the case of $d\geq 4$ no parity odd structures can be constructed. This follows from the simple fact that the Levi-Civita has too many indices and that the contractions of the form $\epsilon^{abcd}x^cx^d$ trivially vanish.

\subsection{Stress-tensor}
\label{sec:T}
Let us now turn our attention to the local stress-tensor $T^{ab}(x)$ totally symmetric in its indices, namely $T^{ab}(x)=T^{ba}(x)$. In a $d$-dimensional non-conformal quantum field theory the stress-tensor transforms in the reducible representation of the rotational group $SO(d)$. It can be decomposed as a direct sum of the trivial and the symmetric traceless representations: $\bullet\oplus\,\bf{d}$. The trivial representation corresponds to the trace of the stress-tensor which we denote by
\begin{equation}
\Theta(x) \equiv T^{aa}(x).
\end{equation}

Logically the discussion in this section will be identical to the one of conserved currents with several minor complications. The most general Euclidean two-point function consistent with the rotational symmetry and the translational invariance has the following form
\begin{align}
\<0|T^{ab}(x_1)T^{cd}(x_2)|0\>_E =
\frac{1}{r^{2d}} \left(\sum_{m} h_m(r) \mathbb{T}_m^{abcd}(x_1,x_2)
+\sum_n i g_n(r) \mathbf{T}_n^{abcd}(x_1,x_2)
\right).
\label{eq:2pt_T}
\end{align}
Here $\mathbb{T}_m^{abcd}$ and $ \mathbf{T}_n^{abcd}$ denote parity even and odd tensor structures respectively. The imaginary unit $i$ is introduced in the parity odd part for the later convenience. The functions $h_m(r)$ and $g_n(r)$ multiplying these structures are dimensionless. Since the Euclidean correlation functions are time-ordered one has the following symmetry condition
\begin{equation}
\label{eq:symmetry_condition_T}
\<0| T^{ab}(x_1)T^{cd}(x_2)|0\>_E = \<0| T^{cd}(x_2)T^{ab}(x_1)|0\>_E.
\end{equation}

In the presence of the conformal symmetry the form of the two-point function \eqref{eq:2pt_T} gets severely restricted and the two-point function becomes
\begin{align}
\nn
\<0|T^{ab}(x_1)T^{cd}(x_2)|0\>_E
&=\frac{C_T}{x_{12}^{2d}}\times\left( \frac{1}{2}\left(\mathcal{I}^{ac}(x_1,x_2)\mathcal{I}^{bd}(x_1,x_2)+
\mathcal{I}^{ad}(x_1,x_2)\mathcal{I}^{bc}(x_1,x_2)\right)
-\frac{1}{d}\,\delta^{ab}\delta^{cd}\right)\\
\nn
&+\frac{i\,C^\prime_T}{x_{12}^{2d}}\times\frac{\delta_{d,2}}{4}\,\bigg(
\mathcal{I}^{ac}(x_1,x_2)\mathcal{E}^{bd}(x_1,x_2)+
\mathcal{I}^{ad}(x_1,x_2)\mathcal{E}^{bc}(x_1,x_2)\\
&+\mathcal{I}^{bc}(x_1,x_2)\mathcal{E}^{ad}(x_1,x_2)+
\mathcal{I}^{bd}(x_1,x_2)\mathcal{E}^{ac}(x_1,x_2)
\bigg).
\label{eq:2pt_T_cft_main}
\end{align}
where the objects $\mathcal{I}^{ab}$ and $\mathcal{E}^{ab}$ were defined in \eqref{eq:objects_IE}. We derive this expression in appendix \ref{app:2pt_CFTs}. The coefficient $C_T$ is called the stress-tensor central charge. It was first introduced in \cite{Osborn:1993cr}.\footnote{The stress-tensor defines the conformal algebra. For instance the dilatation operator is defined as $D \equiv -\int d\Omega_d r^{d-2} x^a x^ b T^{ab}(x)$, where $\Omega_d$ is the sphere in $d$ dimensions, for instance $\Omega_3=4\pi$ and $\Omega_4=2\pi^2$. Under dilatations the primary operators transform as $[D,\cO(0)]=-i\Delta_\cO\cO(x)$, where $\Delta_\cO$ is the scaling dimension. Once the scaling dimensions are fixed there is no possibility to rescale the stress-tensor. Thus, the value of $C_T$ cannot be changed in contrast to the coefficients appearing in two-point functions of generic operators. See also footnote \ref{foot:C_J} for the similar discussion in the case of conserved currents.} It is a universal quantity in any number of dimensions. In unitary theories $C_T>0$. The quantity $C^\prime_T$ is another central charge which is allowed only in $d=2$ dimensions. In unitary theories $-C_T\leq C^\prime_T\leq +C_T$, see \eqref{eq:conditions_CT}.

Given that our quantum field theory is UV complete, namely its UV fixed point is described by a UV CFT (and by an IR CFT in the IR),  we have the following conditions
\begin{align}
\label{eq:limit_1_E}
\lim_{x\rightarrow 0}\,
r^{2d}&\times\left(
\<0|T^{ab}(x) T^{cd}(0)|0\>_E -
\<0|T^{ab}(x) T^{cd}(0)|0\>_{E,\;UV\;CFT}
\right)=0,\\
\label{eq:limit_2_E}
\lim_{x\rightarrow \infty}\,
r^{2d}&\times\left(
\<0|T^{ab}(x) T^{cd}(0)|0\>_E -
\<0|T^{ab}(x) T^{cd}(0)|0\>_{E,\;IR\;CFT}
\right)=0.
\end{align}

\subsection*{Parity even part}
Let us now focus on the parity even part of the two-point function \eqref{eq:2pt_T}. In general number of dimensions one can write five linearly independent tensor structures which read as
\begin{equation}
\label{eq:tensor_structures}
\begin{aligned}
\mathbb{T}_1^{abcd}(x_1,x_2)\equiv &\frac{x_{12}^a x_{12}^b x_{12}^c x_{12}^d}{r^4},\\
\mathbb{T}_2^{abcd}(x_1,x_2)\equiv &\frac{x_{12}^a x_{12}^b \delta^{cd}+x_{12}^c x_{12}^d \delta^{ab}}{r^2},\\
\mathbb{T}_3^{abcd}(x_1,x_2)\equiv&\frac{
	x_{12}^a x_{12}^c \delta^{bd}+
	x_{12}^b x_{12}^c \delta^{ad}+
	x_{12}^a x_{12}^d \delta^{bc}+
	x_{12}^b x_{12}^d \delta^{ac}}{r^2},\\
\mathbb{T}_4^{abcd}(x_1,x_2)\equiv&\delta^{ab}\delta^{cd}\\
\mathbb{T}_5^{abcd}(x_1,x_2)\equiv &
\delta^{ac}\delta^{bd}+\delta^{bc}\delta^{ad}.
\end{aligned}
\end{equation}
In $d=2$ only four tensor structures are linearly independent due to the following relation
\begin{equation}
d=2:\quad
2\,\mathbb{T}_2^{abcd}(x_1,x_2)
-\mathbb{T}_3^{abcd}(x_1,x_2)
-2\,\mathbb{T}_4^{abcd}(x_1,x_2)+
\mathbb{T}_5^{abcd}(x_1,x_2)=0.
\end{equation}
Notice that all the structures in \eqref{eq:tensor_structures} satisfy automatically
the condition \eqref{eq:symmetry_condition_T}. Using these structures we can translate the asymptotic conditions \eqref{eq:limit_1_E} and \eqref{eq:limit_2_E} into the following conditions on the dimensionless functions $h_m$
\begin{equation}
\label{eq:asymptotic_conditions_T}
\begin{aligned}
\lim_{r\rightarrow 0} h_1(r) &= 4\,C_T^{UV},\\
\lim_{r\rightarrow 0} h_2(r) &= 0,\\
\lim_{r\rightarrow 0} h_3(r) &= -C_T^{UV},\\
\lim_{r\rightarrow 0} h_4(r) &= -C_T^{UV}/d,\\
\lim_{r\rightarrow 0} h_5(r) &= C_T^{UV}/2,
\end{aligned}
\qquad\qquad\quad
\begin{aligned}
\lim_{r\rightarrow \infty} h_1(r) &= 4\,C_T^{IR},\\
\lim_{r\rightarrow \infty} h_2(r) &= 0,\\
\lim_{r\rightarrow \infty} h_3(r) &= -C_T^{IR},\\
\lim_{r\rightarrow \infty} h_4(r) &= -C_T^{IR}/d,\\
\lim_{r\rightarrow \infty} h_5(r) &= C_T^{IR}/2.
\end{aligned}
\end{equation}

Similar to section \ref{sec:integral_expressions_J} we can derive the integral expression for the stress-tensor central charge. It reads\footnote{For further details see \cite{Cardy:1988cwa} and section 2.6 in \cite{Karateev:2019ymz}.}
\begin{multline}
\label{eq:integral_expression_T_form1}
C_T^{UV} - C_T^{IR} = (d+1)\times \lim_{r_{min}\rightarrow 0}
\lim_{r_{max}\rightarrow \infty}\int_{r_{min}}^{r_{max}}
\frac{dr}{r}\\
\left(\frac{r^{2d}}{d-1}\,\<0|\Theta(x)\Theta(0)|0\>_E
+\frac{d-2}{2}\,h_2(r)\right).
\end{multline}
Using \eqref{eq:2pt_T} one can express $h_2(r)$ as some contraction of the stress-tensor two-point function. As a result \eqref{eq:integral_expression_T_form1} can be brought to the following equivalent form
\begin{multline}
\label{eq:integral_expression_T_form2}
C_T^{UV} - C_T^{IR} = \frac{1}{2\,(d-1)} \lim_{r_{min}\rightarrow 0}
\lim_{r_{max}\rightarrow \infty}\int_{r_{min}}^{r_{max}}
dr\, r^{2d-1}
R^{abcd}(x)
\<0|T^{ab}(x)T^{cd}(0)|0\>_E,
\end{multline}
where we have defined
\begin{align}
\label{eq:definition_R}
R^{abcd}(x)&\equiv
(4-d^2)\,\frac{x^a x^b x^c x^d}{r^4}+
\frac{d^2+d-2}{2}\,\frac{x^a x^b \delta^{cd}+x^c x^d \delta^{ab}}{r^2}\\
\nn
&-\frac{
	x^a x^c \delta^{bd}+
	x^b x^c \delta^{ad}+
	x^a x^d \delta^{bc}+
	x^b x^d \delta^{ac}}{r^2}+
(d+2)\,\delta^{ab}\delta^{cd}+
\big(\delta^{ac}\delta^{bd}+\delta^{bc}\delta^{ad}\big).
\end{align}

In appendix \ref{app:reflection_positivity} using reflection positivity we show that
\begin{equation}
\forall x:\quad\<0|\Theta(x)\Theta(0)|0\>_E \geq 0,
\end{equation}
in particular see \eqref{eq:reflection_positivity_scalar}. Because of this in $d=2$ the integrand in the right-hand side of \eqref{eq:integral_expression_T_form1} is a non-negative function which is integrated over a positive region. As a result we get a simple inequality
\begin{equation}
\label{eq:c-theorem}
d=2:\quad
C_T^{UV} - C_T^{IR}\geq 0
\end{equation}
known as the Zamolodchikov's c-theorem \cite{Zamolodchikov:1986gt,Cardy:1988tj}. Using the machinery of appendix \ref{app:reflection_positivity} no positivity statement however can be made about $h_2(r)$, thus no statement similar to \eqref{eq:c-theorem} can be made in $d\geq 3$ using these arguments. We will prove the c-theorem one more time but in a different way in section \ref{sec:spectral_densities_and_central_charges}.

\subsection*{Parity odd part} 
As before we need to consider $d=2$, $d=3$ and $d\geq 4$ dimensions separately.

We start with $d=2$. One can naively write six parity odd tensor structures, however only four of them will be linearly independent. Moreover, due to the symmetry condition \eqref{eq:symmetry_condition_T} there exist two additional constraints. Taking them into account we are left only with two structures which reads as
\begin{align}
\nn
\mathbf{T}^{abcd}_1(x_1,x_2) &\equiv
\delta^{ac}\epsilon^{bd} + \delta^{ad}\epsilon^{bc} + \delta^{bc}\epsilon^{ad} + \delta^{bd}\epsilon^{ac} \\
\label{eq:odd_T_d=2}
&+2\,\frac{\delta^{ac}x^b\epsilon^{de} x^e+
	\delta^{ad}x^b\epsilon^{ce} x^e+
	\delta^{bc}x^a\epsilon^{de} x^e+
	\delta^{bd}x^b\epsilon^{ae} x^e
}{r^2},\\
\mathbf{T}^{abcd}_2(x_1,x_2) &\equiv\frac{
	x^ax^c\epsilon^{bd}+x^ax^d\epsilon^{bc}+x^bx^c\epsilon^{ad}+x^bx^d\epsilon^{ac}}{r^2}+4\,\frac{x^ax^bx^c\epsilon^{de}x^e+x^ax^bx^d\epsilon^{ce}x^e}{r^4}.
\nn
\end{align}
Notice that the symmetry property required by \eqref{eq:symmetry_condition_T} is not manifest here. One needs to use relations between different tensor structures in order to show that \eqref{eq:odd_T_d=2} obeys \eqref{eq:symmetry_condition_T}. Conservation of the stress-tensor implies the following differential equations
\begin{equation}
g'_1(r)=0,\qquad
g_1'(r)+g_2'(r)= \frac{2}{r}\times(
2g_1(r)+g_2(r)).
\end{equation}
Solving them and taking into account the asymptotic constraints \eqref{eq:limit_1_E} and \eqref{eq:limit_2_E} we get
\begin{equation}
g_1(r)=\frac{1}{4}\,C_T^{\prime\,IR}=\frac{1}{4}\,C_T^{\prime\,UV},\qquad
g_2(r)=-\frac{1}{2}\,C_T^{\prime\,IR}=-\frac{1}{2}\,C_T^{\prime\,UV}.
\end{equation}
The central charge $C_T^{\prime\,UV}$ remains well defined and invariant along the flow all the way to the IR fixed point. One can identify $C_T^{\prime\,UV}$ with the gravitational anomaly in $d=2$.\footnote{See for example \cite{AlvarezGaume:1983ig} for the discussion on gravitational anomalies.}

In $d=3$ one can construct two parity odd tensor structures which automatically satisfy the condition \eqref{eq:symmetry_condition_T}.  They read as
\begin{equation}
\label{eq:odd_T_d=3}
\begin{aligned}
\mathbf{T}^{abcd}_1(x_1,x_2) &\equiv
\epsilon^{ade}\delta^{bc}\frac{x^e}{r}+
\epsilon^{ace}\delta^{bd}\frac{x^e}{r}+
\epsilon^{bde}\delta^{ac}\frac{x^e}{r}+
\epsilon^{bce}\delta^{ad}\frac{x^e}{r},\\
\mathbf{T}^{abcd}_2(x_1,x_2) &\equiv
\epsilon^{ade}\frac{x^bx^cx^e}{r^3}+
\epsilon^{ace}\frac{x^bx^dx^e}{r^3}+
\epsilon^{bde}\frac{x^ax^cx^e}{r^3}+
\epsilon^{bce}\frac{x^ax^dx^e}{r^3}.
\end{aligned}
\end{equation}
Conservation implies
\begin{equation}
g_1'(r)+g_2'(r) =\frac{1}{r}\times(7g_1(r)+4g_2(r)),
\end{equation}
where due to the asymptotic conditions \eqref{eq:limit_1_E} and \eqref{eq:limit_2_E} one has
\begin{equation}
\lim_{r\rightarrow 0}g_1(r)=
\lim_{r\rightarrow 0}g_2(r)=
\lim_{r\rightarrow \infty}g_1(r)=
\lim_{r\rightarrow \infty}g_2(r)=0.
\end{equation}
We emphasize that even though no parity odd terms in the two-point function of the stress-tensors are allowed at the fixed points, they can be present along the flow.
The parity odd terms in $d=3$ do not contain any information about the UV or IR CFTs and thus will not be studied further in this paper.

In $d\geq 4$ no parity tensor structures can be constructed.

\subsubsection*{Trace of the stress-tensor}
It is useful to make several statements about the trace of the stress-tensor. From \eqref{eq:2pt_T} and the explicit expressions of tensor structures \eqref{eq:tensor_structures}, \eqref{eq:odd_T_d=2} and \eqref{eq:odd_T_d=3} it follows that in any number of dimension one has
\begin{equation}
\label{eq:2pt_trace}
\<0|\Theta(x_1)\Theta(x_2)|0\>_E =
\frac{1}{r^{2d}} \times\left(h_1(r)+2d\,h_2(r)+4\,h_3(r)+d^2\,h_4(r)+2d\,h_5(r)\right).
\end{equation}
Using the asymptotic conditions \eqref{eq:asymptotic_conditions_T} we get then
\begin{align}
\label{eq:requirement_1}
\lim_{r\rightarrow 0} r^{2d}&\times\<0|\Theta(x)\Theta(0)|0\>_E = 0,\\
\label{eq:requirement_2}
\lim_{r\rightarrow \infty} r^{2d}&\times\<0|\Theta(x)\Theta(0)|0\>_E = 0.
\end{align}

We define a particular quantum field theory as a deformation of some UV CFT. In practice it means that we pick a scalar operator $\cO$ with the conformal dimension $\Delta_\cO$ which has the following UV CFT two-point function
\begin{equation}
\label{eq:2pf_relevant}
\< 0|\cO(x) \cO(0)|0\>_{E,\;UV\;CFT} = \frac{1}{r^{2\Delta_\cO}}
\end{equation} 
and introduce a dimensionful parameter with the mass dimension $[g]=d-\Delta_\cO$ which breaks explicitly the scaling invariance and triggers the renormalization group flow. In this process the trace of the stress-tensor is fixed by the deforming operator $\cO$, namely
\begin{equation}
\label{eq:deformation}
\Theta(x) = g \cO(x).
\end{equation}
Notice that $[\Theta]=d$. By plugging \eqref{eq:deformation} and \eqref{eq:2pf_relevant} into \eqref{eq:requirement_1} we obtain the following consistency condition
\begin{equation}
\Delta_\cO< d.
\end{equation}
This is nothing but the requirement that the deforming operator $\cO$ must be relevant.

Analogously, at low energies we can write the trace of the stress-tensor in terms of the local scalar operator deforming the IR conformal field theory. Let us denote this operator by $\cO'$.\footnote{ The operators $\cO$ and $\cO'$ belong to different bases. One basis is more natural for working at hight energies and the other one is more natural for working at low energies.} Applying the above logic we conclude that \eqref{eq:requirement_2} is satisfied only if
\begin{equation}
\Delta_\cO' > d.
\end{equation}
In other words $\cO'$ must be irrelevant.

\section{Lorentzian two-point functions}
\label{sec:spectral_densities}
In this section we discuss two-point functions in the Lorentzian signature.
Contrary to the Euclidean signature where only the time-ordered two-point functions exist, in the Lorentzian signature we can define Wightman, time-ordered, advanced and retarded two-point functions. In what follows we will discuss the first two. The Wightman two-point functions suit best for defining spectral densities. They are automatically well defined at coincident points and do not have any contact terms. Time-ordered Lorentzian two-point functions will be employed in section \ref{sec:spectral_densities_and_central_charges} due to the following property: they simply become the Euclidean two-point functions under the Wick rotation. It will be sufficient to work with time-ordered correlators at non coincident points. This allows to avoid complications due to presence of contact terms.

The discussion presented in this section is completely generic for $d\geq 4$. In $d=2$ and $d=3$ two-point functions are allowed to have a parity odd contribution. Bellow we will completely ignore this possibility.

\subsection{Wightman two-point functions}
\label{sec:wightman}
The scalar Wigthman two-point function in position space is defined as the ordered vacuum expectation value of two real scalar operators  $\cO(x)$ as
\begin{equation}
\label{eq:wightman_function_scalar}
\<0| \cO_1(x) \cO_2(y)|0\>_W \equiv
\lim_{\epsilon\rightarrow 0^+} \<0| \cO_1(\hat x) \cO_2(y)|0\>,
\end{equation}
where we have defined
\begin{equation}
\label{eq:xhat}
\hat x^\mu \equiv \{x^0-i\epsilon,\vec x\},\quad
\epsilon>0.
\end{equation}
See appendix \ref{app:wightman} for further details.
The small imaginary part in the time component is needed to regularize various integrals of the Wightman correlation function. The $\epsilon$ prescription in \eqref{eq:wightman_function_scalar} should be understood as follows: perform all the necessary manipulation with the finite but small $\epsilon$ and then take the limit. The notation $0^+$ indicates that we approach zero from the positive values.

We define the spectral density $\rho_\cO$ of the local operator $\cO(x)$ as the Fourier transform of its Wightman two-point function as\footnote{By definition \eqref{eq:wightman_function_scalar} the first entry in \eqref{eq:spectral_representation_scalar} is equivalent to
\begin{equation}
\nn
\lim_{\epsilon\rightarrow 0^+}\int d^dx\, e^{i p^0 x^0 -i\vec p \cdot \vec x}\<0| \cO(x^0-i\epsilon,\,\vec x) \cO(0)|0\>=
\lim_{\epsilon\rightarrow 0^+}\int d^dx\, e^{i p^0 (x^0+i\epsilon) -i\vec p \cdot \vec x}\<0| \cO(x^0,\,\vec x) \cO(0)|0\>.
\end{equation}	
Here we have simply performed the change of variables. Notice that $\epsilon$ enters the right-hand side of the above equations as $e^{-\epsilon}$. It plays the role of a dumping factor.
}
\begin{equation}
\label{eq:spectral_representation_scalar}
\begin{aligned}
(2\pi)\theta(p^0)\rho_\cO(-p^2) &\equiv \int d^dx\, e^{-i p \cdot x}\<0| \cO(x) \cO(0)|0\>_W,\\
\<0| \cO(x_1) \cO(x_2)|0\>_W  &= \lim_{\epsilon\rightarrow 0^+}\int \frac{d^dp}{(2\pi)^d}\;e^{i p \cdot \hat x_{12}}
(2\pi)\theta(p^0)\rho_\cO(-p^2).
\end{aligned}
\end{equation}
The appearance of the Heaviside step function $\theta(p^0)$ enforces the fact that we work with non-negative energies $p^0\geq 0$ only. For convenience we also define the $s$ Mandelstam variable
\begin{equation}
\label{eq:s_mand}
s\equiv -p^2 \geq 0.
\end{equation}
The reason why $s\geq 0$ will be explained shortly.
It is standard to rewrite the second entry in \eqref{eq:spectral_representation_scalar} by adding a $\delta$-function and integrating over it as
\begin{equation}
\label{eq:spectral_representation_scalar_2}
\begin{aligned}
\<0| \cO(x) \cO(0)|0\>_W &= \int_0^\infty ds \rho_\cO(s) \Delta_W(x;s),\\
\Delta_W(x;s) &\equiv \lim_{\epsilon\rightarrow 0^+}\int \frac{d^dk}{(2\pi)^d}\;
e^{ik\cdot \hat x}\; (2\pi)\theta(k^0)\delta(s+k^2).
\end{aligned}
\end{equation}
We refer to the object $\Delta_W(x;s)$ as the scalar Wightman propagator. Its explicit form can be found in \eqref{eq:scalar_wightman_propagator}.

In unitary poincare invariant QFTs the states transform in the unitary infinite-dimensional representation constructed by Wigner. They are labeled by $-p^2$ and  by the irreducible representation of the Little group to be defined shortly. There are three distinct possibilities, namely
\begin{equation*}
-p^2<0,\quad
-p^2=0,\quad
-p^2>0.
\end{equation*}
In QFTs one deals only with the last two options. The reason for that is the necessity to have a unique vacuum state which is defined to be the lowest energy state in the theory. States with $-p^2<0$ would obviously allow for arbitrary small negative energies. In the case $-p^2>0$ using Lorentz transformations one can obtain any $d$-momentum $p^\mu$ from a standard frame which is conventionally chosen to be
\begin{equation}
\label{eq:standard_frame}
\bar p^\mu \equiv \{M, \vec 0\},
\end{equation}
where $M>0$ is some real constant. The group of transformations leaving invariant \eqref{eq:standard_frame} is called the Little group. Clearly in this case it is $SO(d-1)$. The most universal irreducible representation of the Little group which exists in any dimension is the traceless symmetric representation
\begin{equation}
\label{eq:Little_group}
\ytableausetup{boxsize=0.7em}
\begin{ytableau}
~ & ~ & ~
\end{ytableau}
\ldots
\end{equation}
We refer to \eqref{eq:Little_group} with $\ell$ boxes simply as the spin $\ell$ (Little group) representation. For further details in the $d=4$ case see appendix A in \cite{Hebbar:2020ukp}.  
In any particular QFT model we can choose a basis of states which we denote schematically by $|b\>$. As discussed above these states transform in the unitary representation of the Poincar\'e group. One chooses the basis to diagonalize the generators of translations $P^\mu$, namely $P^\mu|b\>=p^\mu_b|b\>$. The following completeness relation holds
\begin{equation}
\label{eq:completeness}
\mathbb{I} = \sum_{b} |b\>\<b|,
\end{equation} 
where the summation over $b$ is a schematic notation which stands for summing Poincare and all the additional labels characterizing the state.

In the case $-p^2=0$ the standard frame is usually chosen to be $\bar p^{\,\mu}\equiv\{M,0,\ldots,0,M\}$ leading to a different Little group  which is $ISO(d-2)$.  It is usually assumed that ``translation'' generators of this group are realized trivially and the Little group in this case effectively becomes $SO(d-2)$.  This changes the set of labels $b$ needed to describe the state compared to the $-p^2>0$ case. In \eqref{eq:completeness} and below we keep $b$ at a schematic level, thus our discussion applies for both $-p^2=0$ and $-p^2>0$ cases.  In future sections however when we need the explicit structure of the Little group we will restrict our attention to the $-p^2>0$ case only.

Let us inject \eqref{eq:completeness} into \eqref{eq:wightman_function_scalar} we get
\begin{align}
\nn
\<0| \cO(x) \cO(0)|0\>_W
\nn&=\lim_{\epsilon\rightarrow 0^+}\sum_b \<0| \cO(\hat x) |b\>\<b| \cO(0)|0\>\\
\label{eq:expression_scalar_wightman}
&=\lim_{\epsilon\rightarrow 0^+}\sum_b e^{i p_b \cdot \hat x} \; \<0| \cO(0) |b\>\<b| \cO(0)|0\>\\
&=\lim_{\epsilon\rightarrow 0^+}\int \frac{d^dp}{(2\pi)^d}\;e^{i p \cdot \hat x}
\sum_b  (2\pi)^d\delta^{(d)}(p-p_b)
\left|\<b| \cO(0) |0\>\right|^2.
\nn
\end{align}
In the second equality we have used the translation invariance
\begin{equation}
\label{eq:trans_invariance}
\cO(x) = e^{-iP\cdot x} \cO(0) e^{+iP\cdot x}.
\end{equation}
Comparing \eqref{eq:spectral_representation_scalar} with \eqref{eq:expression_scalar_wightman} we get the desired expansion of the spectral density
\begin{align}
\label{eq:spectral_density_scalar}
(2\pi)\theta(p^0)\rho_\cO(-p^2) = \displaystyle\sum_b\,  (2\pi)^d\delta^{(d)}(p-p_b)
\left|\<b| \cO(0) |0\>\right|^2.
\end{align} 
Since for each basis state we have $p_b^0\geq 0$ and $-p_b^2\geq 0$ we conclude from the expression \eqref{eq:spectral_density_scalar} that $p^0\geq 0$ and $-p^2\geq 0$ in accordance with \eqref{eq:s_mand}.

The basis states $|b\>$ at this point are rather abstract. There is a large class of quantum field theories however for which the basis $|b\>$ can be defined in a straightforward constructive way as a tensor product of $n$ free particle states dressed with the M{\o}ller operators, see section 2.1 in \cite{Karateev:2019ymz} for further details. Such basis states are called asymptotic and are denoted here by $|\n\>_{in}$ or $|\n\>_{out}$.\footnote{Asymptotic states can usually be defined in the QFTs with a mass gap. In very special situations one can also define asymptotic states for massless particles such as pions or photons in $d=4$.}
In other words
\begin{equation}
|b\> = |\n\>_{in}
\quad\text{or}\quad
|b\> = |\n\>_{out}.
\end{equation}
Analogously to \eqref{eq:spectral_density_scalar} for massive theories we get
\begin{align}
\label{eq:spectral_density_scalar_as}
(2\pi)\theta(p^0)\rho_\cO(-p^2) = \displaystyle \SumInt_n\,  (2\pi)^d\delta^{(d)}(p-p_\n)
\left|{}_{out}\<\n| \cO(0) |0\>\right|^2,
\end{align} 
where $p_\n^\mu$ is the $d$-momenta of the $|\n\>_{in}$ asymptotic state and $\SumInt_n$ stands for summation over all possible number of particles and integrating over their relative motion. The matrix element ${}_{out}\<\n| \cO(0) |0\>$ is called the form factor. See section 2.4 in \cite{Karateev:2019ymz} for a discussion of from factors and their properties.

We will now define the spectral density of conserved (Abelian) currents and the stress-tensor.

\subsubsection*{Conserved currents}
\label{sec:spectral_J}
The two-point Wightman function of two spin one Lorentz currents $J_\mu(x)$  is defined as
\begin{equation}
\label{eq:wightman_function_vecotr}
\<0| J^\mu(x) J^\nu(y)|0\>_W\equiv
\lim_{\epsilon\rightarrow 0^+}\<0| J^\mu(\hat x) J^\nu(y)|0\>.
\end{equation}
As in the scalar case the spectral density $\rho_J^{\mu \nu}$ (which has two Lorentz indices now) is defined as the Fourier transform of the Wightman two-point function \eqref{eq:wightman_function_vecotr} as

\begin{equation}
\label{eq:spectral_representation}
\begin{aligned}
(2\pi)\theta(p^0)\rho_J^{\mu\nu}(p) &\equiv \int d^dx\, e^{-i p \cdot x}\<0| J^\mu(x) J^\nu(0)|0\>_W,\\
\<0| J^\mu(x_1) J^\nu(x_2)|0\>_W   &= \lim_{\epsilon\rightarrow 0^+}\int \frac{d^dp}{(2\pi)^d}\;e^{i p \cdot \hat x_{12}}
(2\pi)\theta(p^0)\rho_J^{\mu\nu}(p).
\end{aligned}
\end{equation}
In QFTs where the asymptotic states can be defined, the decomposition of the spectral density into the form factors ${}_{out}\<\n| J^\mu(0) |0\>$ reads as
\begin{align}
\label{eq:spectral_density_vector}
(2\pi)\theta(p^0)\rho_J^{\mu\nu}(p) = \displaystyle\SumInt_n\,  (2\pi)^d\delta^{(d)}(p-p_\n)
{}_{out}\<\n| J^\mu(0) |0\>^*\;{}_{out}\<\n| J^\nu(0) |0\>.
\end{align} 

Because of the Lorentz invariance the spectral density $\rho^{\mu\nu}$ can be written in the following most general form
\begin{equation}
\label{eq:decomposition_spectral_density_J}
\rho_J^{\mu\nu}(p) = 
p^2\,\Big(\rho_J^0(-p^2)\,\Pi^{\mu\nu}_{0}(p) - \rho_J^1(-p^2)\,\Pi^{\mu\nu}_{1}(p)\Big),
\end{equation}
where $\rho_J^0$ and $\rho_J^1$ are the (spin 0 and spin 1)\footnote{These names will be explained shortly.} components of the spectral density $\rho^{\mu\nu}$ and the tensor structures are defined as
\begin{equation}
\label{eq:projectors_vector_main_text}
\Pi^{\mu\nu}_{0}(p)\equiv \frac{p^\mu p^\nu}{p^2},\qquad
\Pi^{\mu\nu}_{1}(p)\equiv \eta^{\mu\nu}-\frac{p^\mu p^\nu}{p^2}.
\end{equation}
The overall factor $p^2$ and the minus sign in the second term in \eqref{eq:decomposition_spectral_density_J} were introduced for the later convenience.

The objects \eqref{eq:projectors_vector_main_text} have a more profound meaning than being simply the tensor structures. Let us zoom on this. We are in the situation when $-p^2>0$. Consider the Lorentz spin one operator $J^\mu(p)$ in momentum space. It transforms in the irreducible representation of the Lorentz group, however the states it creates from the vacuum transform in irreducible representations of the Little group $SO(d-1)$. It is thus important to know how to decompose (or project in other words) irreducible representations of the Lorentz group $SO(1,d-1)$ into irreducible representations of the Little group $SO(d-1)$. For Lorentz spin one representation one has
\begin{equation}
\square_{SO(1,d-1)} = \bullet_{SO(d-1)}\oplus \square_{SO(d-1)}.
\end{equation}
It is easy to perform such a decomposition explicitly in the frame \eqref{eq:standard_frame}. One has
\begin{equation}
\label{eq:decomposition_vector}
J^\mu(\bar p) = J^\mu_{0}(\bar p)+J^\mu_{1}(\bar p),
\end{equation}
where we have defined
\begin{equation}
J^\mu_{0}(\bar p)\equiv \{J^0(\bar p),\;\vec 0\},
\quad J^\mu_{1}(\bar p)\equiv\{0,\;\vec J(\bar p)\}.
\end{equation}
In a generic frame this decomposition is achieved by
\begin{equation}
\label{eq:projection_generic_frame}
J_0^{\mu}(p) = \Pi^{\mu\nu}_{0}(p) J_\nu(p),\qquad
J_1^{\mu}(p) = \Pi^{\mu\nu}_{1}(p) J_\nu(p).
\end{equation}
The equivalence of \eqref{eq:decomposition_vector} and \eqref{eq:projection_generic_frame} is trivial to see in the frame \eqref{eq:standard_frame}. Thus, the objects in \eqref{eq:projectors_vector_main_text} are the Little group spin 0 and 1 projectors. From their definitions it is straightforward to check that they satisfy the standard properties of projectors, namely
\begin{equation}
\label{eq:property_projectors}
\Pi^{\mu\nu}_{0}(p) + \Pi^{\mu\nu}_{1}(p) = \eta^{\mu\nu},\qquad
\eta_{\nu\rho}\Pi^{\mu\nu}_{i}(p)\Pi^{\rho\sigma}_{j}(p)= \delta_{ij}\Pi^{\mu\sigma}_{i}(p).
\end{equation}

From \eqref{eq:spectral_density_vector} it is clear that $\rho_J^{\mu\nu}(p)$ is a $d\times d$ hermitian semi-positive definite matrix for any value of $p^\mu$ satisfying $-p^2>0$. We then can evaluate this matrix in the standard frame \eqref{eq:standard_frame}. The semi-positivity then translates into non-negativity of the spectral density components
\begin{equation}
\label{eq:positivity_J}
\rho_J^{0}(-\bar p^2)\geq 0,\qquad
\rho_J^{1}(-\bar p^2)\geq 0.
\end{equation}
Since the components of the spectral densities are scalar quantities, they remain invariant under any Lorentz transformation, thus the inequalities \eqref{eq:positivity_J} hold true in any frame.
It is also useful to deduce the mass dimensions of the components of the spectral density. Since the Heaviside step function is dimensionless from \eqref{eq:spectral_representation} we get
\begin{equation}
\label{eq:dimensions_J}
[J^\mu(x)]=d-1
\quad\Rightarrow\quad
[\rho_J^{\mu\nu}]=d-2
\quad\Rightarrow\quad
[\rho_J^0]=
[\rho_J^1]= d-4.
\end{equation}

Analogously to the scalar case using the definition of the components of the current spectral density we can bring the second entry in \eqref{eq:spectral_representation} to a very convenient form
\begin{equation}
\label{eq:spectral_representation_2}
\begin{aligned}
\<0| J^\mu(x) J^\nu(0)|0\>_W &= \int_0^\infty ds\,
\Big(
-\rho_J^0(s) \Delta^{\mu\nu}_{W,\,0}(x;s)+
\rho_J^1(s) \Delta^{\mu\nu}_{W,\,1}(x;s)
\Big),\\
\Delta^{\mu\nu}_{W,\,i}(x;s) &\equiv \lim_{\epsilon\rightarrow 0^+}s\int \frac{d^dp}{(2\pi)^d}\;
e^{ip\cdot \hat x}\; (2\pi)\theta(p^0)\delta(p^2+s)\Pi^{\mu\nu}_{i}(p),
\end{aligned}
\end{equation}
where $\Delta^{\mu\nu}_i$ are the Lorentz spin one Wightman propagators. By using the explicit expressions for the projectors \eqref{eq:projectors_vector_main_text} and the integration by parts procedure, these two propagators can be written in terms of the Wightman scalar propagator \eqref{eq:spectral_representation_scalar_2} as
\begin{equation}
\label{eq:wightman_propagator_J}
\Delta^{\mu\nu}_{W,\,0}(x;s) = \partial^\mu \partial^\nu\; \Delta_{W}(x;s),\quad
\Delta^{\mu\nu}_{W,\,1}(x;s) = \left(s\,\eta^{\mu\nu}-
\partial^\mu \partial^\nu
\right) \Delta_{W}(x;s).
\end{equation}

The scalar Wightman propagator $\Delta_{W}$ in the limit $s\rightarrow 0$ (for space-like separated points $x^2>0$) is given in \eqref{eq:finiteness_wightman_propagator} . 
It remains finite and depends only on $x^2$. Consequently the Wightman propagators in \eqref{eq:wightman_propagator_J} remain finite in the limit $s\rightarrow 0$. This is the reason for introducing the overall factor $p^2$ in \eqref{eq:decomposition_spectral_density_J}.

Finally, conservation of the current implies
\begin{equation}
\partial_\mu J^\mu(x)=0
\quad\Rightarrow\quad
p_\mu \rho_J^{\mu\nu}(p) =0,\quad
p_\nu \rho_J^{\mu\nu}(p) =0.
\end{equation}
Using \eqref{eq:decomposition_spectral_density_J} and \eqref{eq:projectors_vector_main_text} these conditions in turn imply
\begin{equation}
\label{eq:implication_conservation}
p^2\rho_J^{0}(-p^2)=0
\quad\Rightarrow\quad
\rho_J^{0}(-p^2)=A\times\delta(-p^2),\qquad
[A]=d-2,
\end{equation}
where $A$ is some dimensionful constant whose mass dimension follows from \eqref{eq:dimensions_J}.

\subsubsection*{Stress-tensor}
\label{sec:spectral_T}
Let us consider an operator transforming in the two-index symmetric reducible representation $T^{\mu\nu}(x)$ and consider its Wightman two-point function
\begin{equation}
\label{eq:wightman_function_tensor}
\<0| T^{\mu\nu}(x) T^{\rho\sigma}(y)|0\>_W\equiv
\lim_{\epsilon\rightarrow 0^+}\<0| T^{\mu\nu}(\hat x) T^{\rho\sigma}(y)|0\>.
\end{equation}
The spectral density $\rho_T^{\mu\nu;\,\rho\sigma}$ of the operators $T^{\mu\nu}(x)$ is the Fourier transform of \eqref{eq:wightman_function_tensor}, namely
\begin{equation}
\label{eq:spectral_decomposition_T}
\begin{aligned}
(2\pi)\theta(p^0)\rho_{T}^{\mu\nu;\rho\sigma}(p) &\equiv \int d^dx\, e^{-i p \cdot x}\<0| T^{\mu\nu}(x) T^{\rho\sigma}(0)|0\>_W,\\
\langle 0| T^{\mu\nu}(x_1) T^{\rho\sigma}(x_2)|0\rangle_W &=
\lim_{\epsilon\rightarrow 0^+}
\int \frac{d^dp}{(2\pi)^d}\;e^{i p \cdot \hat x_{12}}
(2\pi)\theta(p^0)\rho_{T}^{\mu\nu;\,\rho\sigma}(p).
\end{aligned}
\end{equation}
In QFTs which can be described in terms of asymptotic states the spectral density can be written as a sum of the form factors ${}_{out}\<\n| T_{\mu\nu}(0) |0\>$ as
\begin{align}
\label{eq:spectral_density_T}
(2\pi)\theta(p^0)\rho_{T}^{\mu\nu;\,\rho\sigma}(p) \equiv \displaystyle\SumInt_n\,  (2\pi)^d\delta^{(d)}(p-p_\n)
{}_{out}\<\n| T^{\mu\nu}(0) |0\>^*\;{}_{out}\<\n| T^{\rho\sigma}(0) |0\>.
\end{align}

The operator $T^{\mu\nu}$ is in the reducible representation, one can decompose it into two irreducible representations as
\begin{equation}
\bullet_{SO(1,d-1)} +\;\,
\ytableausetup{boxsize=0.7em}
\begin{ytableau}
~ & ~ 
\end{ytableau}_{SO(1,d-1)}.
\end{equation}
The operators transforming in theses two irreducible representations are
\begin{equation}
\label{eq:That}
\Theta(x)\equiv \eta_{\mu\nu} T^{\mu\nu}(x),\qquad
\hat T^{\mu\nu} \equiv T^{\mu\nu}(x)-\frac{1}{d}\,\eta^{\mu\nu} \Theta(x).
\end{equation}
They are the trace and the traceless-symmetric part of $T^{\mu\nu}$.
Now instead of \eqref{eq:wightman_function_tensor} one should consider the following three (generically independent) Wightman two-point functions
\begin{equation}
\label{eq:wightman_function_tensor_splin}
\<0| \Theta(x) \Theta(0)|0\>_W,\quad
\<0| \Theta(x) \hat T^{\rho\sigma}(0)|0\>_W,\quad
\<0| \hat T^{\mu\nu}(x) \hat T^{\rho\sigma}(0)|0\>_W.
\end{equation}
In the case when $T^{\mu\nu}$ is the stress-tensor, the conservation condition 
\begin{equation}
\label{eq:conservation_T}
\partial_\mu T^{\mu \nu}(x) = 0
\end{equation}
however mixes all three correlators in \eqref{eq:wightman_function_tensor_splin}. 
Using the splitting \eqref{eq:wightman_function_tensor_splin} one can define the following spectral densities
\begin{equation}
\label{eq:spectral_decomposition_traceT}
\begin{aligned}
(2\pi)\theta(p^0)\rho_{\Theta}(-p^2) &\equiv \int d^dx\, e^{-i p \cdot x}\<0| \Theta(x) \Theta(0)|0\>_W,\\
(2\pi)\theta(p^0)\rho^{\mu\nu}_{\Theta\hat T}(p) &\equiv \int d^dx\, e^{-i p \cdot x}\<0| \Theta(x) \hat T^{\mu\nu}(0)|0\>_W,\\
(2\pi)\theta(p^0)\rho^{\mu\nu;\,\rho\sigma}_{\hat T}(p) &\equiv \int d^dx\, e^{-i p \cdot x}\<0| \hat T^{\mu\nu}(x) \hat T^{\rho\sigma}(0)|0\>_W.
\end{aligned}
\end{equation}
Analogously to $\rho^{\mu\nu}_{\Theta\hat T}(p)$ one can define the spectral density $\rho^{\mu\nu}_{\hat T\Theta}(p)$. One can show however that the latter is identical to the former. Using these one can write
\begin{equation}
\label{eq:spectral_T}
\rho_{T}^{\mu\nu;\,\rho\sigma}(p) =
\rho_{\hat T}^{\mu\nu;\,\rho\sigma}(p) +
\frac{1}{d}\,\left(
\eta^{\mu\nu}\rho_{\Theta \hat T}^{\rho\sigma}(p) +
\eta^{\rho\sigma}\rho_{\Theta\hat T}^{\mu\nu}(p)
\right)
+\frac{1}{d^2}\,\eta^{\mu\nu}\eta^{\rho\sigma}\rho_{\Theta}(-p^2).
\end{equation}

The decomposition of the Lorentz spin 2 operator into the irreducible representations of the Little group $SO(d-1)$ reads as
\begin{equation}
\label{eq:Little_group_spin2}
\ytableausetup{boxsize=0.7em}
\begin{ytableau}
~ & ~ 
\end{ytableau}_{SO(1,d-1)} =
\bullet_{SO(d-1)} +
\ytableausetup{boxsize=0.7em}
\begin{ytableau}
~
\end{ytableau}_{SO(d-1)} +
\ytableausetup{boxsize=0.7em}
\begin{ytableau}
~ & ~ 
\end{ytableau}_{SO(d-1)}.
\end{equation}
The decomposition \eqref{eq:Little_group_spin2} can be done by using three projectors constructed out of \eqref{eq:projectors_vector_main_text}. They read
\begin{equation}
\label{eq:projectors}
\begin{aligned}
\Pi_{0}^{\mu\nu;\,\rho\sigma}(p) &\equiv \frac{d}{d-1}\,
\widetilde\Pi_{0}^{\mu\nu}(p)\widetilde\Pi_{0}^{\rho\sigma}(p),\\
\Pi_1^{\mu\nu;\,\rho\sigma}(p) &\equiv \frac{1}{2}\,
\Big(
\Pi_{0}^{\mu\rho}(p)   \Pi_{1}^{\nu\sigma}(p)+
\Pi_{0}^{\mu\sigma}(p) \Pi_{1}^{\nu\rho}(p)+
\Pi_{1}^{\mu\rho}(p)   \Pi_{0}^{\nu\sigma}(p)+
\Pi_{1}^{\mu\sigma}(p) \Pi_{0}^{\nu\rho}(p)
\Big),\\
\Pi_2^{\mu\nu;\,\rho\sigma}(p) &\equiv 
-\frac{1}{d-1}\Pi_{1}^{\mu\nu}(p)   \Pi_{1}^{\rho\sigma}(p)
+\frac{1}{2}\,\Pi_{1}^{\mu\rho}(p)  \Pi_{1}^{\nu\sigma}(p)
+\frac{1}{2}\,\Pi_{1}^{\mu\sigma}(p)\Pi_{1}^{\nu\rho}(p),
\end{aligned}
\end{equation}
where we have defined
\begin{equation}
\label{eq:traceless_spin0}
\widetilde\Pi^{\mu\nu}_{0}(p)\equiv \frac{p^\mu p^\nu}{p^2}-\frac{1}{d}\,\eta^{\mu\nu}.
\end{equation} 
The projectors \eqref{eq:projectors} are required to be symmetric and traceless in both pairs of indices $(\mu\nu)$ and $(\rho\sigma)$. They also satisfy the following relations
\begin{equation}
\label{eq:T_projectors_properties}
\begin{aligned}
\sum_{i=0}^2\Pi_i^{\mu\nu;\,\rho\sigma}(p) &= \frac{1}{2}\left(\eta^{\mu\rho}\eta^{\nu\sigma}
+\eta^{\nu\rho}\eta^{\mu\sigma}\right)-\frac{1}{d}\,\eta_{\mu\nu}\eta_{\rho\sigma},\\
\Pi_i^{\mu\nu;\,\rho\sigma}(p)\Pi_j{}_{\rho\sigma}{}^{\alpha\beta}(p) &=
\delta_{ij}\, \Pi_i^{\mu\nu;\,\alpha\beta}(p).
\end{aligned}
\end{equation}
Using the projectors \eqref{eq:projectors} we can write the decomposition \eqref{eq:Little_group_spin2} explicitly as
\begin{equation}
\label{eq:T_splitting}
\hat T^{\mu\nu}(p) = \hat T_0^{\mu\nu}(p) + \hat T_1^{\mu\nu}(p) + \hat T_2^{\mu\nu}(p),
\end{equation}
where the Little group spin 0, 1 and 2 representations, analogously to \eqref{eq:projection_generic_frame}, read as
\begin{equation}
\label{eq:components_T_splitting}
\begin{aligned}
\hat T_0^{\mu\nu}(p) \equiv \Pi_{0}^{\mu\nu}{}_{\rho\sigma}(p)\hat T^{\rho\sigma}(p),\\
\hat T_1^{\mu\nu}(p) \equiv \Pi_{1}^{\mu\nu}{}_{\rho\sigma}(p)\hat T^{\rho\sigma}(p),\\
\hat T_2^{\mu\nu}(p) \equiv \Pi_{2}^{\mu\nu}{}_{\rho\sigma}(p)\hat T^{\rho\sigma}(p).
\end{aligned}
\end{equation}

Using the Lorentz invariance one can write the decomposition of the spectral densities into components as
\begin{equation}
\label{eq:spectral_ThetaT}
\rho^{\mu\nu}_{\Theta\hat T}(p) = \rho_{\Theta\hat T}(-p^2)\times p^2\widetilde\Pi^{\mu\nu}_{0}(p),
\end{equation}
together with
\begin{equation}
\label{eq:decomposition_density_irreps}
\rho_{\hat T}^{\mu\nu;\,\rho\sigma}(p) = 
p^4\times\Big(\rho_{\hat T}^{0}(-p^2) \Pi^{\mu\nu;\,\rho\sigma}_{0}(p)-\\
\rho_{\hat T}^{1}(-p^2) \Pi^{\mu\nu;\,\rho\sigma}_{1}(p) + 
\rho_{\hat T}^{2}(-p^2) \Pi^{\mu\nu;\,\rho\sigma}_{2}(p)\Big).
\end{equation}
It is also useful to deduce the mass dimensions of the components of the spectral density. It simply reads $[\rho_T^{\mu\nu}]=d$.

Conservation condition \eqref{eq:conservation_T} implies
\begin{equation}
p_\mu \rho_T^{\mu\nu;\,\rho\sigma}(p)=0,\qquad
p_\rho \rho_T^{\mu\nu;\,\rho\sigma}(p)=0.
\end{equation}
Using \eqref{eq:spectral_T}, \eqref{eq:decomposition_density_irreps} and \eqref{eq:spectral_ThetaT} we obtain the following constraints
\begin{equation}
\label{eq:conservation_components}
\rho_\Theta(s) = d(d-1)\,s^2\rho_{\hat T}^0(s),\qquad
\rho_{\Theta \hat T}(s) = d\,s\rho_{\hat T}^0(s),\qquad
\rho_{\hat T}^1(s) = 0.
\end{equation}
Apart from some singular points at $p^2=0$, the condition \eqref{eq:conservation_components} leaves us with two components of the stress-tensor spectral density, namely $\rho_{\Theta}$ and $\rho_{\hat T}^2$ and we can compactly write
\begin{equation}
\label{eq:spectral_T_conserved}
\rho_{T}^{\mu\nu;\,\rho\sigma}(p) = \frac{1}{(d-1)^2}\,\rho_\Theta(s) \Pi_1^{\mu\nu}\Pi_1^{\rho\sigma}+s^2\rho_{\hat T}^2(s)\Pi^{\mu\nu;\,\rho\sigma}_2(p).
\end{equation}
Since the spectral density \eqref{eq:spectral_density_T} is a hermitian matrix we conclude that
\begin{equation}
\rho_{\Theta}(-p^2)\geq 0,\qquad
\rho_{\hat T}^2(-p^2)\geq 0.
\end{equation}

Plugging \eqref{eq:spectral_T_conserved} into \eqref{eq:spectral_decomposition_T}, analogously to section \ref{sec:spectral_J}, we obtain the spectral representation for the conserved stress-tensor. It reads
\begin{equation}
\label{eq:spectral_representation_T_final1}
\begin{aligned}
\<0| T^{\mu\nu}(x) T^{\rho\sigma}(0)|0\>_W
= \frac{1}{(d-1)^2}\int_0^\infty ds \rho_{\Theta}(s) \Delta_{W,\Theta}^{\mu\nu;\rho\sigma}(x;s)+
\int_0^\infty ds \rho_{\hat T}^2(s) \Delta^{\mu\nu;\rho\sigma}_{W,\,2}(x;s),
\end{aligned}
\end{equation}
where we have defined
\begin{equation}
\label{eq:Wightman_propagators_T}
\begin{aligned}
\Delta^{\mu\nu}_{W,\Theta}(x;s) &\equiv \lim_{\epsilon\rightarrow 0^+} s^2 \int \frac{d^dp}{(2\pi)^d}\;
e^{ip\cdot \hat x}\; (2\pi)\theta(p^0)\delta(p^2+s)
\Pi_1^{\mu\nu}(p)
\Pi_1^{\rho\sigma}(p),\\
\Delta^{\mu\nu;\,\rho\sigma}_{W,\,2}(x;s) &\equiv \lim_{\epsilon\rightarrow 0^+}s^2 \int \frac{d^dp}{(2\pi)^d}\;
e^{ip\cdot \hat x}\; (2\pi)\theta(p^0)\delta(p^2+s)\Pi^{\mu\nu;\,\rho\sigma}_2(p).
\end{aligned}
\end{equation}
Up to an overall constant the expression \eqref{eq:spectral_representation_T_final1} matches precisely the equation (3.1) of \cite{Cappelli:1990yc}.

To conclude let us express the Wightman propagators of the stress-tensor in terms of the scalar Wightman propagator \eqref{eq:spectral_representation_scalar_2}. Taking \eqref{eq:Wightman_propagators_T} and expressing the momenta as derivatives one can write straightforwardly
\begin{equation}
\label{eq:wightman_propagator_T_final}
\begin{aligned}
\Delta^{\mu\nu;\,\rho\sigma}_{W,\,\Theta}(x;s) &=
(s\,\eta^{\mu\nu}-\partial^\mu \partial^\nu)
(s\,\eta^{\rho\sigma}-\partial^\rho \partial^\sigma)
\Delta_{W}(x;s),\\
\Delta^{\mu\nu;\,\rho\sigma}_{W,\,i}(x;s) &= s^2\, 
\Pi^{\mu\nu;\,\rho\sigma}_{i}(p^\alpha\rightarrow i\partial^\alpha,\;p^2\rightarrow - s)
\Delta_{W}(x;s).
\end{aligned}
\end{equation}

\subsection{Time-ordered two-point functions}
\label{sec:time-ordered}
Time-ordered correlators are widely used because of several reasons. First they can be straightforwardly computed in perturbation theory. Second, they appear in the LSZ reduction formula and third they can be easily mapped to Euclidean correlators using the Wick rotation. In this section we will discuss two-point time-ordered correlaotrs.

Given two real scalar operators $\cO_1(x)$ and $\cO_2(x)$ the time-ordered two-point function is defined as
\begin{equation}
\<0| \cO_1(x_1) \cO_2(x_2)|0\>_T \equiv
\theta(x_{12}^0) \<0| \cO_1(x_1) \cO_2(x_2) |0\>_W+
\theta(x_{21}^0) \<0| \cO_2(x_2) \cO_1(x_1) |0\>_W.
\end{equation}
Plugging the expression for the Wightman two-point functions in terms of the spectral density \eqref{eq:spectral_representation_scalar_2} we get
\begin{equation}
\label{eq:KL_representation_scalar}
\begin{aligned}
\<0| \cO(x) \cO(0)|0\>_T &= -i\int_0^\infty ds \rho_\cO(s) \Delta_F(x;s),\\
-i\Delta_F(x;s) &\equiv \theta(x^0)\Delta_W(x;s) + \theta(-x^0)\Delta_W(-x;s)\\
 &= \lim_{\epsilon\rightarrow 0^+} \int \frac{d^dp}{(2\pi)^d}e^{ip\cdot x} \frac{-i}{p^2+s-i\epsilon},
\end{aligned}
\end{equation}
where $\Delta_F$ is called the Feynman propagator.\footnote{
In order to rewrite the Feynman propagator defined in the second line of \eqref{eq:KL_representation_scalar} in a conventional form given by the last line of \eqref{eq:KL_representation_scalar}, one uses the integral representation of the step function
\begin{equation*}
\theta(t) =-\frac{1}{2\pi i} \int_{-\infty}^{+\infty} ds\,\frac{e^{-ist}}{s+i\epsilon}
\end{equation*}
and performs a change of variables. In most textbooks the equivalence of the second and third lines of \eqref{eq:KL_representation_scalar} is shown backwards by integrating the last line of \eqref{eq:KL_representation_scalar} in $p^0$ using the residue theorem.}
Its explicit expression can be found in \eqref{eq:feynman_propagator_explicit}.
The representation of the time-ordered two-point correlation function in terms of the spectral density \eqref{eq:KL_representation_scalar} is called the K\"{a}ll\'en-Lehmann representation. We also exclude $x^\mu=0$ point from the discussion to avoid talking about contact terms.

Using \eqref{eq:KL_representation_scalar} one can also express the spectral density in terms of the real part of the time-ordered two-point function. To show that we use the relation
\begin{equation}
\lim_{\epsilon\rightarrow 0^+}\frac{1}{p^2+s-i\epsilon}=
P \frac{1}{p^2+s}+i\pi \delta(p^2+s),
\end{equation}
inside the Feynman propagator, where $P$ stands for the principal value. Performing the inverse Fourier transformation we obtain
\begin{equation}
\theta(p^0) \rho_\cO(-p^2) = \frac{1}{\pi}\,\int d^d x e^{-i p \cdot x} \text{Re} \<0| \cO(x) \cO(0)|0\>_T.
\end{equation}

In what follows we derive the analogs of \eqref{eq:KL_representation_scalar} for conserved currents and the stress-tensor.

\subsubsection*{Conserved currents}
Analogously to the scalar case the time-ordered two-point function of two Lorentz spin one operators is defined as
\begin{equation}
\label{eq:T_J}
\<0| J^\mu(x_1) J^\nu(x_2)|0\>_T =
\theta(x_{12}^0)  \<0| J^\mu(x_1) J^\nu(x_2)|0\>_W+
\theta(x_{21}^0) \<0| J^\nu(x_2) J^\mu(x_1)|0\>_W.
\end{equation}
Plugging here the spectral representation of the current Wightman function \eqref{eq:spectral_representation_2} we get the  K\"{a}ll\'en-Lehmann representation for the currents. It reads
\begin{equation}
\label{eq:spectral_J}
\<0| J^\mu(x) J^\nu(0)|0\>_T = -i \int_0^\infty ds 
\left(
-\rho_J^0(s) \Delta_{F,\,0}^{\mu\nu}(x;s) +
\rho_J^1(s) \Delta_{F,\,1}^{\mu\nu}(x;s)
\right),
\end{equation}
where the Feynman propagators are defined as
\begin{equation}
\label{eq:feynman_J}
-i\Delta_{F,\,i}^{\mu\nu}(x;s) = \theta(x^0)\Delta_{W,\,i}^{\mu\nu} (x;s) +
\theta(-x^0)\Delta_{W,\,i}^{\nu\mu} (-x;s).
\end{equation}
Using \eqref{eq:wightman_propagator_J}, the derivative of the step function
\begin{equation}
\frac{\partial }{\partial t} \theta(t)=\delta(t)
\end{equation}
and the following property of the scalar Wightman propagator\footnote{The left-hand side of \eqref{eq:property} can be potentially non-zero only for $x^0=0$. In the latter case however the scalar propagator \eqref{eq:spectral_representation_scalar_2} is symmetric under the exchange $\vec x \leftrightarrow -\vec x$.}
\begin{equation}
\label{eq:property}
\delta(x^0) \times \big(\Delta_W(x;s)-\Delta_W(-x;s)\big) = 0,
\end{equation}
one obtains the following simple expressions for the Feynman propagator
\begin{equation}
\label{eq:feynman_propagator_J}
\Delta_{F,\,0}^{\mu\nu}(x;s) = \partial_\mu \partial_\nu\; \Delta_F(x;s),\quad
\Delta_{F,\,1}^{\mu\nu}(x;s) = \left(s\,\eta_{\mu\nu}-\partial_\mu \partial_\nu\right) \Delta_F(x;s).
\end{equation}

\subsubsection*{Stress-tensor}
The identical discussion holds for the time-ordered two-point correlation function of the stress-tensors. In what follows we will only state its K\"{a}ll\'en-Lehmann representation in terms of the components of the stress-tensor spectral density. It reads
\begin{equation}
\label{eq:spectral_TT}
\begin{aligned}
\<0| T^{\mu\nu}(x) T^{\rho\sigma}(0)|0\>_T 
=&-\frac{i}{(d-1)^2}\int_0^\infty ds \rho_{\Theta}(s) \Delta_{F,\Theta}^{\mu\nu;\,\rho\sigma}(x;s)\\
& -i \int_0^\infty ds \rho_{\hat T}^2(s) \Delta_{F,\,2}^{\mu\nu;\,\rho\sigma}(x;s),
\end{aligned}
\end{equation}
where the Feynman propagators read as
\begin{equation}
\label{eq:feynman_propagator_T}
\begin{aligned}
\Delta_{F,\Theta}^{\mu\nu;\rho\sigma}(x;s) &=
(s\,\eta^{\mu\nu}-\partial^\mu \partial^\nu)
(s\,\eta^{\rho\sigma}-\partial^\rho \partial^\sigma)
\Delta_F(x;s),\\
\Delta^{\mu\nu;\,\rho\sigma}_{F,\,2}(x;s) &=
s^2\Pi^{\mu\nu;\,\rho\sigma}_2(p^\alpha\rightarrow i\partial^\alpha,\;p^2\rightarrow - s)
\Delta_{F}(x;s)
\end{aligned}
\end{equation}

\section{Spectral densities in Lorentzian CFTs}
\label{sec:conformal_spectral}
In the previous section we defined spectral densities as the Fourier transform of Wightman two-point functions. In the presence of conformal symmetry the two-point functions are purely kinematic objects. In other words their form is completely fixed by the conformal symmetry. As a consequence we can straightforwardly compute the CFT spectral densities.

Let us start from the very well known case of a real scalar operator $\cO$ with the scaling dimension $\Delta_\cO$. In unitary theories there is a lower bound on this scaling dimension which reads as $\Delta_\cO\geq \frac{d-2}{2}$. As in the previous section we will work in the Lorentzian metric here. The Wightman two-point function of the operator $\cO$ reads
\begin{equation}
\label{eq:scalar_2pf_conformal}
\<0|\cO(x)\cO(0)|0\>_{W,\;CFT} =
\lim_{\epsilon\rightarrow 0^+} \frac{\mathcal{N}_\mathcal{O}}{\left(\hat x^2\right)^{\Delta_\cO}},
\end{equation}
where $\mathcal{N}_\mathcal{O}$ is the normalization constant which can be set to one. Plugging \eqref{eq:scalar_2pf_conformal} into the definition \eqref{eq:spectral_representation_scalar} one gets
\begin{align}
\label{eq:result_FT_scalar}
(2\pi)\theta(q^0)\rho_{\cO}(-q^2) = 
\lim_{\epsilon\rightarrow 0^+}
\int d^dx\,\frac{ e^{-i q \cdot \hat x}}{(\hat x^{2})^{\Delta_\cO}}.
\end{align}
Performing the integration and taking the limit we arrive at the following expression for the spectral density\footnote{This result can be found for example in equations (2.22) in \cite{Bautista:2019qxj}, see also \cite{Gillioz:2018mto}. Notice that for some special values of $\Delta_\cO$ the integration procedure leads to additional terms in \eqref{eq:scalar_spectral_CFT}. These terms are not present however for all the cases relevant to this section due to unitarity bounds on scaling dimensions. }
\begin{equation}
\label{eq:scalar_spectral_CFT}
2\pi\rho_{\cO}(s) =
\mathcal{N}_\mathcal{O}\;
\kappa(d,\Delta_\cO)\theta(s)s^{\Delta_\cO-d/2},
\end{equation}
where the coefficient $\kappa(d,\Delta)$ is defined as
\begin{equation}
\label{eq:kappa}
\kappa(d,\Delta)\equiv\frac{\pi^{d/2+1}}{2^{2\Delta-d-1}\Gamma(\Delta)\Gamma(\Delta-\frac{d-2}{2})}.
\end{equation}

In what follows we will derive the spectral densities of a generic Lorentz spin one and Lorentz spin two operators. Notice, that we will completely ignore the parity odd part in $d=2$.

\subsection*{Lorentz spin one operator}
Consider a generic Lorentz spin one operator $J^\mu$ with the scaling dimension $\Delta_J\geq d-1$. In the Euclidean signature the two-point function of such operators was already given in \eqref{eq:2pt_J_conformal}. Analogously in the Lorentzian signature we have 
\begin{equation}
\label{eq:2pt_J_conformal_initial}
\<0|J^\mu(\hat x) J^\nu(0)|0\>_{W,\;CFT} = 
\lim_{\epsilon\rightarrow 0^+}
\frac{C_J}{(\hat x^{2})^{\Delta_J}}\times
\left(\eta^{\mu\nu} - 2\,\frac{\hat x^\mu \hat x^\nu}{\hat x^2}\right).
\end{equation}
For a generic Lorentz spin one operator $C_J$ is a normalization constant. When $J^\mu$ is a conserved current instead, $C_J$ becomes the central charge and the scaling dimension $\Delta_J$ saturates the unitarity bound $\Delta_J=d-1$.

Plugging  \eqref{eq:2pt_J_conformal_initial} into the first equation in \eqref{eq:spectral_representation} we get
\begin{align}
\nn
(2\pi)\theta(q^0)\rho_{J}^{\mu\nu}(q) &=\int d^dx\, e^{-i q \cdot x}
\frac{C_J}{(\hat x^{2})^{\Delta_J}}\times
\left(\eta^{\mu\nu} - 2\,\frac{\hat x^\mu \hat x^\nu}{\hat x^2}\right)\\
&=C_J \times\left(\eta^{\mu\nu}\int d^dx\,\frac{ e^{-i q \cdot \hat x}}{(\hat x^{2})^{\Delta_J}}
+2\partial_q^\mu\partial_q^\nu\,\int d^dx\,\frac{ e^{-i q \cdot \hat x}}{(\hat x^{2})^{{\Delta_J}+1}}\right).
\label{eq:spectral_J_CFT}
\end{align}
The integrals in this expression have already been evaluated in \eqref{eq:result_FT_scalar} and we are only left with taking derivatives. Using the properties\footnote{To see that the last entry in \eqref{eq:properties_del} indeed vanishes, notice that because of the $\delta$-function it can be non-zero only if $\mu\neq 0$ and $q^0=0$. This leads however to the step function of a negative argument which vanishes unless all $q^i=0$, with $i=1,2,\ldots,d$. The latter case also gives zero because of the $q^\mu$ factor.
}
\begin{equation}
\label{eq:properties_del}
\begin{aligned}
f(x)\delta'(x) &= -\delta(x) \frac{d}{dx}f(x),\\
x \delta(x) &= 0,\\
q^\mu\delta(q^0)\theta(-q^2) & = q^\mu\delta(q^0)\theta(-\vec q\,^2) =0,
\end{aligned}
\end{equation}
and \eqref{eq:result_FT_scalar} together with \eqref{eq:scalar_spectral_CFT} one can show that
\begin{equation}
\label{eq:technical expression}
\begin{aligned}
\partial_q^\mu\partial_q^\nu\,\int d^dx\,\frac{ e^{-i q \cdot x}}{(x^{2})^{{\Delta_J}+1}} &= -2a \left(\eta^{\mu\nu}
+2(a-1) \frac{q^\mu q^\nu}{q^2}\right)\theta(q^0) \theta(-q^2) (-q^2)^{a-1}\\
&+4aq^\mu q^\nu \theta(q^0) \delta(-q^2) (-q^2)^{a-1},
\end{aligned}
\end{equation}
where we have defined the parameter 
\begin{equation}
a \equiv \Delta_J-d/2+1\geq d/2.
\end{equation}
Notice that the last term in \eqref{eq:technical expression} vanishes unless $a=1$. This is possible only in a very special case of conserved currents (with $\Delta_J=1$) in $d=2$ dimensions. 
Plugging \eqref{eq:result_FT_scalar}, \eqref{eq:scalar_spectral_CFT} and \eqref{eq:technical expression} into \eqref{eq:spectral_J_CFT} and bringing the result into the form \eqref{eq:decomposition_spectral_density_J} we can read off the expressions of the components of the Lorentz spin one spectral density
\begin{equation}
\label{eq:J_cft_density}
\begin{aligned}
2\pi\rho_J^0(s) &= C_J\times\frac{\Delta_J-d+1}{\Delta_J}\times\kappa(d,\Delta_J)\theta(s)s^{\Delta_J-d/2-1}\\
&+C_J\times\frac{2}{\Delta_J}\times\kappa(d,\Delta_J)\delta(s)s^{\Delta_J-d/2},\\
2\pi\rho_J^1(s) &= C_J\times\frac{\Delta_J-1}{\Delta_J}\times\kappa(d,\Delta_J)\theta(s)s^{\Delta_J-d/2-1}.
\end{aligned}
\end{equation}

Let us now focus on the case when $J^\mu$ is a conserved current. The expressions \eqref{eq:J_cft_density} then simplify and read
\begin{align}
\label{eq:J_spin0_cft_cons}
&2\pi\rho_{J}^0(s) = 4\pi^2\,\delta_{d,2}\,C_J\times\delta(s),\\
\label{eq:J_spin1_cft_cons}
&2\pi\rho_{J}^1(s)= \frac{d-2}{d-1}\,\kappa(d,d-1)C_J\times \theta(s)s^{d/2-2}.
\end{align}
In $d=2$ the spin 1 component vanishes and the spin 0 component is proportional to $\delta(s)$. This is in agreement with \eqref{eq:implication_conservation}. By comparing \eqref{eq:J_spin0_cft_cons} and \eqref{eq:implication_conservation} we can even determine the coefficient $A$ introduced in \eqref{eq:implication_conservation}, it reads $A=2\pi\,C_J.$
In $d\geq 3$ the spin 1 component instead is always non-zero whereas the spin 0 component always vanishes. This is again in agreement with \eqref{eq:implication_conservation} since the coefficient $A$ introduced there is a dimensionful quantity and thus must vanish in CFTs, $A=0$.

\subsection*{Lorentz spin two operator}
Using the identical logic we can  derive the components of spectral densities for the Lorentz spin two operator $\hat T^{\mu\nu}$. We keep the hat in order to indicate explicitly that the operator is trace-less. Skipping all the details we provide only the final answer which reads
\begin{equation}
\label{eq:T_density}
\begin{aligned}
2\pi\rho_{\hat T}^0(s) &= C_T\times\frac{(\Delta_T-d)(\Delta_T-d+1)}{\Delta_T\,(\Delta_T+1)}\times\kappa(d,\Delta_T)\theta(s)s^{\Delta_T-d/2-2}\\
2\pi\rho_{\hat T}^1(s) &= C_T\times\frac{(\Delta_T-d)(\Delta_T-1)}{\Delta_T\,(\Delta_T+1)}\times\kappa(d,\Delta_T)\theta(s)s^{\Delta_T-d/2-2}+\frac{\pi^2}{6}\,C_T\,\delta(s)\delta_{d,2}\delta_{\Delta_{\hat T},2},\\
2\pi\rho_{\hat T}^2(s) &= C_T\times\frac{\Delta_T-1}{\Delta_T+1}\times\kappa(d,\Delta_T)\theta(s)s^{\Delta_T-d/2-2},
\end{aligned}
\end{equation}
where the stress-tensor central charge $C_T$ is defined in \eqref{eq:2pt_T_cft_main}.\footnote{The expression \eqref{eq:2pt_T_cft_main} is given in the Euclidean signature. Its parity even part in the Lorentzian signature is obtained straightforwardly by simply replacing the Kronecker delta with the Lorentzian metric.}

\section{Spectral densities and central charges}
\label{sec:spectral_densities_and_central_charges}

As explained in section \ref{sec:spectral_densities}, in the case of a conserved current $J^\mu(x)$ there is a single (Little group spin one) component of the spectral density denoted by $\rho_J^1(s)$. Analogously, in the case of the stress-tensor $T^{\mu\nu}(x)$ there are two components of the spectral density, namely the trace part $\rho_\Theta(s)$ and the Little group spin two part $\rho_{\hat T}^2(s)$.\footnote{Spectral densities give an alternative description of the two-point functions to the position space functions $h_i(r)$ introduced in section \ref{sec:two-point_functions}, see \eqref{eq:2pt_J} and \eqref{eq:2pt_T}. For instance in the case of conserved currents we have two functions $h_1(r)$ and $h_2(r)$ related by a single differential equation. In the case of the stress-tensor we have five functions $h_i(r)$ with three differential constraints.}

In what follows we will explain how the information about the UV and IR central charges (for their precise definition see either section \ref{sec:conformal_spectral} or section \ref{sec:two-point_functions}) are encoded in the components of spectral densities. We will see that $d=2$ and $d\geq 3$ are drastically different. We will consider only the continuous part of the spectral densities excluding the $s=0$ point from the discussion.\footnote{We are allowed however to be infinitesimally close to $s=0$.}

\section*{Conserved currents}
Taking into account \eqref{eq:J_spin0_cft_cons}, the requirement that the quantum field theory under consideration has the UV and IR fixed points described by the UV  and IR conformal field theories respectively at the level of spectral densities is imposed by the conditions\footnote{We remind that $s$ plays the role of energy squared. At very small and very large energies we expect to restore conformal invariance since we approach the IR and UV fixed points.}
\begin{equation}
\label{eq:asymptotics}
\begin{aligned}
\lim_{s\rightarrow 0} \,
 s^{2-d/2}&\times\left(
\rho_J^1(s) - \rho_{J}^1(s)\big|_{IR\;CFT}\right)=0,\\
\lim_{s\rightarrow \infty} \,
s^{2-d/2}&\times\left(
\rho_J^1(s) - \rho_{J}^1(s)\big|_{UV\;CFT}\right)=0.
\end{aligned}
\end{equation}
These are completely equivalent to the position space conditions \eqref{eq:limit_1_J} and \eqref{eq:limit_2_J}. Due to \eqref{eq:J_spin0_cft_cons} the above requirement can also be written as
\begin{equation}
\label{eq:asymptotics_elaborated}
\begin{aligned}
\lim_{s\rightarrow 0} s^{2-d/2}\times\rho_J^1(s) &= C^{IR}_J\times(2\pi)^{-1}\frac{d-2}{d-1}\,\kappa(d,d-1),\\
\lim_{s\rightarrow \infty} s^{2-d/2}\times\rho_J^1(s) &= C^{UV}_J\times(2\pi)^{-1}\frac{d-2}{d-1}\,\kappa(d,d-1),
\end{aligned}
\end{equation}
where $C_J^{IR}$ and $C_J^{UV}$ are the usual IR an UV conserved current central charges and the numerical coefficient $\kappa$ is given by \eqref{eq:kappa}.
We see that in $d\geq 3$ the central charges govern the asymptotics of $\rho_J^1(s)$. In $d=2$ instead the right-hand side of \eqref{eq:asymptotics_elaborated} simply vanishes and surprisingly the dependence on the UV and IR central charges disappears.

In order to understand what is happening in $d=2$ dimensions let us re-derive \eqref{eq:asymptotics_elaborated} in a different way. Consider the integral expression for the difference of $UV$ and $IR$ central charges \eqref{eq:integral_J} valid in $d\geq 2$. It contains the Euclidean two-point function of conserved currents. We then write its spectral (K\"{a}ll\'en-Lehmann) decomposition in terms of $\rho_J^1(s)$. This is done by applying the Wick rotation to the Lorentzian spectral decomposition \eqref{eq:spectral_J}. We discuss all the technical details in appendix \ref{app:technical_dentails} and state here only the final answer: in $d\geq 3$ the sum-rule \eqref{eq:integral_J} reduces to the asymptotic conditions \eqref{eq:asymptotics_elaborated}, instead in $d=2$ one gets the following integral expression
\begin{equation}
\label{eq:sum_rule_J_2d_main_text}
C_J^{UV}-C_J^{IR} = \frac{1}{2\pi}
\lim_{s_{min}\rightarrow 0}
\int_{s_{min}}^{\infty} \frac{ds}{s}\,\rho_J^1(s).
\end{equation}
This result is not well known in the literature, nevertheless it was obtained long before this paper, see \cite{VilasisCardona:1994ri}.

In $d=2$ another central charge $C^\prime_J$ also exists, see \eqref{eq:2pt_J_conformal}, which is actually the global anomaly and according to the discussion of section \ref{sec:J} remains invariant along the RG flow. In other words
\begin{equation}
\label{eq:condition_CJbar}
C_J^{\prime\,UV}=C_J^{\prime\,IR}.
\end{equation}
For more details see section \ref{sec:J}.
In order to write \eqref{eq:sum_rule_J_2d_main_text} in a canonical form we define the holomorphic and the anti-holomorphic parts of the conserved currents. The associated central charges are denoted by $k$ and $\bar k$ respectively and are related to $C_J$ and $C^\prime_J$ in the following way
\begin{equation}
k\equiv (2\pi)^2 \times \frac{C_J+C^\prime_J}{2},\qquad
\bar k\equiv (2\pi)^2 \times \frac{C_J-C^\prime_J}{2}.
\end{equation}
For details see the end of appendix \ref{app:2pt_CFTs}. In terms of $k$ and $\bar k$ the sum-rule \eqref{eq:sum_rule_J_2d_main_text} due to the condition \eqref{eq:condition_CJbar} reads as
\begin{equation}
\label{eq:sum_rule_J_2d_main_text_kkbar}
\begin{aligned}
k_{UV}-k_{IR} =
\bar k_{UV}-\bar k_{IR}= \pi
\lim_{s_{min}\rightarrow 0}
\int_{s_{min}}^{\infty} \frac{ds}{s}\,\rho_J^1(s).
\end{aligned}
\end{equation}

In section \eqref{sec:wightman} we proved that $\rho_J^1(s)\geq0$ for all the energies. As a result from \eqref{eq:sum_rule_J_2d_main_text} we conclude that
\begin{equation}
\label{eq:k_theorem_1}
C_J^{UV}-C_J^{IR} \geq 0.
\end{equation}
Alternatively from \eqref{eq:sum_rule_J_2d_main_text_kkbar} we conclude that
\begin{equation}
\label{eq:k_theorem_2}
k_{UV}-k_{IR} \geq 0,\qquad
\bar k_{UV}-\bar k_{IR} \geq 0.
\end{equation}
The inequalities \eqref{eq:k_theorem_1} and \eqref{eq:k_theorem_2} are referred to as the ``c-theorem'' for conserved currents or the ``k-theorem''. Notice that the equal sign can appear only if the theory is conformal where $\rho_J^1(s)=0$ for all the energies according to \eqref{eq:J_spin1_cft_cons}.
In $d\geq 3$ the situation is very different. Due to unitarity  $C_J^{UV}\geq 0$ and $C_J^{IR}\geq 0$. However from \eqref{eq:asymptotics_elaborated} one cannot deduce further relations between them, in other words both options   $C_J^{UV}\geq C_J^{IR}$ and $C_J^{UV}< C_J^{IR}$ are perfectly viable.

\section*{Stress-tensor}
The identical discussion holds for the stress-tenors. The requirement that the UV and IR fixed points are governed by the UV and IR conformal field theories translates into the conditions on the components of the stress-tensor spectral density.

We start with the spectral density of the trace of the stress-tensor $\rho_\Theta(s)$. In a conformal theory we strictly have $\Theta(x)=0$. In a quantum field theory instead the trace operator is given by the relevant scalar operator $\cO$ deforming the UV CFT, namely
\begin{equation}
\Theta(x)=g \cO(x).
\end{equation}
The operator $\cO$ has the scaling dimension $\Delta_\cO<d$ and the coupling constant $g$ has the mass dimension $[g]=d-\Delta_\cO$. As a result $\rho_\Theta(s)=g^2 \rho_\cO(s)$. Using \eqref{eq:scalar_spectral_CFT} we conclude that
\begin{equation}
\lim_{s\rightarrow \infty} \,
s^{d/2-\Delta_\cO}\times
\left(\rho_\Theta(s) - g^2 \rho_\cO(s)\big|_{UV\;CFT}\right)=0
\end{equation}
or equivalently
\begin{equation}
\label{eq:limit_Tr}
\lim_{s\rightarrow \infty}
s^{d/2-\Delta_\cO}\times \rho_\Theta(s) =
g^2(2\pi)^{-1}\mathcal{N}_\mathcal{O}\;
\kappa(d,\Delta_\cO).
\end{equation}
The coefficient $\kappa$ is given by \eqref{eq:kappa} and $\mathcal{N}_\mathcal{O}$ is the normalization constant of the operator $\cO$, see \eqref{eq:scalar_2pf_conformal}. Analogous we can write
\begin{equation}
\Theta(x)=g' \cO'(x),
\end{equation}
where $\cO'$ is an irrelevant operator describing the deformation of the IR conformal field theory. We have then
\begin{equation}
\label{eq:asymptotics_Theta}
\lim_{s\rightarrow 0}
s^{d/2-\Delta_\cO'}\times \rho_\Theta(s) =
g^{\prime\, 2}(2\pi)^{-1}\mathcal{N}_{\cO'}\;
\kappa(d,\Delta_\cO).
\end{equation}
The operators $\cO$ and $\cO'$ are related by some change of basis. One basis is more convenient for working at high energies, the other one is more convenient for working at low energies. 
For some extra details on the trace of the stress-tensor see the last paragraph of section \ref{sec:T}.

Let us address now the Little group spin two component of the spectral density $\rho^2_{\hat T}(s)$. Its asymptotic behavior due to \eqref{eq:T_density} reads as
\begin{equation}
\label{eq:asymptotics_T}
\begin{aligned}
\lim_{s\rightarrow 0} \,
s^{2-d/2}&\times\left(
\rho_{\hat T}^2(s) - \rho_{\hat T}^2(s)\big|_{IR\;CFT}\right)=0,\\
\lim_{s\rightarrow \infty} \,
s^{2-d/2}&\times\left(
\rho_{\hat T}^2(s) - \rho_{\hat T}^2(s)\big|_{UV\;CFT}\right)=0.
\end{aligned}
\end{equation}
These are equivalent to
\begin{equation}
\label{eq:asymptotics_T2hat}
\begin{aligned}
\lim_{s\rightarrow 0}
s^{2-d/2}\times\rho_{\hat T}^2(s) &=C_T^{IR}\,\times(2\pi)^{-1} \frac{d-1}{d+1}\kappa(d,d),\\
\lim_{s\rightarrow \infty}
s^{2-d/2}\times\rho_{\hat T}^2(s) &=C_T^{UV}\times(2\pi)^{-1} \frac{d-1}{d+1}\kappa(d,d).
\end{aligned}
\end{equation}

As we can see, the asymptotic behavior of $\rho_\Theta(s)$ is governed by the properties of the UV and IR ``deforming'' operators, instead the asymptotic behavior of $\rho_{\hat T}^2(s)$ is governed by the central charges. We remind however that $\rho_{\hat T}^2(s)$ exists only in $d\geq 3$. In $d=2$ the central charge information is encoded instead into $\rho_\Theta(s)$ in a very non-trivial way. To understand how, we use \eqref{eq:integral_expression_T_form2} and plug there the Wick rotated spectral decomposition of the two-point function of the stress-tensor \eqref{eq:spectral_TT}. We recover \eqref{eq:asymptotics_T2hat} in $d\geq 3$ and in $d=2$ dimensions obtain instead
\begin{equation}
\label{eq:sum_rule_T_2d_main_text}
C_T^{UV} - C_T^{IR} = \frac{6}{\pi}
\lim_{s_{min}\rightarrow 0}
\int_{s_{min}}^{\infty} \frac{ds}{s^2}\rho_\Theta(s).
\end{equation}
For more details see appendix \ref{app:technical_dentails}.
The sum-rule \eqref{eq:sum_rule_T_2d_main_text} was derived in \cite{Cappelli:1990yc}.

In $d=2$ there is also another central charge $C^\prime_T$, see \eqref{eq:2pt_T_cft_main}, which is actually the gravitational anomaly and thus remains invariant along the RG flow, in other words
\begin{equation}
\label{eq:gravitational_anomaly}
C_T^{\prime\,UV}=C_T^{\prime\,IR}.
\end{equation}
For more details see section \ref{sec:T}. In order to write \eqref{eq:sum_rule_T_2d_main_text} in a canonical form we define the holomorphic and the anti-holomorphic parts of the stress-tensor. The associated central charges are denoted by $c$ and $\bar c$ respectively and are related to $C_T$ and $C^\prime_T$ in the following way
\begin{equation}
c\equiv (2\pi)^2\times\frac{C_T + C^\prime_T}{2},\qquad
\bar c\equiv (2\pi)^2\times\frac{C_T - C^\prime_T}{2}.
\end{equation}
For details see the end of appendix \ref{app:2pt_CFTs}. Taking into account \eqref{eq:gravitational_anomaly}, the sum-rule \eqref{eq:sum_rule_T_2d_main_text} can be written as
\begin{equation}
\label{eq:sum_rule_T_2d_main_text_kkbar}
c_{UV}-c_{IR}=\bar c_{UV}-\bar c_{IR}=
12\pi
\lim_{s_{min}\rightarrow 0}
\int_{s_{min}}^{\infty} \frac{ds}{s^2}\rho_\Theta(s).
\end{equation}

In section \eqref{sec:wightman} we proved that $\rho_\Theta(s)\geq0$ for all the energies. As a result from \eqref{eq:sum_rule_T_2d_main_text} we conclude that
\begin{equation}
\label{eq:c_theorem_1}
C_T^{UV}-C_T^{IR} \geq 0.
\end{equation}
Alternatively from \eqref{eq:sum_rule_T_2d_main_text_kkbar} we conclude that
\begin{equation}
\label{eq:c_theorem_2}
c_{UV}-c_{IR} \geq 0,\qquad
\bar c_{UV}-\bar c_{IR} \geq 0.
\end{equation}
The inequalities \eqref{eq:c_theorem_1} and \eqref{eq:c_theorem_2} were found by A. Zamolodchikov \cite{Zamolodchikov:1986gt}, see also \cite{Cardy:1988tj}. They are referred to as the ``c-theorem''.
Notice that the equal sign can appear only if the theory is conformal where $\rho_\Theta(s)=0$ for all the energies according to \eqref{eq:T_density}.
In $d\geq 3$ the situation is very different. Due to unitarity  $C_T^{UV}\geq 0$ and $C_T^{IR}\geq 0$. However from \eqref{eq:asymptotics_T2hat} one cannot deduce further relations between them, in other words both options   $C_T^{UV}\geq C_T^{IR}$ and $C_T^{UV}< C_T^{IR}$ are perfectly viable.

\section{Applications to bootstrap}
\label{sec:discussion}

In section 3 of \cite{Karateev:2019ymz} it was shown how to use unitarity to construct non-trivial constraints on partial amplitudes, form factors and spectral densities. This was done in a presence of a single scalar local operator. Here we extend the analysis of section 3 in \cite{Karateev:2019ymz} to include the full stress-tensor. We will conclude this section by defining concrete bootstrap problems.

We will focus on quantum field theory with a mass gap (or equivalently on the QFTs with an empty IR fixed point). The spectrum of such theories is described by one-particle asymptotic in and out states. We will work here with identical scalar particles for simplicity. For precise definitions of asymptotic states see section 2.1 in \cite{Karateev:2019ymz}. One can build the two-particle asymptotic in and out states by taking the symmetrized tensor product of two one-particle states.\footnote{Symmetrization is required for identical particles in order to make the state invariant under the exchange of two particles.} We denote such two-particle states by
\begin{equation}
|m,\vec p_1;m,\vec p_2\>_{in},\qquad
|m,\vec p_1;m,\vec p_2\>_{out}.
\end{equation}
The four-momenta of the one-particle asymptotic states by definition obey
\begin{equation}
\label{eq:on-shell}
p_1^2=-m^2,\qquad
p_2^2=-m^2.
\end{equation}
We also define
\begin{equation}
\label{eq:p_def}
p^\mu \equiv p_1^\mu + p_2^\mu,\qquad
(p_1+p_2)^2=-s,\qquad
(p_1-p_2)^2=s-4m^2,
\end{equation}
where $s$ is the squared total energy of the two-particle state.

\subsection{Stress-tensor form factor}
\label{sec:form_factor}

Let us start by recalling the definitions of the form-factor and its properties in the case of the stress-tensor. (See also sections 2.4 and 2.6 of \cite{Karateev:2019ymz}.) The trace of the stress-tensor two-particle form factor is defined as\footnote{Compared to \cite{Karateev:2019ymz}, in all the formulas here we drop the subscript $2$ for the form factors in order to simplify the notation. Originally this subscript was introduced to stress that we deal with two-particle form factors.}
\begin{equation}
\label{eq:ff_trace}
\mathcal{F}_{\Theta}(s) \equiv  {}_{out}\<m,\vec p_1;m,\vec p_2|\Theta(0)|0\>.
\end{equation}
The two-particle form factor of the full stress-tensor is defined as
\begin{equation}
\label{eq:ff_T}
\mathcal{F}^{\mu\nu}_T(p_1,p_2) \equiv  {}_{out}\<m,\vec p_1;m,\vec p_2|T^{\mu\nu}(0)|0\>.
\end{equation}
Analogously, one can define the stress-tensor form factors with the in asymptotic states. They are however related in a simple way to the ones here due to the CPT invariance.
One can decompose it in the basis of tensor structures 
\begin{equation}
p_1^\mu p_1^\nu,\quad
p_1^\mu p_2^\nu+p_2^\mu p_1^\nu,\quad
p_2^\mu p_2^\nu,\quad
(p_1+p_2)^2\, \eta^{\mu\nu},
\end{equation}
which are totally symmetric in $\mu$ and $\nu$ indices. The conservation of the stress-tensor leads to the following condition
\begin{equation}
(p_1+p_2)_\mu\, \mathcal{F}^{\mu\nu}_T(p_1,p_2)=0.
\end{equation}
As a result, the most general form of the stress-tensor form factor in $d\geq 3$ reads as
\begin{equation}
\label{eq:stress-tensor_form_factor}
\begin{aligned}
\mathcal{F}^{\mu\nu}_T(p_1,p_2) =
&-\mathcal{F}_{(0)}(s)
\times\left(\eta^{\mu\nu}-\frac{(p_1+p_2)^\mu(p_1+p_2)^\nu}{(p_1+p_2)^2}\right)\\
&+\mathcal{F}_{(2)}(s)
\times\frac{(p_1-p_2)^\mu(p_1-p_2)^\nu}{(p_1-p_2)^2},
\end{aligned}
\end{equation}
where the functions $\mathcal{F}_{(0)}$ and $\mathcal{F}_{(2)}$ are the coefficients in the tensor structure decomposition.\footnote{Since we work in the Lorentizian signature the two tensor structures introduced in \eqref{eq:stress-tensor_form_factor} have poles when $p_1=\pm p_2$. The appearance of these poles is completely artificial. As a result they must be removed in the full expression of the stress-tensor form factor by the presence of appropriate zeros in the components of the stress-tensor form factor $\mathcal{F}_{(0)}(s)$ and $\mathcal{F}_{(2)}(s)$.} We notice also that the first tensor structure in \eqref{eq:stress-tensor_form_factor} due to \eqref{eq:p_def} is precisely the $\Pi^{\mu\nu}_1(p)$ projector defined in \eqref{eq:projectors_vector_main_text}. In $d=2$ the two tensor structures in \eqref{eq:stress-tensor_form_factor} are equal to each other, thus we should keep only one of them in the decomposition. The simplest way to proceed is to leave the second tensor structure which means that we effectively set $\mathcal{F}_{(0)}(s)=0$ in $d=2$. Contracting both sides of \eqref{eq:stress-tensor_form_factor} with the metric $\eta_{\mu\nu}$ and comparing with \eqref{eq:ff_trace} we conclude that
\begin{align}
d=2:\qquad \mathcal{F}_{\Theta}(s) &= \mathcal{F}_{(2)}(s),\\
d\geq 3:\qquad \mathcal{F}_{\Theta}(s) &= \mathcal{F}_{(2)}(s)-(d-1)\,\mathcal{F}_{(0)}(s).
\label{eq:trace_FF}
\end{align}
Due to the fact that the stress-tensor enters the definition of the Poincar\'e generators, one can derive the following normalization conditions
\begin{equation}
\label{eq:normalization_2FF_2d}
\lim_{s\rightarrow 0}s^{-1}\mathcal{F}_{(0)}(s) = \text{const},\qquad
\lim_{s\rightarrow 0}\mathcal{F}_{(2)}(s) = -2m^2,\qquad
\lim_{s\rightarrow 0}\mathcal{F}_{\Theta}(s) = -2m^2,
\end{equation}
where const is some undetermined constant. For the detailed derivation see appendix \ref{app:FF_normalization}.

Consider now the Fourier transformed stress-tensor $T^{\mu\nu}$. Using \eqref{eq:That}, \eqref{eq:T_splitting} and \eqref{eq:components_T_splitting} we can split the stress-tensor form factor into three pieces with the Little group spin 0, 1 and 2 as follows
\begin{equation}
\label{eq:decomposition_FF}
\mathcal{F}^{\mu\nu}_T(p_1,p_2) = 
\left(
(\Pi_0^{\mu\nu;\rho\sigma}(p)+\frac{1}{d}\,\eta^{\mu\nu}\eta^{\rho\sigma})+
\Pi_1^{\mu\nu;\rho\sigma}(p)+\Pi_2^{\mu\nu;\rho\sigma}(p)
\right)\mathcal{F}^{\mu\nu}_T(p_1,p_2),
\end{equation}
where $p^\mu$ was defined in \eqref{eq:p_def}. We notice immediately that the Little group spin 1 term vanishes identically leaving us only with the first and the last terms.  Let us now introduce the center of mass (COM) frame for two-particle states
\begin{equation}
\label{eq:com}
p_1^\text{com} \equiv\{\sqrt s/2, +\vec k\},\qquad
p_2^\text{com} \equiv\{\sqrt s/2, -\vec k\},
\end{equation}
where due to the conditions \eqref{eq:on-shell} we have $s=4m^2+\vec k\,^2$. Plugging \eqref{eq:stress-tensor_form_factor} into \eqref{eq:decomposition_FF} and going to the center of mass frame one gets
\begin{align}
\label{eq:stress-tensor_form_factor_COM}
\mathcal{F}^{\mu\nu}_T(p_1^\text{com},p_2^\text{com}) =
\left(\frac{\mathcal{F}_{(2)}(s)}{d-1}-\mathcal{F}_{(0)}(s)\right)
\times\delta^{mn}+
\mathcal{F}_{(2)}(s)\times
\left(\frac{4\, k^m k^n}{s-4m^2}
-\frac{\delta^{mn}}{d-1}\right),
\end{align}
where the expression in the left-hand side of  \eqref{eq:stress-tensor_form_factor_COM} vanishes if $\mu=0$ or $\nu=0$ and the indices $m$ and $n$ are defined as $\mu=\{0,m\}$ and $\nu=\{0,n\}$. The first term in \eqref{eq:stress-tensor_form_factor_COM} corresponds to the Little group spin 0. Taking into account \eqref{eq:trace_FF} we see that it is simply driven by the trace of the stress-tensor form factor. The second term in \eqref{eq:stress-tensor_form_factor_COM} corresponds to the Little group spin 2.

\section*{Relation with the spectral density}
The stress-tensor spectral density in terms of its components $\rho_\Theta(s)$ and $\rho_{\hat T}^2(s)$ is given in \eqref{eq:spectral_T_conserved}. We reproduce it here again for convenience
\begin{equation}
\label{eq:spectral_T_conserved_repeat}
\rho_{T}^{\mu\nu;\,\rho\sigma}(p) = \frac{1}{(d-1)^2}\,\rho_\Theta(s) \Pi_1^{\mu\nu}\Pi_1^{\rho\sigma}+s^2\rho_{\hat T}^2(s)\Pi^{\mu\nu;\,\rho\sigma}_2(p).
\end{equation}
In what follows we compute the components of the spectral density in terms of the components $\mathcal{F}_\Theta$ and $\mathcal{F}_{(2)}$ of the form factor defined in \eqref{eq:stress-tensor_form_factor} and \eqref{eq:trace_FF}. To do this we use \eqref{eq:spectral_density_T}. Writing explicitly the contribution of two-particle states and denoting by $\ldots$ the contribution of multiparticle states, we can write
\begin{multline}
2\pi\rho_{T}^{\mu\nu;\,\rho\sigma}(p) =\frac{1}{2}\int \frac{d^{d-1}p_1}{(2\pi)^{d-1}}\frac{1}{2p^0_1}
\int \frac{d^{d-1}p_2}{(2\pi)^{d-1}}\frac{1}{2p^0_2}\\
(2\pi)^d\delta^{(d)}(p-p_1-p_2)
\mathcal{F}^{*\mu\nu}_T(p_1,p_2)\mathcal{F}^{\rho\sigma}_T(p_1,p_2)+\ldots.
\end{multline}
The overall $1/2$ factor here appears because we deal with identical particles.
Going to the center of mass frame \eqref{eq:com} and switching to the spherical coordinates according to (A.17) in \cite{Karateev:2019ymz} one gets
\begin{equation}
\label{eq:integral_spectral}
2\pi\rho_{T}^{\mu\nu;\,\rho\sigma}(p) =\frac{1}{2\,\mathcal{N}_d}\,
\int\frac{d\Omega_{d-1}}{(2\pi)^{d-2}}
\mathcal{F}^{*\mu\nu}_T(p_1^\text{com},\,p_2^\text{com})\mathcal{F}^{\rho\sigma}_T(p_1^\text{com},\,p_2^\text{com})+\ldots.
\end{equation}
Using the properties of the projectors we can write
\begin{equation}
\label{eq:components}
\begin{aligned}
\rho_\Theta(s)&=\eta_{\mu\nu}\eta_{\rho\sigma}\rho_{T}^{\mu\nu;\,\rho\sigma}(p),\\
\rho_{\hat T}^2(s)&=\frac{2}{(d-2)(d+1)}
\Pi_{2\,\mu\nu;\,\rho\sigma}
\rho_{T}^{\mu\nu;\,\rho\sigma}(p).
\end{aligned}
\end{equation}
We can now perform the integration in \eqref{eq:integral_spectral} and plug the result into \eqref{eq:components}. One then obtains
\begin{equation}
\label{eq:components_spectral_ff}
\begin{aligned}
2\pi\rho_\Theta(s) &= \omega^2\, |\mathcal{F}_\Theta(s)|^2+\ldots,\\
2\pi s^2\rho_{\hat T}^2(s) &= \frac{2\,\omega^2}{d^2-1}\, |\mathcal{F}_{(2)}(s)|^2+\ldots,
\end{aligned}
\end{equation}
where the coefficient $\omega$ reads as
\begin{equation}
\label{eq:omega}
\omega^2 = \frac{1}{\mathcal{N}_d}\frac{\Omega_{d-1}}{2(2\pi)^{d-2}}.
\end{equation}
The spherical angle $\Omega_n$ is defined in \eqref{eq:spherical_angle} and the coefficient $\mathcal{N}_d$ reads as
\begin{equation}
\label{eq:Nd}
\mathcal{N}_d \equiv 2^{d-1} \sqrt{s} \,\left(s-4m^2\right)^{(3-d)/2}.
\end{equation}

\section*{Projection to definite spin}
Let us introduce now the two-particle in and out asymptotic states in the center of mass frame projected to a definite $SO(d-1)$ Little group total spin $j$,
\begin{equation}
\label{eq:state_12}
|\psi_1\>_j \equiv \mathbf{\Pi}_j|m,+\vec k;m, -\vec k\,\>_{in},\qquad
|\psi_2\>_j \equiv \mathbf{\Pi}_j|m,+\vec k;m, -\vec k\,\>_{out}.
\end{equation}
The projector $\mathbf{\Pi}_j$ was defined in equation (2.14) of \cite{Karateev:2019ymz}, it reads
\begin{align}
\label{eq:irreps}
\mathbf{\Pi}_j &\equiv \gamma_j\times\int d\Omega_{d-1}C_{j}^{(d-3)/2}(\cos\theta_1),
\end{align}
where $C_j^k$ is the Gegenbauer polynomial, $\theta_1$ is the angle of the $(d-1)$-dimensional vector $\vec k$ with respect to the $x^{d-1}$ spatial axis.
The coefficient $\gamma_j$ is defined as
\begin{equation}
\label{eq:coefficient_gamma}
\gamma_j \equiv
\omega\, \Gamma\left(\frac{d-3}{2}\right)\times
\left(
\frac{2^{d-4}}{2\pi\Omega_{d-1}\Omega_{d-2}}
\frac{j!\,(d-3+2j)}{\Gamma(d-3+j)}
\right)^{1/2}.
\end{equation}
We will also need to introduce the following state
\begin{align}
\label{eq:T_state}
|\psi_3\>^{\mu\nu} &\equiv \lim_{\epsilon\rightarrow 0}m^{-d/2}
\int d^d x e^{+ip\cdot x}T^{\mu\nu}(\hat x^*)|0\>,
\end{align}
where the coordinate $\hat x^\mu$ has a small imaginary part in the time direction according to \eqref{eq:xhat}.
The factor $m^{-d/2}$ is introduced to match the dimensions of the states \eqref{eq:state_12}.

Let us study now the inner product of the states $|\psi_2\>$ and $|\psi_3\>^{\mu\nu}$. One has
\begin{align}
\<\psi_2|\psi_3\>^{\mu\nu} &= m^{-d/2}\; \mathbf{\Pi}_j
\int d^d x e^{+ip\cdot x}
{}_{out}\<m,+\vec k;m, -\vec k|
T^{\mu\nu}(x)|0\>\\
&= m^{-d/2}(2\pi)^d\delta^d(p-p_1^\text{com}-p_2^\text{com}) \times \mathbf{\Pi}_j
\mathcal{F}^{\mu\nu}_T(p_1^\text{com},\,p_2^\text{com}).
\end{align}
In the second line we used \eqref{eq:trans_invariance} and the definition \eqref{eq:ff_T}. Applying \eqref{eq:irreps} to \eqref{eq:stress-tensor_form_factor_COM} we obtain\footnote{In practice we performed the integration explicitly using Mathematica for several values of $d$ and then guessed the general result. For the definition of spherical coordinates see appendix A in \cite{Karateev:2019ymz}.}
\begin{equation}
\mathbf{\Pi}_j\mathcal{F}^{\mu\nu}_T(p_1^\text{com},p_2^\text{com}) = 0,\qquad \forall j\neq 0,2.
\end{equation}
The only non-zero result appears for $j=0$ and $j=2$. In the former case only the first term in \eqref{eq:stress-tensor_form_factor_COM} gives a non-zero contribution. In the latter case only the second term in \eqref{eq:stress-tensor_form_factor_COM} gives a non-zero contribution. More precisely
\begin{equation}
\label{eq:projected_FF}
\begin{aligned}
\mathbf{\Pi}_0\mathcal{F}^{\mu\nu}_T(p_1^\text{com},p_2^\text{com}) &=
\mathcal{F}_\Theta(s)\times \frac{\omega}{d-1}\,\Pi_1^{\mu\nu}(p^\text{com}),\\
\mathbf{\Pi}_2\mathcal{F}^{\mu\nu}_T(p_1^\text{com},p_2^\text{com}) &=
\mathcal{F}_{(2)}(s)
\times 
\sqrt{\frac{2\,\omega^2}{(d-2)(d+1)}}\,\Pi_2^{\mu\nu;11}(p^\text{com}),
\end{aligned}
\end{equation}

\subsection{Unitarity constraints}
\label{sec:unitarity_constraints}

Having set up all the necessary ingredients, let us finally address unitarity. We start by taking all possible inner products of the states \eqref{eq:state_12}.  Skipping the detail, which were explained in section 3 in \cite{Karateev:2019ymz}, we arrive at the following matrix
\begin{equation}
\label{eq:positivity_S}
\begin{pmatrix}
1 & \mathcal{S}^*_j(s)\\
\mathcal{S}_j(s) & 1
\end{pmatrix}\succeq 0,\quad
\forall j=0,2,4\ldots\quad\text{and}\quad\forall s\geq 4m^2
\end{equation}
which must be semi-positive definite in unitary theories according to the discussion of appendix \ref{app:unitarity_appendix}. Here $\mathcal{S}_j(s)$ is the partial amplitude related to the full scattering amplitude $S(s,t,u)$ (the amplitude containing the disconnected piece) as
\begin{equation}
\mathcal{S}_j(s) = \frac{\kappa_j}{\gamma_j \Omega_{d-2}} \mathbf{\Pi}_j S(s,t(s,\cos\theta_1),u(s,\cos\theta_1)),
\end{equation}
where the Mandelstam variables can be explicitly expressed in terms of the scattering angle $\theta_1$ as follows
\begin{equation}
\label{eq:mandelstam_variables_angle}
t = - \frac{s-4m^2}{2}\,(1-\cos\theta_1),\quad
u = - \frac{s-4m^2}{2}\,(1+\cos\theta_1).
\end{equation}
The coefficient $\kappa_j$ was computed in equation (2.41) in \cite{Karateev:2019ymz}. It reads
\begin{equation}
\kappa_j\equiv \frac{j!}{(d-3)\Omega_{d-1}\mathcal{N}_d\Gamma(d-3+j)}.
\end{equation}
In order to proceed we also need to consider the inner product of the state \eqref{eq:T_state} with itself
\begin{equation}
\begin{aligned}
{}^{\mu\nu}\<\psi_3|\psi_3\>^{\rho\sigma} &=
m^{-d}
\int d^d x
\int d^d y
e^{-ip'\cdot x}
e^{+ip\cdot y}
\<0|T^{\mu\nu}(x)T^{\rho\sigma}(y)|0\>_W\\
&=(2\pi)^d\delta^d(p'-p)\times 2\pi\rho_{T}^{\mu\nu;\,\rho\sigma}(p).
\end{aligned}
\end{equation}
Here we used \eqref{eq:trans_invariance}, performed the change of variables and employed \eqref{eq:spectral_decomposition_T}.

Let us now consider the following three states
\begin{equation}
|\psi_1\>_{j=0},\qquad
|\psi_2\>_{j=0},\qquad
\eta_{\mu\nu}|\psi_3\>^{\mu\nu}.
\end{equation}
Taking all possible inner products of these states we obtain a 3x3 hermitian matrix which components were carefully derived in section 3 of \cite{Karateev:2019ymz}. Using the unitarity requirement, as explained in appendix \ref{app:unitarity_appendix},
we obtain the following semi-positive definite constraint
\begin{equation}
\label{eq:positivity_theta}
\begin{pmatrix}
1 & \mathcal{S}^*_0(s)
& \omega\,m^{-d/2} \mathcal{F}^*_{\Theta}(s)\\
\mathcal{S}_0(s) & 1
& \omega\,m^{-d/2} \mathcal{F}_{\Theta}(s)\\
\omega\,m^{-d/2} \mathcal{F}_{\Theta}(s) 
& \omega\,m^{-d/2} \mathcal{F}^*_{\Theta}(s) &
2\pi\,m^{-d}\rho_\Theta(s)
\end{pmatrix}\succeq 0.
\end{equation}
This condition should be satisfied for all the energies $s\geq 4m^2$.
We can also consider the following three states instead
\begin{equation}
|\psi_1\>_{j=0},\qquad
|\psi_2\>_{j=0},\qquad
X_{\mu\nu}\Pi_2^{\mu\nu}{}_{\rho\sigma}|\psi_3\>^{\rho\sigma},
\end{equation}
where we have defined
\begin{equation}
X^{\mu\nu}\equiv\sqrt\frac{d-1}{d-2}\frac{(p_1-p_2)^\mu(p_1-p_2)^\nu}{s-4m^2}.
\end{equation}
Taking all possible inner product of these states, removing the overall $\delta$-function and using \eqref{eq:projected_FF} we obtain the following semi-positive definite condition
\begin{equation}
\label{eq:positivity_2}
\begin{pmatrix}
1 & \mathcal{S}^*_2(s)
& f(\theta_1)\omega\,m^{-d/2} \mathcal{F}^*_{(2)}(s)\\
\mathcal{S}_2(s) & 1
& f(\theta_1)\omega\,m^{-d/2} \mathcal{F}_{(2)}(s)\\
f(\theta_1)\omega\,m^{-d/2} \mathcal{F}_{(2)}(s) 
& f(\theta_1)\omega\,m^{-d/2} \mathcal{F}^*_{(2)}(s) &
2\pi\,m^{-d}s^2\,\rho_{\hat T}^2(s)
\end{pmatrix}\succeq 0,
\end{equation}
where we have defined
\begin{equation}
f(\theta_1) \equiv \sqrt\frac{2}{d^2-1}\times
\frac{(d-1)\cos(2\theta_1)-1}{d-2}.
\end{equation}
The condition \eqref{eq:positivity_2} should be satisfies for all the energies $s\geq 4m^2$ and angles $\theta_1\in[0,\pi]$. Notice the appearance of the angle $\theta_1$ compared to the trace case. As we will see shortly, the strongest bounds come from $\theta_1=0$ configuration. Thus, the semi-positive constraint \eqref{eq:positivity_2} simply reduces to the following form
\begin{equation}
\label{eq:positivity_2_final}
\begin{pmatrix}
1 & \mathcal{S}^*_2(s)
& \varepsilon\,m^{-d/2} \mathcal{F}^*_{(2)}(s)\\
\mathcal{S}_2(s) & 1
& \varepsilon\,m^{-d/2} \mathcal{F}_{(2)}(s)\\
\varepsilon\,m^{-d/2} \mathcal{F}_{(2)}(s) 
& \varepsilon\,m^{-d/2} \mathcal{F}^*_{(2)}(s) &
2\pi\,m^{-d}s^2\,\rho_{\hat T}^2(s)
\end{pmatrix}\succeq 0,
\end{equation}
where we have defined
\begin{equation}
\varepsilon \equiv \omega\,\sqrt\frac{2}{d^2-1}.
\end{equation}

The semi-definite positive conditions \eqref{eq:positivity_theta} and \eqref{eq:positivity_2_final} are the main results of this section.

\section*{Sylvester's criterion}
A semi-positive definite matrix $M\succeq 0$ has only non-negative eigenvalues. In order to address semi-positive constraints in practice we use the Sylvester's criterion: a matrix $M$ is semi-positive definite if and only if all its principal minors (including the determinant) are non-negative.

Let us now analyze the constraints \eqref{eq:positivity_S}, \eqref{eq:positivity_theta} and \eqref{eq:positivity_2} using the Sylvester's criterion. First of all one recovers the standard unitary constraint on the partial amplitudes
\begin{equation}
\label{eq:part_constraints}
|S_j(s)|^2 \leq 1,\qquad \forall j=0,2,4,\ldots
\end{equation}
Here and below all the inequalities are given in the physical domain of squared energies $s\geq 4m^2$.
Second we recover the non-negativity of the components of the stress-tensor spectral densities
\begin{equation}
\rho_\Theta(s)\geq 0,\qquad
\rho_{\hat T}^2(s)\geq 0
\end{equation}
already derived in section \ref{sec:wightman}. Third, we derive the inequalities
\begin{equation}
\label{eq:ff_spectral_inequalities}
|\mathcal{F}_\Theta(s)|^2 \leq 2\pi \omega^{-2}\rho_\Theta(s),\qquad
|\mathcal{F}_{(2)}(s)|^2 \leq 2\pi \omega^{-2}f^{-2}(\theta_1)s^2\,\rho_{\hat T}^2(s).
\end{equation}
We notice that the strongest constraint in \eqref{eq:ff_spectral_inequalities} comes from $\theta_1=0$ (or equivalently $\theta_1=\pi$) configuration since the function $f^{-2}(\theta_1)$ has its minimum there. These bounds are in a perfect agreement with \eqref{eq:components_spectral_ff}.

Finally the determinants ot the matrices \eqref{eq:positivity_theta} and \eqref{eq:positivity_2} lead to the following set of constraints
\begin{equation}
\label{eq:det_constraints}
\begin{aligned}
2\pi \omega^{-2}\rho_\Theta(s) \left(1-|\mathcal{S}_0(s)|^2\right)
&-2|\mathcal{F}_\Theta(s)|^2 +
\mathcal{F}_\Theta^{*2}(s)\mathcal{S}_0(s) +
\mathcal{F}_\Theta^2(s)\mathcal{S}_0^*(s)\geq 0,\\
2\pi \omega^{-2}f^{-2}(\theta_1)s^2\,\rho_{\hat T}^2(s) \left(1-|\mathcal{S}_2(s)|^2\right)
&-2|\mathcal{F}_{(2)}(s)|^2 +
\mathcal{F}_{(2)}^{*2}(s)\mathcal{S}_2(s) +
\mathcal{F}_{(2)}^2(s)\mathcal{S}_2^*(s)\geq 0.
\end{aligned}
\end{equation}
We notice now that the first term in both equalities is non-negative. Thus, the strongest bound happens at the minimum of the function $f^{-2}(\theta_1)$ which is at $\theta_1=0$, since it becomes harder to compensate for the negative second term and for the potentially negative third and fourth terms.

\section*{Elastic unitarity}
In the special region of energies
\begin{equation}
\label{eq:elastic_range}
s\in[4m^2,\,9m^2],
\end{equation}
called the elastic regime, the inequality \eqref{eq:part_constraints} become saturated,\footnote{In $d=2$ this can also happen for $s>9m^2$ in the case of integrable models.} namely
\begin{equation}
|S_j(s)|^2 = 1,\qquad \forall j=0,2,4,\ldots
\end{equation}
Using this fact we can rewrite the equations \eqref{eq:det_constraints} as
\begin{equation}
\label{eq:watson}
\mathcal{F}_\Theta^*(s)\mathcal{S}_0(s)=
\mathcal{F}_\Theta(s),\qquad
\mathcal{F}_{(2)}^*(s)\mathcal{S}_2(s)=
\mathcal{F}_{(2)}(s).
\end{equation}
These are known as the Watson's equations. They allow to express the partial amplitudes in terms of the components of the (two-particle) stress-tensor form factor in the ``elastic'' range of energies \eqref{eq:elastic_range}.

\section*{Asymptotic behavior}
Let us now study the inequalities \eqref{eq:ff_spectral_inequalities} in the $s\rightarrow \infty$ limit. Using \eqref{eq:limit_Tr} and \eqref{eq:asymptotics_T2hat} together with \eqref{eq:omega} and \eqref{eq:Nd} we obtain
\begin{align}
\label{eq:inequality_1}
\lim_{s\rightarrow \infty}
|s^{\frac{d-\Delta_\cO}{2}-1}\mathcal{F}_\Theta(s)|^2 &\leq
2^d(2\pi)^{d-2}\Omega_{d-1}^{-1}\times g^2\mathcal{N}_\cO \kappa(d,\Delta_\cO),\\
\label{eq:inequality_2}
\lim_{s\rightarrow \infty}
|s^{-1}\mathcal{F}_{(2)}(s)|^2 &\leq
2^d(2\pi)^{d-2}\Omega_{d-1}^{-1}f^{-2}(0)\times
C_T^{UV}\frac{d-1}{d+1}\kappa(d,d).
\end{align}
We remind that the trace of the stress tensor at high energy is given by the relevant scalar operator $\cO$ (with the scaling dimension $\Delta_\cO$ and the two-point normalization $\mathcal{N}_\cO$) which deforms the UV CFT and $g$ is the dimensionful coupling governing the deformation. The numerical constant $\kappa$ was defined in \eqref{eq:kappa}. In 2d the inequality \eqref{eq:inequality_1} was first derived in \cite{Delfino:2003yr}, see formulas (3.33) and (3.34).

It is interesting to notice that even if one constructs a scattering amplitude such that all its partial amplitudes obey the unitarity condition \eqref{eq:part_constraints} at all the energies, it is not clear if one can read off any UV CFT data from it.\footnote{Notice however that in $d\ge 3$ using holography one can argue that the regime of hard scattering (high energy and fixed angle) should be directly related to the UV CFT \cite{Polchinski:2001tt}.} The conditions \eqref{eq:det_constraints} in the limit $s \rightarrow \infty$ together with \eqref{eq:limit_Tr}, \eqref{eq:asymptotics_T2hat} and \eqref{eq:inequality_1}, \eqref{eq:inequality_2} could in principle provide this connection. Under closer investigation it does not seem however that one can draw from them any generic statements.

\subsection{Bootstrap problems}
\label{sec:bootstrap_problems}
One can use the semi-positive definite constraints \eqref{eq:positivity_S}, \eqref{eq:positivity_theta} and \eqref{eq:positivity_2_final} to define several bootstrap problems.\footnote{Notice that these constraints are already written in the form which is straightforward to implement into the semi-definite problem solver SDPB \cite{Simmons-Duffin:2015qma,Landry:2019qug}.} There are at least two distinct possibilities.

Let us start with the first one.
The constraint \eqref{eq:positivity_S} allows to bound various non-perturbative  S-matrix coupling constants using the numerical procedure of \cite{Paulos:2016but, Paulos:2017fhb}, see also section 1 of \cite{Hebbar:2020ukp} for a concise summary. One can now re-run this procedure in the presence of \eqref{eq:positivity_theta} and \eqref{eq:positivity_2} where we inject some known numerical\footnote{One could obtain some numerical data using Hamiltonian truncation methods, see for instance \cite{Anand:2020gnn,Anand:2020qnp,EliasMiro:2020uvk}.} data about the stress-tensor form factors and the spectral density. This provides a more restrictive setup and injects model specific information in the numerical procedure.

The second possibility in $d\geq 3$ is to apply the numerical procedure \cite{Paulos:2016but, Paulos:2017fhb} to \eqref{eq:positivity_S}, \eqref{eq:positivity_theta} and \eqref{eq:positivity_2_final}, writing an ansatz for the components of the stress-tensor form factor and the spectral density, in order to minimize the following quantities
\begin{equation}
\label{eq:chi}
\psi_m \equiv \int_{4m^2}^{\infty} ds\;s^{-m}\rho_{\Theta}(s)
\quad\text{or}\quad
\chi_n \equiv \int_{4m^2}^{\infty} ds\;s^{-n}\rho_{\hat T}^2(s),
\end{equation}
where $m$ and $n$ are some real parameters. Their allowed range is constrained by the convergence of the integral at large values of $s$ due to the asymptotic behaviour \eqref{eq:asymptotics_Theta} and  \eqref{eq:asymptotics_T2hat} of the components of the stress-tensor spectral density.  For example one concludes that $n>d/2-1$. The only disadvantage of this procedure is that the quantities in \eqref{eq:chi} do not have any clear physical meaning. Notice that the non-trivial result of such a minimization procedure is guaranteed by the presence of the form factor normalization conditions \eqref{eq:normalization_2FF_2d}.

In $d=2$ instead of \eqref{eq:chi} we can minimize the UV central charge $c_{UV}$ given by the integral expression \eqref{eq:sum_rule_T_2d_main_text_kkbar}. This was already employed in \cite{Karateev:2019ymz}. In the presence of global symmetries one can also minimize the conserved current central charge $k_{UV}$ given by the integral expression \eqref{eq:sum_rule_J_2d_main_text_kkbar}. This for example can be employed to study further the $O(N)$ models in $d=2$.

\section*{Acknowledgements}
I am grateful to Jo\~ao Penedones for the initial collaboration on the project and for the numerous discussions later on.
I also thank Liam Fitzpatrick, Mark Gillioz, Andrei Khmelnitsky, Alexander Monin, Riccardo Rattazzi, Sylvain Ribault, Slava Rychkov, and Matt Walters for useful discussions.

The work of DK is supported by the Simons Foundation grant 488649 (Simons Collaboration on the Nonperturbative Bootstrap) 
and by the Swiss National Science Foundation through the project
200021-169132 and through the National Centre of Competence in Research SwissMAP.

\appendix

\section{Correlation functions in Euclidean CFTs}
\label{app:2pt_CFTs}
The conformal group in $d$-dimensional Euclidean space is $SO(1,d+1)$. We will consider local operators with spin $\ell$, namely the ones transforming in the traceless-symmetric representation of the $SO(d)$ subgroup.\footnote{In $d=3$ all the bosonic representations are traceless-symmetric. In $d\geq 4$ even bosonic representation can be non-traceless symmetric.} Such operators can be encoded in the following index-free objects
\begin{equation}
\label{eq:encoding}
\cO(x,z)\equiv \cO(x)_{a_1\ldots a_\ell} z^{a_1}\ldots z^{a_\ell},
\end{equation}
where $z^a$ are real vectors called polarizations. One can invert \eqref{eq:encoding} as
\begin{equation}
\label{eq:Index_full}
\cO(x)_{a_1\ldots a_\ell} = \frac{1}{\ell!\,(d/2-1)_\ell}\,D_{a_1}\ldots D_{a_\ell}\, \cO(x,z),
\end{equation}
where $x_\ell\equiv x(x+1)\ldots(x+\ell-1)$ is the Pochhammer symbol and $D_a$ is the Todorov differential operator defined as
\begin{equation}
\label{eq:todorov}
D^a\equiv \left(d/2-1+z\cdot \partial_z\right)\partial_z^a
-\frac{1}{2}\,z^a\,\partial^2_z.
\end{equation}
The Todorov operator is strictly defined for $d\geq 3$. One can still use it in $d=2$ by keeping $d$ generic and taking the limit $d\rightarrow 2$ in the very end of the computation. 

The conformal group can be realized linearly in $D\equiv d+2$ dimensional embedding space. Using the formalism developed in \cite{Costa:2011mg} one can represent the traceless-symmetric local operator \eqref{eq:encoding} as a function of $D$-dimensional light-cone coordinates $X^A\equiv \{x^a,\;X^+,\; X^-\}$ and polarizations $Z^A\equiv \{z^a,\;Z^+,\;Z^- \}$.\footnote{In order to work with more general representations other embedding formalisms are required. For general representations (bosonic and fermionic) in $d=3$ see \cite{Iliesiu:2015qra}. For general representations in $d=4$ see \cite{SimmonsDuffin:2012uy,Elkhidir:2014woa,Cuomo:2017wme}. For general bosonic operators in $d\geq 4$ see \cite{Costa:2014rya}.} The metric in the light-cone coordinates is\footnote{\label{foot:cartesian}The Cartesian coordinates in $D$-dimensions read as $X^2=X^aX^a-(X^{d})^2+(X^{d+1})^2$. The light-cone coordinates are then defined as $X^+\equiv X^{d}+X^{d+1}$ and $X^-\equiv X^{d}-X^{d+1}$.}
\begin{equation}
\eta^{AB}=\begin{pmatrix}
\delta^{ab} & 0 & 0\\ 
0 & 0 & -2\\
0 & -2 & 0\\

\end{pmatrix},\qquad
\eta_{AB}=\begin{pmatrix}
\delta_{ab} & 0 & 0 \\
0 & 0 & -\frac{1}{2} \\
0 & -\frac{1}{2} & 0\\ 
\end{pmatrix}.
\end{equation}
The map between the embedding space and the original space is given by
\begin{align}
\label{eq:map}
\cO(x,z) = \cO\left(
X^A\rightarrow \{x^a,\,1,\,x^2\},\; Z^A \rightarrow \{z^a,\,0,\,2x\cdot z \}
\right).
\end{align}

It is straightforward to construct $n$-point functions in the embedding formalism. They read
\begin{equation}
\<0|\cO_1(X_1,Z_1)\ldots \cO_n(X_n,Z_n)|0\> = \sum_I g_I(u,v,\ldots)T_I(X_i,Z_i),
\end{equation}
where $g_I$ are some undetermined functions of the conformally invariant variables $u,v,\ldots$ also known as the cross-ratios and $T_I$ are the tensor structures. For $n=2$ and $n=3$ there are no cross-ratios, thus the functions $g_I$ can only be constants. Tensor structures are built as products of the following conformally invariant objects
\begin{equation}
\label{eq:invariants_1}
\begin{aligned}
X_{ij} &\equiv -2 (X_i\cdot X_j),\\
H_{ij} &\equiv -2 \left((Z_i\cdot Z_j)(X_i\cdot X_j)-(Z_i\cdot X_j)(X_i\cdot Z_j)\right),\\
V_{k,ij} &\equiv X_{ij}^{-1}\left((Z_k\cdot X_i)(X_k\cdot X_j)-(Z_k\cdot X_j)(X_k\cdot X_i)\right).
\end{aligned}
\end{equation}
These are parity even objects. One can also construct various parity odd conformally invariant objects which contain a single $D$-dimensional Levi-Civita symbol. The number and the structure of such objects depend on the number of dimensions. For instance in $d=2$ ($D=4$) and for $n=2$ one can write a single object
\begin{equation}
\label{eq:invariant_odd}
F_{ij} \equiv -2\,\epsilon_{ABCD} X_i^A X_j^B Z_i^C Z_j^D,
\end{equation}
where the Levi-Civita symbol in $D=4$ Lorentzian space with one time direction is  $\epsilon^{0123}=-\epsilon_{0123}=+1$.
Here the indices $A, B=0,1,\ldots,d+1$ denote the Cartesian and not the light-cone embedding coordinates as in the footnote \ref{foot:cartesian}. The powers of $X_{ij}$ are fixed by the homogeneity requirement $\cO(\lambda X,Z) = \lambda^{-\Delta_\cO}\cO(X,Z)$, where $\Delta_\cO$ is the usual scaling dimension of the operator $\cO$.

Using the map \eqref{eq:map} we can write the projection of the invariants to the original $d$-dimensional space. One gets
\begin{equation}
\label{eq:projection_1}
\begin{aligned}
X_{ij} & \rightarrow  x_{ij}^2,\\
H_{ij} & \rightarrow  x_{ij}^2 \left(z_i\cdot z_j-2\frac{(z_i\cdot x_{ij})(z_j\cdot x_{ij})}{x_{ij}^2}\right),\\
V_{k,ij} & \rightarrow  \frac{x_{ki}^2 x_{kj}^2}{x_{ij}^2}\times \left(
\frac{z_k \cdot x_{ki}}{x_{ki}^2}-\frac{z_k \cdot x_{kj}}{x_{kj}^2}
\right).
\end{aligned}
\end{equation}
Analogously for the parity odd invariant \eqref{eq:invariant_odd} we have
\begin{equation}
\label{eq:projection_2}
F_{ij} \rightarrow 
(x_i^2-x_j^2)(\epsilon_{ab}\,z_{i}^a z_j^b)
-2\,(x_i\cdot z_i)(\epsilon_{ab}\,x_{ij}^a z_j^b)
-2\,(x_j\cdot z_j)(\epsilon_{ab}\,x_{ji}^a z_i^b),
\end{equation}
where the Levi-Civita symbol in Euclidean $d=2$ space is $\epsilon^{01}=\epsilon_{01}=+1$.

\subsection*{Examples}
As the first application consider the two-point functions of Abelian conserved currents
\begin{equation}
\<0|J(X_1,Z_1) J(X_2,Z_2)|0\> = 
\frac{C_J H_{12}+i\,\delta_{d,2}C^\prime_J F_{12}}{X_{12}^{d}},
\end{equation}
where $C_J$ and $C^\prime_J$ are some constants undetermined by the conformal symmetry. They are called the current central charges. The imaginary unit $i$ was introduced in the second term for the later convenience.
Using the projections \eqref{eq:projection_1}, \eqref{eq:projection_2} and the Todorov operator \eqref{eq:todorov} we get the following indexful expression for the Euclidean two-point function
\begin{equation}
\label{eq:2pt_J_cft_full}
\<0|J^a(x_1) J^b(x_2)|0\>_E = 
\frac{C_J}{(x_{12}^2)^{d-1}}\times
\mathcal{I}^{ab}(x_1,x_2)+
\frac{i\,C^\prime_J}{(x_{12}^2)^{d-1}}\times\delta_{d,2}\,\mathcal{E}^{ab}(x_1,x_2),
\end{equation}
where we have introduced the auxiliary objects
\begin{align}
\label{eq:obj_IE}
\mathcal{I}^{ab}(x_i,x_j) \equiv \delta^{ab} - 2\,\frac{x_{ij}^a x_{ij}^b}{x_{ij}^2},\qquad
\mathcal{E}^{ab}(x_i,x_j) \equiv \epsilon^{ab} +
2 \frac{ x_{ij}^a\epsilon^{bc} x_{ij}^c}{x_{ij}^2}.
\end{align}
Notice that both of these objects are translation invariant as they should be. Moreover they are also invariant under the the transformation $a\leftrightarrow b$ and $x_i\leftrightarrow x_j$.\footnote{This is obvious for $\mathcal{I}^{ab}$. In order to show it for $\mathcal{E}^{ab}$ one needs to use \eqref{eq:condition_2d}.}
It is straightforward to check that the two-point functions \eqref{eq:2pt_J_cft_full} is automatically conserved.

As will be discussed in appendix \ref{app:reflection_positivity}, Euclidean two-point functions in unitary theories must obey reflection positivity. For Lorentz spin one current this condition is given in \eqref{eq:reflection_positivity_current}. Plugging \eqref{eq:2pt_J_cft_full} into \eqref{eq:reflection_positivity_current} and using \eqref{eq:objIE_refpos} we get
\begin{equation}
\label{eq:hermitian_J}
\begin{pmatrix}
C_J & -i\,C^\prime_J \\
i\,C^\prime_J & \;\;C_J
\end{pmatrix}\succeq 0.
\end{equation}
The semi-positive condition \eqref{eq:hermitian_J} can be satisfied only if the matrix in \eqref{eq:hermitian_J} is hermitian. As a result both $C_J$ and $C^\prime_J$ must be real. Furthermore using the Sylvester's criterion, the condition \eqref{eq:hermitian_J} leads to the following constraints
\begin{equation}
\label{eq:conditions_CJ}
C_J\geq 0,\quad
-C_J\leq C^\prime_J \leq +C_J.
\end{equation}

As the second example let us consider the two-point function of the stress-tensor
\begin{equation}
\<0|T(X_1,Z_1) T(X_2,Z_2)|0\> = \frac{C_T H_{12}^2+i\,\delta_{d,2}C^\prime_T H_{12}F_{12}}{X_{12}^{d+1}},
\end{equation}
where as before $C_T$ andc $C^\prime_T$ are some constants undetermined by the conformal symmetry coefficients referred to as the central charges.
Again using the projections \eqref{eq:projection_1}, \eqref{eq:projection_2} and the Todorov operator \eqref{eq:todorov} we get the following indexful expression for the Euclidean two-point function
\begin{align}
\nn
\<0|T^{ab}(x_1)T^{cd}(x_2)|0\>_E
&=\frac{C_T}{x_{12}^{2d}}\times\left( \frac{1}{2}\left(\mathcal{I}^{ac}(x_1,x_2)\mathcal{I}^{bd}(x_1,x_2)+
\mathcal{I}^{ad}(x_1,x_2)\mathcal{I}^{bc}(x_1,x_2)\right)
-\frac{1}{d}\,\delta^{ab}\delta^{cd}\right)\\
\nn
&+\frac{i\,C^\prime_T}{x_{12}^{2d}}\times\frac{\delta_{d,2}}{4}\,\bigg(
\mathcal{I}^{ac}(x_1,x_2)\mathcal{E}^{bd}(x_1,x_2)+
\mathcal{I}^{ad}(x_1,x_2)\mathcal{E}^{bc}(x_1,x_2)\\
&+\mathcal{I}^{bc}(x_1,x_2)\mathcal{E}^{ad}(x_1,x_2)+
\mathcal{I}^{bd}(x_1,x_2)\mathcal{E}^{ac}(x_1,x_2)
\bigg).
\label{eq:2pt_T_cft_full}
\end{align}
This expression is automatically conserved. As in the case of conserved currents reflection-positivity imposes constraints on the central charges $C_T$ and $C^\prime_T$. Plugging \eqref{eq:2pt_T_cft_full} into \eqref{eq:condition_T_refpos} and using \eqref{eq:objIE_refpos} we get that both must be real and obey the following inequalities
\begin{equation}
\label{eq:conditions_CT}
C_T\geq 0,\quad
-C_T\leq C^\prime_T \leq +C_T.
\end{equation}

Notice the presence of parity odd terms in $d=2$ both in \eqref{eq:2pt_J_cft_full} and \eqref{eq:2pt_T_cft_full}. No such terms can be constructed in $d\geq 3$. In general it can be shown that two-point functions of local primary operators transforming in the irreducible Lorentz representation have a single tensor structure, see for example \cite{Kravchuk:2016qvl}. In $d=2$ both $J^a$ and $T^{ab}$ transform however in a reducible representation of the Lorentz group.\footnote{In $d=2$ the Lorentz group is $SO(2)$ with $U(1)$ being its double cover. As a result the $SO(2)$ representations can also be labeled by the $U(1)$ charges. For instance the spin one $SO(2)$ representation is the direct sum of $\pm 1$ $U(1)$ charges. Analogously the spin two representation is the direct sum of $\pm 2$ $U(1)$ charges.} As we will shortly see they can be split into irreducible representations which have a single tensor structure in their two-point functions.

\subsection*{Conventions in d=2}
Let us summarize now the standard $d=2$ notation. One defines the complex coordinates
\begin{equation}
\label{eq:2d_coordinates}
z\equiv x^1+i x^2,\quad
\bar z\equiv x_1-i x_2.
\end{equation}

In theses coordinates one can define the following components of the spin one Lorentz operators
\begin{equation}
\label{eq:J_holomorphic}
\begin{aligned}
J(z,\bar z) &\equiv (2\pi )\times J^z(z,\bar z) =
\pi \times \left(J^1(x) - i J^2(x) \right),\\
\overline J(z,\bar z) &\equiv (2\pi)\times J^{\bar z}(z,\bar z) =
\pi \times \left(J^1(x) + iJ^2(x) \right).
\end{aligned}
\end{equation}
Conservation implies
\begin{equation}
\partial_{\bar z} J(z,\bar z) +
\partial_{z} \overline J(z,\bar z) = 0.
\end{equation}
Using \eqref{eq:2pt_J_cft_full} and \eqref{eq:J_holomorphic} one simply gets
\begin{equation}
\langle 0 | J(z) J(0) |0\rangle = -\frac{k}{z^2},\quad
\langle 0 | J(z) \overline J(0) |0\rangle = 0,\quad
\langle 0 | \overline J(\bar z) \overline J(0) |0\rangle = -\frac{\bar k}{\bar z^2},
\end{equation}
where the coefficients $k$ and $\bar k$ read as
\begin{equation}
\label{eq:relation_J}
k\equiv (2\pi)^2 \times \frac{C_J+C^\prime_J}{2},\qquad
\bar k\equiv (2\pi)^2 \times \frac{C_J-C^\prime_J}{2}.
\end{equation}
From \eqref{eq:conditions_CJ} it follows that $k\geq 0$ and $\bar k\geq 0$.
As an example one can use free theory of a massless Dirac fermions which has $k=\bar k$. Using \eqref{eq:relation_J} these values can be read off from equation (5.6) in \cite{Osborn:1993cr}.\footnote{Notice that $C_V$ in \cite{Osborn:1993cr} is identical to our $C_J$. The values of $C_J$ depends on the normalization of the $U(1)$symmetry generators.}

Analogously for the stress-tensor we can define the following components
\begin{equation}
\label{eq:rel_2d}
\begin{aligned}
T(z,\bar z) &\equiv (2\pi)\times T_{zz}(z,\bar z)=(2\pi)\times \frac{1}{4}\,\left(T_{11}(x)-T_{22}(x)-2i\, T_{12}(x)\right),\\
\Theta(z,\bar z) &\equiv 4 T_{z\bar z}(z,\bar z)=T_{11}(x)+T_{22}(x),\\
\overline T(z,\bar z) &\equiv (2\pi)\times T_{\bar z \bar z}(z,\bar z)=(2\pi)\times \frac{1}{4}\,\left(T_{11}(x)-T_{22}(x)+2i\, T_{12}(x)\right).
\end{aligned}
\end{equation}
Conservation implies
\begin{equation}
\partial_{\bar z} T(z,\bar z)+\frac{\pi}{2}\,\partial_{z}\Theta(z,\bar z)= \partial_z \overline T(z,\bar z) +\frac{\pi}{2}\,\partial_{\bar z}\Theta(z,\bar z) = 0.
\end{equation}
The conformal invariance implies
\begin{equation}
\Theta(z,\bar z)=0.
\end{equation}
Using \eqref{eq:2pt_T_cft_full} and \eqref{eq:rel_2d} one simply gets
\begin{equation}
\langle 0 | T(z) T(0) |0\rangle = \frac{c/2}{z^4},\quad
\langle 0 | T(z) \overline T(0) |0\rangle = 0,\quad
\langle 0 | \overline T(\bar z) \overline T(0) |0\rangle = \frac{\bar c/2}{\bar z^4},
\end{equation}
where the coefficients $c$ and $\bar c$ read as
\begin{equation}
\label{eq:relation_c_conventions}
c\equiv (2\pi)^2\times\frac{C_T + C^\prime_T}{2},\qquad
\bar c\equiv (2\pi)^2\times\frac{C_T - C^\prime_T}{2}.
\end{equation}
From \eqref{eq:conditions_CT} we conclude that $c>0$ and $\bar c>0$.
In a free theory of a single real scalar and also in a free theory of a single Dirac fermion $c=\bar c = 1$, see equations (5.5) and (5.6) in \cite{Osborn:1993cr}. For an alternative derivation see also appendix C of \cite{Karateev:2019ymz}.

\section{Unitarity}
\label{app:unitarity_appendix}
Unitary quantum field theories are defined to have non-negative norms of all its states. Consider some state $|\psi\>$. The unitarity then requires
\begin{equation}
\label{eq:unitarity_simple}
\<\psi | \psi\> \geq 0.
\end{equation} 
In a more complicated situation when we have $N$ states $|\psi\>^I$ with the label $I=1,\ldots N$, the above condition becomes the semi-positive requirement on the $N\times N$ hermitian matrix
\begin{equation}
\label{eq:unitarity}
{}^I\<\psi | \psi\>^{J} \succeq 0.
\end{equation} 
In what follows we will use \eqref{eq:unitarity_simple} and \eqref{eq:unitarity} to derive some concrete constraints on two-point functions. We will give the discussion in the Euclidean and in the Lorentzian signature separately.

\subsection{Implications in Euclidean signature}
\label{app:reflection_positivity}
We start with the Euclidean signature. As indicated in the main text we pick the first coordinate and assign to it the role of Euclidean time
\begin{equation}
t_E\equiv x^0,\quad
\vec x \equiv \{x^1,\ldots , x^{d-1}\}.
\end{equation}
The hermitian conjugation of local operators contrary to the Lorentzian signature is very non-trivial in the Euclidean signature. With the choice of the Euclidean time made above the hermitian conjugation of a generic real\footnote{One defines real and complex operators in the Lorentzian signature and then analytically continues to the Euclidean signature. See appendix \ref{app:unitarity_lorentzian} for some details.} operator with spin reads as\footnote{For the derivation of \eqref{eq:conjugation} see section 7.1 of \cite{Simmons-Duffin:2016gjk}.}
\begin{equation}
\label{eq:conjugation}
\left(\cO^{a b \ldots}(t_E, \vec x) \right)^\dagger = 
\kappa^{a a'}\kappa^{b b'}\ldots\;\cO^{a' b' \ldots}(-t_E, \vec x),
\end{equation}
where the prefactors $\kappa$ are defined as
\begin{equation}
\kappa^{a a'} \equiv \delta^{a a'} - 2 \delta^{a 0} \delta^{a' 0}.
\end{equation}

Let us give special names for the following coordinates
\begin{equation}
x_+^a\equiv \{t_E,\vec x \},\qquad
x_-^a\equiv \{-t_E,\vec x \}.
\end{equation}
We then choose the following state
\begin{equation}
\label{eq:state}
|\psi\>^{I} \equiv \cO^{ab\ldots}(x_-) |0\>,
\end{equation}
where $I$ is a collective index for the indices $a,b,\ldots$
Using \eqref{eq:conjugation}, the condition \eqref{eq:unitarity} becomes the following property of the two-point function\footnote{\label{foot:smearing}The choice of the state \eqref{eq:state} is very particular. More generally one should define a state by smearing the operator with some ``test'' function. By changing the test function one changes the state. As a result one gets an infinite number of smeared constraints \eqref{eq:reflection_positivity}.}
\begin{equation}
\label{eq:reflection_positivity}
\kappa^{a a'}\kappa^{b b'}\ldots\;
\< 0 | \cO^{a'b'\ldots}(x_+) \cO^{cd\ldots}(x_-) |0\>_E \succeq 0,
\end{equation}
where the semi-poisitivity is imposed on the square matrix $(ab\ldots)\times(cd\ldots)$. The property \eqref{eq:reflection_positivity} is called reflection-positivity of the two-point function.

To be concrete let us consider two simple examples. First, if we deal with a scalar operator the condition \eqref{eq:reflection_positivity} simply reads
\begin{equation}
\label{eq:reflection_positivity_scalar}
\< 0 | \cO(x_+) \cO(x_-) |0\>_E \geq 0.
\end{equation}
Second, if the operator is a vector, the condition \eqref{eq:reflection_positivity} reads as
\begin{equation}
\label{eq:reflection_positivity_current}
\begin{pmatrix}
\;-\< 0 | J^0(x_+) J^0(x_-) |0\>_E &
\;-\< 0 | J^0(x_+) \vec J(x_-) |0\>_E\\
+\< 0 | \vec J(x_+) J^0(x_-) |0\>_E & 
+\< 0 | \vec J(x_+) \vec J(x_-) |0\>_E
\end{pmatrix}
\succeq 0.
\end{equation}
As an application consider the parity even part of the two-point function \eqref{eq:2pt_J_conformal}, We have then
\begin{align}
\<0|J^a(x_+) J^b(x_-)|0\>_E &= 
\frac{1}{(2t_E)^{2(d-1)}}\times
\left(
h_1(r) \delta^{ab} + h_2(r)\delta^{a0}\delta^{b0}
\right).
\end{align}
Plugging this expression into \eqref{eq:reflection_positivity_current} and using the Sylvester's criterion for semi-positive definiteness of a real matrix we get the following conditions
\begin{equation}
\forall r:\quad
h_1(r)\geq 0,\quad
h_1(r)+h_2(r)\leq 0.
\end{equation}
When we are concerned with two-point functions in conformal field theories it is convenient to write explicitly their tensor structures in a ``reflection-positive'' frame
\begin{equation}
\label{eq:objIE_refpos}
\mathcal{I}^{ab}(x_+, x_-)=\delta^{ab}-2\delta^{a0}\delta^{b0},\qquad
\mathcal{E}^{ab}(x_+, x_-)=
\begin{pmatrix}
0 & -1 \\
-1 & 0
\end{pmatrix}.
\end{equation}

Finally consider the case of the stress-tensor. The reflection-positivity condition \eqref{eq:reflection_positivity} becomes a $4\times 4$ block matrix spanned by the collective indixed $I$ and $J$, where $I\equiv ab = \{00,0i,j0,ij\}$ and $J\equiv cd = \{00,0k,l0,kl\}$ and $i,j,k,l=1,\ldots,d-1$. In order to write a compact formula we also define
\begin{equation}
\label{eq:definition}
\frac{1}{r^{2d}}\times K^{ab;\,cd}\equiv \< 0 | T^{ab}_E(x_+) J^{cd}_E(x_-) |0\>.
\end{equation}
Then the condition \eqref{eq:reflection_positivity} in terms of \eqref{eq:definition} reads
\begin{equation}
\label{eq:condition_T_refpos}
\begin{pmatrix}
+K^{00;\,00} &
+K^{00;\,0l} &
+K^{00;\,k0} &
+K^{00;\,kl} \\
-K^{0i;\,00} &
-K^{0i;\,0l} &
-K^{0i;\,k0} &
-K^{0i;\,kl} \\
-K^{j0;\,00} &
-K^{j0;\,0l} &
-K^{j0;\,k0} &
-K^{j0;\,kl} \\
+K^{ij;\,00} &
+K^{ij;\,0l} &
+K^{ij;\,k0} &
+K^{ij;\,kl}
\end{pmatrix}\succeq 0.
\end{equation}

\subsection{Implications in Lorentzian signature}
\label{app:unitarity_lorentzian}
Consider the Lorentzian space. We denote the Lorentzian time and the spacial coordinates in the following way
\begin{equation}
t_L\equiv x^0,\quad
\vec x \equiv \{x^1,\ldots , x^{d-1}\}.
\end{equation}
Consider now some real local operator with spin. The hermitian conjugation has a very straightforward action on such an operator in the Lorentzian space. It reads
\begin{equation}
\label{eq:hermitian_conjugation}
\left( \cO^{\mu_1\mu_2\ldots}(x) \right)^\dagger =
\cO^{\mu_1\mu_2\ldots}(x^*).
\end{equation}
The coordinates $x^\mu$ are mostly real, however we often include a small imaginary part in the time component in order to regularize two-point functions, see appendix \ref{app:wightman}. This is the reason why we kept $x^{*}$ in the right-hand side of \eqref{eq:hermitian_conjugation}.

Consider now the following state\footnote{The same comment as in the footnote \ref{foot:smearing} applies here.}
\begin{equation}
|\psi\> \equiv \cO(\hat x^*)|0\>,
\end{equation}
where $\cO$ is some real scalar local operator and as in section \ref{sec:spectral_densities} we have defined
\begin{equation}
\hat x^\mu \equiv \{x^0-i\epsilon,\vec x\},\quad
\epsilon>0.
\end{equation}
Unitarity condition \eqref{eq:unitarity_simple} together with \eqref{eq:hermitian_conjugation} then implies the following condition on the ordered two-point function of the local operator $\cO$
\begin{equation}
\label{eq:positivity_scalar}
\<0|\cO(\hat x)
\cO(\hat x^*)|0\> \geq 0.
\end{equation}
Similarly for the Lorentz spin one operator we can construct the following  states
\begin{equation}
|\psi\>^\mu \equiv J^\mu(\hat x^*)|0\>.
\end{equation}
Unitarity condition \eqref{eq:unitarity} together with \eqref{eq:hermitian_conjugation} imply then the following constraint on the ordered two-point function
\begin{align}
\label{eq:positivity_currents}
\<0| J^\mu(\hat x) J^\nu(\hat x^*)|0\> \succeq 0.
\end{align}

It is also important to note that the  reality condition \eqref{eq:hermitian_conjugation} poses further constraints on ordered two-point functions of real operators. Consider for example the case of conserved currents. One has
\begin{equation}
\label{eq:condition}
\<0|J^\mu(x_1)J^\nu(x_2)|0\>^*=\<0|\left(J^\mu(x_1)J^\nu(x_2)\right)^\dagger|0\>=
\<0|J^\nu(x_2^*)J^\mu(x_1^*)|0\>.
\end{equation}

As an example let us consider the ordered two-point function of conserved currents in Lorentzian conformal field theory. One has
\begin{equation}
\label{eq:2pt_J_cft_full_W}
\<0|J^\mu(x_1) J^\nu(x_2)|0\> = 
\frac{C_J}{(x_{12}^2)^{d-1}}\times
\mathcal{I}^{\mu\nu}(x_1,x_2)-
\frac{C^\prime_J}{(x_{12}^2)^{d-1}}\times\delta_{d,2}\,\mathcal{E}^{\mu\nu}(x_1, x_2),
\end{equation}
where we have defined
\begin{align}
\label{eq:obj_IE_W}
\mathcal{I}^{\mu\nu}(x_i,x_j) \equiv \eta^{\mu\nu} - 2\,\frac{x_{ij}^\mu x_{ij}^\nu}{x_{ij}^2},\qquad
\mathcal{E}^{\mu\nu}(x_i,x_j) \equiv \epsilon^{\mu\nu} +
2 \frac{ x_{ij}^\mu\epsilon^{\nu\rho} x_{ij}^\rho}{x_{ij}^2}.
\end{align}
The Levi-Civita symbol obeys $\epsilon^{01}=-\epsilon_{01}=+1$.
The expression \eqref{eq:2pt_J_cft_full_W} can either be derived from scratch adapting appendix \ref{app:2pt_CFTs} to Lorentzian signature or can be simply translated from the Euclidean expression \eqref{eq:2pt_J_cft_full} using \eqref{eq:J_connection}.
Plugging \eqref{eq:2pt_J_cft_full_W} in \eqref{eq:condition} and taking into account the following properties
\begin{equation}
\mathcal{I}^{\mu\nu}(x_1,x_2)=+\mathcal{I}^{\nu\mu}(x_2,x_1),\quad
\mathcal{E}^{\mu\nu}(x_1,x_2)=+\mathcal{I}^{\nu\mu}(x_2,x_1),
\end{equation}
which simply follow from the definitions \eqref{eq:obj_IE_W}, one concludes that $C_J$ and $C^\prime_J$ are purely real.
Plugging \eqref{eq:2pt_J_cft_full_W} into \eqref{eq:positivity_currents} we obtain the condition
\begin{equation}
\label{eq:positivity}
\begin{pmatrix}
 C_J & C^\prime_J \\
 C^\prime_J & C_J
\end{pmatrix}\succeq 0,
\end{equation}
where we have used
\begin{equation}
\mathcal{I}^{\mu\nu}(\hat x, \hat x^*)=\eta^{\mu\nu}+2\delta^\mu_0\delta^\nu_0,\qquad
\mathcal{E}^{\mu\nu}(\hat x, \hat x^*)=\begin{pmatrix}
0 & -1 \\
-1 & 0
\end{pmatrix}.
\end{equation}
From \eqref{eq:positivity} we get the following conditions
\begin{equation}
C_J\geq 0,\quad
-C_J\leq C^\prime_J \leq +C_J,
\end{equation}
which are identical to the ones obtained in the Euclidean metric and given in \eqref{eq:conditions_CJ}.

\section{Euclidean vs. Lorentzian operators}
\label{app:wightman}
Here we will discuss Euclidean and Lorentzian correlators. We then provide a formal way to define the latter as various analytic continuations of the former.
Part of the discussion here is based on section 7 and appendix B of \cite{Simmons-Duffin:2016gjk}.


\paragraph{Euclidean correlators}
In the Euclidean space two- (and higher-) point correlation functions are computed using the path integral approach. They are denoted by
\begin{equation}
\label{eq:stat_sum}
\< \cO_{E}(x_E) \cO_{E}(y_E)\>.
\end{equation}
We introduced the subscript $E$ for both the coordinates and the operators in order to emphasise that we work in the Euclidean metric. By construction the correlation function \eqref{eq:stat_sum} is time-ordered with respect to Euclidean time, namely
\begin{equation}
\label{eq:property}
\< \cO_{E}(x_E) \cO_{E}(y_E)\> = \< \cO_{E}(y_E) \cO_{E}(x_E)\>.
\end{equation}

We can also reinterpret the correlator \eqref{eq:stat_sum} as the vacuum expectation value of local operators in some Hilbert space.\footnote{One can think about states and operators as vectors and matrices in the infinite-dimensional space. The vacuum state is the state with the lowest energy.} This is done as follows. The vacuum expectation value of two local operators is denote by
\begin{equation}
\label{eq:2pt_Euclidean}
\<0| \cO_{E}(x_E) \cO_{E}(y_E) |0\>.
\end{equation}
The order of operators in this expression is important.
The correlator \eqref{eq:2pt_Euclidean} makes sense only if $x_E^0>y_E^0$. This is easy to see by rewriting \eqref{eq:2pt_Euclidean} in the following way
\begin{equation}
\label{eq:2pt_Euclidean_modified}
\<0| \cO_{E}(x_E) \cO_{E}(y_E) |0\> =
\<0| \cO_{E}(0) e^{-H (x_E^0-y_E^0)}\cO_{E}(0) |0\>,
\end{equation}
where $H$ is the Hamiltonian of the system. Here we simply used the translation invariance
\begin{equation}
\cO_E(x_E) = e^{+P\cdot x_E} \cO_E(0) e^{-P\cdot x_E},
\end{equation}
where $P^a$ are the generators of translations and $H\equiv P^0$. The Hamiltonian $H$ is an infinite-dimensional matrix with non-negative eigenvalues. In other words the eigenvalues of $H$ are bounded from below. The operator $e^{-H (x_E^0-y_E^0)}$ has all finite eigenvalues only if $x_E^0>y_E^0$. It becomes unbounded from above if $x_E^0<y_E^0$. In this case \eqref{eq:2pt_Euclidean_modified} formally diverges. The only option to avoid this and to define the two-point correlation function for any values of $x_E$ and $y_E$ is as follows
\begin{multline}
\label{eq:euclidean_tordered}
\<0| \cO_{E}(x_E) \cO_{E}(y_E) |0\>_E\equiv
\theta(x_E^0-y_E^0)\<0| \cO_{E}(x_E) \cO_{E}(y_E) |0\>+\\
\theta(y_E^0-x_E^0)\<0| \cO_{E}(y_E) \cO_{E}(x_E) |0\>
\end{multline}
By construction this is the time-ordered (with respect to Euclidean time) correlation function. We refer to it as the Euclidean correlator. The equivalence between the path integral formulation \eqref{eq:stat_sum} and the operator formulation \eqref{eq:euclidean_tordered} leads to 
\begin{equation}
\< \cO_{E}(x_E) \cO_{E}(y_E)\> =
\<0| \cO_{E}(x_E) \cO_{E}(y_E) |0\>_E.
\end{equation}

\paragraph{Lorentzian correlators}
Let us now consider the vacuum expectation value of the local operators in the Lorentzian signature
\begin{equation}
\label{eq:vev_Lorentzian}
\<0| \cO_{L}(x_L) \cO_{L}(y_L) |0\>.
\end{equation}
This quantity is not well-defined since it generically contains poles when $(x_L-y_L)^2=0$ at $x_L^\mu \neq y_L^\mu$. In order to define the above vacuum expectation value correctly one should specify how to deal with these poles. In practice we allow for a small imaginary part for the Lorentzian time and then send it to zero. Several different ways (prescriptions) exist. They define different types of correlators, namely Wightman, time-ordered (Feynman), advanced and retarded correlators. 
For instance the Wightman function is defined as follows
\begin{equation}
\label{eq:wightman_def_scalar}
\<0| \cO_{L}(x_L) \cO_{L}(y_L) |0\>_W \equiv\\
\lim_{\epsilon_1\rightarrow 0}\lim_{\epsilon_2\rightarrow 0}
\<0| \cO_{L}(x^0_L-i\epsilon_1,\vec x) \cO_{L}(y^0_L-i\epsilon_2,\vec x_2) |0\>,
\end{equation}
where $\epsilon_1>\epsilon_2$. This is the simplest possible prescription. One can use now the Wightman correlator to define all the other types of correlators (instead of going though various prescriptions). For example the time-ordered (Feynman) correlator is defined as
\begin{multline}
\label{eq:defynman_def_scalar_from_w}
\<0| \cO_{L}(x_L) \cO_{L}(y_L) |0\>_F \equiv
\theta(x^0_L-y^0_L) \<0| \cO_{L}(x_L) \cO_{L}(y_L) |0\>_W +\\
\theta(y_L-x_L) \<0| \cO_{L}(y_L) \cO_{L}(x_L) |0\>_W.
\end{multline}

By definition the Wightman two-point function \eqref{eq:wightman_def_scalar} is a distribution. When integrated with a test-function the $i\epsilon$ prescription leads to an unambiguous result. The time-ordered two-point function \eqref{eq:defynman_def_scalar_from_w} instead is not a distribution due to the presence of the step functions.

\paragraph{Relation between Euclidean and Lorentzian correlators}
We can formalise the discussion of Lorentzian two-point functions by defining them as various analytic continuations of the Euclidean two-point function in time. Let us denote the Euclidean time by $t_E\equiv x_E^0$ and the Lorentzian time by $t_L\equiv x_L^0$.

Let us start from the following Euclidean correlator
\begin{equation}
\<0| \cO_{E}(\epsilon_1,\vec x) \cO_{E}(\epsilon_2,\vec x_2) |0\>_E,
\end{equation}
where $\epsilon_1$ and $\epsilon_2$ are real Euclidean times which obey $\epsilon_1>\epsilon_2$. For $\epsilon_1>\epsilon_2$ only the first term in \eqref{eq:euclidean_tordered} survives and we can thus drop the subscript $E$. We then analytically continue this object to complex times and send both $\epsilon_1$ and $\epsilon_2$ to zero. The resulting object formally defines the Wightman two-point function, namely
\begin{align}
\nn
\<0| \cO_{L}(x^0_L,\, \vec x) \cO_{L}(y^0_L,\, \vec y) |0\>_W &\equiv
\lim_{\epsilon_1\rightarrow 0}\lim_{\epsilon_2\rightarrow 0}
\<0| \cO_{E}(\epsilon_1+ix^0_L,\vec x) \cO_{E}(\epsilon_2+iy^0_L,\vec x_2) |0\>\\
&=
\lim_{\epsilon_1\rightarrow 0}\lim_{\epsilon_2\rightarrow 0}
\<0| \cO_{E}\left(i(x^0_L-i\epsilon_1),\,\vec x\right) \cO_{E}\left(i(y^0_L-i\epsilon_2),\,\vec x_2\right) |0\>\nn\\
&=
\lim_{\epsilon_1\rightarrow 0}\lim_{\epsilon_2\rightarrow 0}
\<0| \cO_{L}(x^0_L-i\epsilon_1,\,\vec x) \cO_{L}(y^0_L-i\epsilon_2,\,\vec x_2) |0\>.
\label{eq:wightman_def}
\end{align}
Here we have decided to relate the Euclidean and Lorentzian times as 
\begin{equation}
\label{eq:times}
t_E=+i t_L,\qquad
t_L=-i t_E
\end{equation}
The equality between the last two entries in \eqref{eq:wightman_def} lead to the formal relation between the Euclidean and Lorentian scalar local operators
\begin{equation}
\label{eq:scalar_connection}
\cO_E(t_E,\vec x) = \cO_L(-i t_E,\vec x),\qquad
\cO_L(t_L,\vec x) = \cO_E(it_L,\vec x).
\end{equation}
For local operators with spin the relation between Euclidean and Lorentzian operators is more complicated. For instance for the vector operators we have
\begin{equation}
\label{eq:J_connection}
J^0_E(t_E,\vec x) = -iJ^0_L(-i t_E,\vec x),\qquad
\vec J_E(t_E,\vec x) = \vec J_L(-i t_E,\vec x).
\end{equation}

The analytic continuation which follows the path
\begin{equation}
\label{eq:approx_wick}
z \rightarrow z'\equiv z e^{i(\pi/2-\epsilon)}\approx z(i+\epsilon)
\end{equation}
is known as the Wick rotation. Without loss of generality let us set $y=0$ by using translation invariance.
Using the Wick rotation one can define the (Feynman) time-ordered two-point function as
\begin{align}
\label{eq:feynman_def_scalar}
\<0| \cO_{L}(x^0_L,\, \vec x) \cO_{L}(0) |0\>_F &\equiv
\lim_{\epsilon\rightarrow 0}
\<0| \cO_{E}(x^0_L e^{i(\pi/2-\epsilon_1)},\,\vec x)\; \cO_{E}(0) |0\>_E\\
&= \lim_{\epsilon\rightarrow 0}
\<0| \cO_{E}(x^0_L (i+\epsilon),\,\vec x)\; \cO_{E}(0) |0\>_E.
\label{eq:feynman_def_scalar_2}
\end{align}
Using the definition of the Euclidean propagator in the right-hand side of \eqref{eq:feynman_def_scalar_2} the last equality can be written as
\begin{equation}
\theta(+x^0_L)
\<0| \cO_{E}(ix^0_L+\epsilon,\,\vec x)\; \cO_{E}(0) |0\>_E +
\theta(-x^0_L)
\<0| \cO_{E}(ix^0_L-\epsilon,\,\vec x)\; \cO_{E}(0) |0\>_E,
\end{equation}
where we have used the fact that $\epsilon\, t_L\approx +\epsilon$ if $t_L>0$ and $\epsilon\, t_L\approx -\epsilon$ if $t_L<0$.
Using the first line of \eqref{eq:wightman_def} one sees the equivalence between the above expression and \eqref{eq:defynman_def_scalar_from_w}.

\section{K\"{a}ll\'en-Lehmann representation in Euclidean signature}
\label{app:KL}
The K\"{a}ll\'en-Lehmann representation of the Lorentzian time-ordered two-point functions was derived in section \ref{sec:time-ordered}. In this section we would like to translate those result to the Euclidean signature. For that we apply the following change of variables
\begin{equation}
x_L^0 \rightarrow -ix^0_E,\qquad
p_L^0 \rightarrow -ip^0_E.
\end{equation}
Which is in agreement with \eqref{eq:times}.

Let us start with the K\"{a}ll\'en-Lehmann representation for the scalar operators given by  \eqref{eq:KL_representation_scalar}. Performing the above change of variables we get
\begin{equation}
\label{eq:Euclidean_propagator}
\begin{aligned}
\<0| \cO(x) \cO(0)|0\>_E &= \int_0^\infty ds \rho_\cO(s) \Delta_E(x;s),\\
\Delta_E(x;s) 
 &= \lim_{\epsilon\rightarrow 0^+} \int \frac{d^dp_E}{(2\pi)^d}e^{ip_E\cdot x_E} \frac{1}{p_E^2+s}.
\end{aligned}
\end{equation}
See appendix B of \cite{Karateev:2019ymz} for some additional details. Notice the absence of the $i\epsilon$ since there are no poles for real $p^2$ to be regularized.

The K\"{a}ll\'en-Lehmann for the conserved currents and the stress-tensor were derived in \eqref{eq:spectral_J} and \eqref{eq:spectral_TT}.  Analogously to the scalar case one gets
\begin{align}
\label{eq:euclidean_J}
\<0| J^a(x) J^b(0)|0\>_T &= \int_0^\infty ds \rho_J^1(s) \Delta_{E,\,1}^{ab}(x;s),\\
\label{eq:euclidean_T}
\<0| T^{ab}(x) T^{cd}(0)|0\>_T &= \int_0^\infty ds 
\left(
\frac{\rho_{\Theta}(s)}{(d-1)^2}\,\Delta_{E,\Theta}^{ab;\,cd}(x;s)
+\rho_{\hat T}^2(s) \Delta_{E,\,2}^{ab;\,cd}(x;s)
\right),
\end{align}
where the Euclidean propagators $\Delta_E$ are obtained from \eqref{eq:feynman_propagator_J} and \eqref{eq:feynman_propagator_T}. The Euclidean correlator for the conserved current is obtained from \eqref{eq:feynman_propagator_J}, it reads
\begin{equation}
\label{eq:feynman_propagator_J_Euclidean}
\Delta_{E,\,1}^{ab}(x;s) = (s\,\delta^{ab}-\partial^a\partial^b)
\Delta_E(x;s)
\end{equation}
and the Euclidean Feynman propagators for the stress-tensor read as
\begin{equation}
\label{eq:feynman_propagator_T_Euclidean}
\begin{aligned}
\Delta_{E,\Theta}^{ab;cd}(x;s) &=
(s\,\delta^{ab}-\partial^a\partial^b)
(s\,\delta^{cd}-\partial^c\partial^d)
\Delta_E(x;s),\\
\Delta^{ab;\,cd}_{E,\,2}(x;s) &= s^2\,
\Pi^{ab;\,cd}_2(\partial;s)
\Delta_{E}(x;s),
\end{aligned}
\end{equation}
where we have defined the Euclidean projectors in the coordinate space
\begin{align}
\Pi_1^{ab}(\partial;s) &\equiv \delta^{ab}-s^{-1}\partial^a \partial^b,\\
\Pi_2^{ab;\,cd}(\partial;s) &\equiv 
-\frac{1}{d-1}\Pi_{1}^{ab}(\partial;s)   \Pi_{1}^{cd}(\partial;s)
+\frac{1}{2}\,\Pi_{1}^{ac}(\partial;s)  \Pi_{1}^{bd}(\partial;s)
+\frac{1}{2}\,\Pi_{1}^{ad}(\partial;s)\Pi_{1}^{bc}(\partial;s).
\nn
\end{align}

\section{Scalar propagators}
\label{app:propagators}

In this appendix we compute the explicit form of the scalar Euclidean, Wightman, Feynman, retarded and advanced propagators.  In what follows we completely ignore the case of coincident points, in other words $x\neq 0$.

\subsection*{Euclidean propagator}
Let us start with the scalar Euclidean propagator in position space. It was defined in \eqref{eq:Euclidean_propagator}. Let us repeat its definition here again for convenience
\begin{equation}
\label{eq:feynman_scalar}
\Delta_E(x;s)\equiv \int \frac{d^dp}{(2\pi)^d}\frac{e^{ip\cdot x}}{p^2+s}.
\end{equation}
We emphasise that in the Euclidean signature $x^2>0$ and $p^2\geq 0$.
Switching to spherical coordinates we get
\begin{align}
\nn
\Delta_E(x;s) &= \frac{\Omega_{d-1}}{(2\pi)^d}\int_0^\infty \frac{p^{d-1}dp}{p^2+s}\int_{-1}^{+1} d\xi (1-\xi^2)^{\frac{d-3}{2}}\,e^{ipr\xi}\\
\label{eq:result_scalar_feynman}
&=\frac{1}{(2\pi)^{d/2}}\int_0^\infty \frac{p^{d-1}dp}{p^2+s}\;
\frac{J_{\frac{d-2}{2}}(pr)}{(pr)^{\frac{d-2}{2}}}\\
&=\frac{1}{(2\pi)^{d/2}}\left(\frac{\sqrt s}{r}\right)^{\frac{d-2}{2}}
\int_0^\infty \frac{k^{\frac{d}{2}}dk}{k^2+1}
J_{\frac{d-2}{2}}(rk\sqrt s).
\nn
\end{align}
In the second line we have defined $r\equiv\sqrt{x^2}$ and used the result 3.387 2. from \cite{Gradshteyn:1702455}. The symbol $J_n(x)$ stands for the Bessel function of the first kind. The last integral in \eqref{eq:result_scalar_feynman} is given by 6.565 4. in \cite{Gradshteyn:1702455}. It  converges only for $d<5$. Performing it we obtain the final expression
\begin{equation}
\label{eq:final_scalar_feynman}
\Delta_E(x;s)=\frac{s^{\frac{d-2}{2}}}{(2\pi)^{d/2}}\times
\frac{K_{\frac{d-2}{2}}(\sqrt{s x^2})}{\left(\sqrt{s x^2}\right)^{\frac{d-2}{2}}},
\end{equation}
where $K_n(x)$ denotes the Bessel function of the second kind. The result \eqref{eq:final_scalar_feynman} is analytic in $d$. Thus, even though the integral in \eqref{eq:result_scalar_feynman} does not converge, we can define it via the analytic continuation of \eqref{eq:final_scalar_feynman}.

\subsection*{Wightman propagator}

The Wightman propagator is defined in \eqref{eq:spectral_representation_scalar_2}. We remind that $s>0$. Integrating over the $\delta$-function we get
\begin{align}
\label{eq:scalar_wightman_start}
\Delta_W(x;s) &= \lim_{\epsilon\rightarrow 0^+}\int \frac{d^{d-1}\vec k}{(2\pi)^{d-1}}\;
\frac{e^{-i\hat x^0\sqrt{s+\vec k^2}}e^{i\vec k\cdot \vec x}}{2\sqrt{s+\vec k^2}},
\end{align}
where as usual time $x^0$ has a small imaginary part according to
\begin{equation}
\hat x^0 \equiv x^0 - i\epsilon.
\end{equation}
To perform the rest of integrals in \eqref{eq:scalar_wightman_start} we split the discussion into two distinct situations: when $x^2>0$ (space-like separation) and when $x^2<0$ (time-like separation). As indicated in the beginning of the section we will completely exclude the light-like separation $x^2=0$ from the discussion.

Let us consider first the space-like separation of points. In this situation we can perform the Lorentz transformation to set $x^0=0$. We can then perform the integral \eqref{eq:scalar_wightman_start} as follows
\begin{equation}
\label{eq:space-like-derivation}
\begin{aligned}
\Delta_W(x;s) &= 
\frac{\Omega_{d-2}}{2\,(2\pi)^{d-1}}
\int_0^\infty\frac{dk\,k^{d-2}}{\sqrt{s+k^2}}
\int_{-1}^{+1}d\eta\, (1-\eta^2)^\frac{d-4}{2}e^{i k \chi \eta}\\
&= \sqrt{\frac{\pi}{2}}\;(2\pi)^{-d/2}
\int_0^\infty \frac{dk\,k^{d-2}}{\sqrt{s+k^2}}
\frac{J_\frac{d-3}{2}(k \chi)}{(k\chi)^\frac{d-3}{2}}\\
&=(2\pi)^{-d/2}\,s^\frac{d-2}{2}\frac{K_\frac{d-2}{2}(\chi\sqrt{s})}{(\chi\sqrt{s})^\frac{d-2}{2}}.
\end{aligned}
\end{equation}
In the first line of \eqref{eq:space-like-derivation} we have switched to the spherical coordinates in $d-1$ dimensions and defined $k\equiv|\vec k|$ together with $\chi\equiv |\vec x|$. We performed the integration over $d-3$ angles to get the spherical angle $\Omega_{d-2}$, where
\begin{equation}
\label{eq:spherical_angle}
\Omega_n \equiv \frac{n\, \pi^{n/2}}{\Gamma(n/2+1)}.
\end{equation}
The variable $\eta$ reflects the integration over the last remaining angle. For details see formulas (A.3) - (A.6) in \cite{Karateev:2019ymz}.
In the second line of \eqref{eq:space-like-derivation} we use the result 3.387 2. from \cite{Gradshteyn:1702455}. The function $J_n(x)$ stands for the Bessel function of the first kind. Finally in the third line of \eqref{eq:space-like-derivation} we use formula 6.564 1. from \cite{Gradshteyn:1702455}. The function $K_n$ is called the Bessel function of the second kind.

Let us now consider the time-like separation of points, namely when $x^2<0$. In this situation we can perform the Lorentz transformation to set $\vec x=0$. We can then perform the integral \eqref{eq:scalar_wightman_start} as follows
\begin{equation}
\label{eq:time-like-derivation}
\begin{aligned}
\Delta_W(x;s) &= \frac{\Omega_{d-1}}{2\,(2\pi)^{d-1}}\lim_{\epsilon\rightarrow 0^+}\int_0^\infty dk\,k^{d-2}
\frac{e^{-i\hat x^0\sqrt{s+k^2}}}{\sqrt{s+k^2}},\\
&=\frac{\Omega_{d-1}\,s^\frac{d-2}{2}}{2\,(2\pi)^{d-1}}\lim_{\epsilon\rightarrow 0^+}\int_1^\infty d\xi\,(\xi^2-1)^\frac{d-3}{2}
\,e^{-i\hat x^0\sqrt{s}\,\xi}\\
&=\frac{i\pi}{2}(2\pi)^{-d/2}s^\frac{d-2}{2}\lim_{\epsilon\rightarrow 0^+}
\frac{H^{(1)}_{\frac{2-d}{2}}(i\epsilon-x^0\sqrt{s})}{(i\epsilon-x^0\sqrt{s})^\frac{d-2}{2}}.
\end{aligned}
\end{equation}
In the first line of \eqref{eq:time-like-derivation} we switch to the spherical coordinates in $d-1$ dimensions and defined $k\equiv |\vec k|$. 
In the second line of \eqref{eq:time-like-derivation} we performed yet another change of variables where $\xi \equiv \sqrt{1-s^{-1}k^2}$. Finally in the last line of \eqref{eq:time-like-derivation} we use formula 3.387 4. from \cite{Gradshteyn:1702455}. Notice that it was crucial to have a small imaginary part in order to define the integral properly. The function $H_n^{(1)}$ is called the Hankel function of the first kind. Notice also that the time component $x^0$ does not have a definite sign.

We still need to perform some work to bring the result \eqref{eq:time-like-derivation} to its final form. To do that we split \eqref{eq:time-like-derivation} into two parts: one with $x^0>0$ and one with $x^0<0$. Using properties of the Hankel function we get then
\begin{equation}
\label{eq:hankel}
\lim_{\epsilon\rightarrow 0^+}
\frac{H^{(1)}_{\frac{2-d}{2}}(i\epsilon-x^0\sqrt{s})}{(i\epsilon-x^0\sqrt{s})^\frac{d-2}{2}}=
-\theta(x^0)
\frac{H^{(2)}_{\frac{2-d}{2}}(x^0\sqrt{s})}{(x^0\sqrt{s})^\frac{d-2}{2}}+
\theta(-x^0)
\frac{H^{(1)}_{\frac{2-d}{2}}(-x^0\sqrt{s})}{(-x^0\sqrt{s})^\frac{d-2}{2}}.
\end{equation}
We can now perform the Lorentz transformation in order to write the expressions \eqref{eq:space-like-derivation} and \eqref{eq:time-like-derivation} in a generic frame. Effectively this is done by replacing $\chi \rightarrow \sqrt{x^2}$ and $x^0 \rightarrow \pm\sqrt{-x^2}$. Taking this and \eqref{eq:hankel} into account we arrive at the final expression for the scalar Wightman propagator
\begin{align}
\label{eq:scalar_wightman_propagator}
\Delta_W&(x;s) = (2\pi)^{-d/2}\,s^\frac{d-2}{2} \times\\
&\left(
\theta(+x^2)\,\frac{K_\frac{d-2}{2}(\sqrt{sx^2})}{(\sqrt{sx^2})^\frac{d-2}{2}}
-\frac{i\pi}{2}\theta(-x^2)
\frac{
\theta(+x^0)H^{(2)}_{\frac{2-d}{2}}(\sqrt{-sx^2})-
\theta(-x^0)H^{(1)}_{\frac{2-d}{2}}(\sqrt{-sx^2})
}{(\sqrt{-sx^2})^\frac{d-2}{2}}
\right).
\nn
\end{align}

\subsection*{Feynman propagator}
The scalar  Feynman propagator was defined in \eqref{eq:KL_representation_scalar}. Using \eqref{eq:scalar_wightman_propagator} we can get its explicit expression which reads as
\begin{align}
\label{eq:feynman_propagator_explicit}
-i\Delta_F(x;s) &\equiv \theta(x^0)\Delta_W(x;s) + \theta(-x^0)\Delta_W(-x;s)\\
&=(2\pi)^{-d/2}\,s^\frac{d-2}{2} \times
\Bigg(
\theta(+x^2)\,\frac{K_\frac{d-2}{2}(\sqrt{sx^2})}{(\sqrt{sx^2})^\frac{d-2}{2}}-\frac{i\pi}{2}\theta(-x^2)
\frac{H^{(2)}_{\frac{2-d}{2}}(\sqrt{-sx^2})}{(\sqrt{-sx^2})^\frac{d-2}{2}}
\Bigg).
\nn
\end{align}
For completeness let us also introduce the retarded and advanced propagators
\begin{equation}
\begin{aligned}
-i\Delta_R(x;s)\equiv +\theta(+x^0) D(x;s),\\
-i\Delta_A(x;s)\equiv -\theta(-x^0) D(x;s),
\end{aligned}
\end{equation}
where we have defined
\begin{equation}
\label{eq:D-object}
\begin{aligned}
D(x;s) &\equiv \Delta_W(x;s)-\Delta_W(-x;s)\\
&=-(2\pi)^{-d/2}\,s^\frac{d-2}{2}
\theta(-x^2)\times i\pi
\Big(\theta(x^0)-\theta(-x^0)\Big)\,
\frac{J_\frac{2-d}{2}(\sqrt{-sx^2})}{\sqrt{-sx^2}^\frac{d-2}{2}}.
\end{aligned}
\end{equation}
To obtain this result we have plugged \eqref{eq:scalar_wightman_propagator} into the first line of \eqref{eq:D-object} and rewrote the sum of two Hankel functions as a Bessel function of the first kind.

The results for Feynman, retarded and advance propagators are well known in $d=2$ and $d=4$ dimensions, see for example \cite{Johnston:2009fr} for the clean summary. The scalar Feynman propagator in general dimensions was also computed in \cite{Zhang:2008jy}, see formula (27). Our results match the ones present in the literature. In order to perform the comparison note the following properties of the special functions
\begin{equation}
\label{eq:relation_0}
d\geq 2:\quad
H^{(1)}_{\frac{2-d}{2}}(z) = (+i)^{d-2}H^{(1)}_{\frac{d-2}{2}}(z),\qquad
H^{(2)}_{\frac{2-d}{2}}(z) = (-i)^{d-2}H^{(2)}_{\frac{d-2}{2}}(z),
\end{equation}
together with
\begin{align}
d\geq 2\;\;\text{(even)}:\quad
&J_{\frac{2-d}{2}}(z) = i^{d-2}J_{\frac{d-2}{2}}(z),\\
d\geq 3\;\;\text{(odd)}:\quad
&J_{\frac{2-d}{2}}(z) = i^{d-1}Y_{\frac{d-2}{2}}(z),
\end{align}
where $Y_n$ is the Neumann function and $z$ is real.

\subsection*{Consistency check}
For the space-like separation of points $x^2>0 $ the Euclidean, the Wightman and the Feynman propagators are simply related as
\begin{equation}
\label{eq:space-like_relation}
x^2>0:\qquad
\Delta_E(x;s) = \Delta_W(x;s) = -i\Delta_F(x;s).
\end{equation}
For the time-like separation of points $x^2<0$ one should be able to obtain the expressions \eqref{eq:scalar_wightman_propagator} and \eqref{eq:feynman_propagator_explicit} using the analytic continuation of the Euclidean result. This re-derivation should be seen as the consistency check of the above computations. For simplicity let us work in the frame where $\vec x =0$, in other words $x^2=-t_L^2$, where $t_L$ is the Lorentzian time. 

Let us start from the Wightman propagator. Using the first line of \eqref{eq:wightman_def} we can write
\begin{equation}
\Delta_W(t_L;s) = \lim_{\epsilon\rightarrow 0} \Delta_E(\epsilon+i t_L;s).
\end{equation}
Using the explicit expression \eqref{eq:final_scalar_feynman} we get
\begin{equation}
\label{eq:euclidean_ac}
\Delta_W(t_L;s) = \frac{s^{\frac{d-2}{2}}}{(2\pi)^{d/2}}\times
\lim_{\epsilon\rightarrow 0}
\frac{K_{\frac{d-2}{2}}(\sqrt{s (\epsilon+i t_L)^2})}{\left(\sqrt{s (\epsilon+i t_L)^2}\right)^{\frac{d-2}{2}}}.
\end{equation}
By splitting this expression in two distinct cases of $t_L<0$ and $t_L>0$ we can take the limit explicitly
\begin{equation}
\label{eq:relation_1_wick}
\lim_{\epsilon\rightarrow 0}
\sqrt{s (\epsilon+i t_L)^2} =
\begin{cases}
+i \sqrt{s\, t_L^2},\quad t_L>0\\
-i \sqrt{s\, t_L^2},\quad t_L<0.
\end{cases}
\end{equation}
The Bessel functions of the second kind are related to the Hankel functions via the following relations
\begin{align}
\label{eq:relation_2_wick}
K_n(+i z) = - \frac{i \pi }{2} (-i)^n H_n^{(2)}(z),\qquad
K_n(-i z) = + \frac{i \pi }{2} (+i)^n H_n^{(1)}(z),
\end{align}
where $z>0$. The equivalence between the expressions \eqref{eq:euclidean_ac} and \eqref{eq:scalar_wightman_propagator} for $x^2<0$ becomes obvious once the relations \eqref{eq:relation_1_wick} and \eqref{eq:relation_2_wick} together with \eqref{eq:relation_0} are used.

Due to \eqref{eq:feynman_def_scalar}, \eqref{eq:feynman_scalar} \eqref{eq:KL_representation_scalar} the Feynman propagator can be obtained from the Euclidean one by means of the Wick rotation as
\begin{equation}
\label{eq:euclidean_wick}
-i\Delta_F(t_L;s) = \lim_{\epsilon\rightarrow 0} \Delta_E(t_L e^{i(\pi/2-\epsilon)};s).
\end{equation}
Using the explicit expression \eqref{eq:final_scalar_feynman} and \eqref{eq:approx_wick} we get
\begin{equation}
\label{eq:euclidean_ac_f}
-i\Delta_F(t_L;s) = \frac{s^{\frac{d-2}{2}}}{(2\pi)^{d/2}}\times
\lim_{\epsilon\rightarrow 0}
\frac{K_{\frac{d-2}{2}}(\sqrt{-s t_L^2+i\epsilon})}{\left(\sqrt{-s t_L^2+i\epsilon}\right)^{\frac{d-2}{2}}}.
\end{equation}
As before using the fact that $\lim_{\epsilon\rightarrow 0}\sqrt{-s t_L^2+i\epsilon}=+i\sqrt{s t_L^2}$ and the first entry in \eqref{eq:relation_2_wick} we conclude that
\begin{equation}
\label{eq:result_ac_ferynman}
-i\Delta_F(t_L;s) = 
\frac{s^{\frac{d-2}{2}}}{(2\pi)^{d/2}}\times
\left(- \frac{i \pi }{2}\right)
\frac{H^{(2)}_{\frac{2-d}{2}}(\sqrt{s t_L^2})}{\left(\sqrt{s t_L^2}\right)^{\frac{d-2}{2}}}.
\end{equation}
Here we have also used the second entry in \eqref{eq:relation_0}. This expression is equivalent to \eqref{eq:feynman_propagator_explicit} when $\vec x=0$ or in other words when $-x^2=t_L^2$.

\subsection*{Massless limit}
Let us study the massless limit $s\rightarrow 0$ of the explicit expressions for the Euclidean, Wightman and Feynman propagators given by \eqref{eq:final_scalar_feynman},  \eqref{eq:scalar_wightman_propagator} and \eqref{eq:feynman_propagator_explicit} respectively. We start with the Wightman propagator \eqref{eq:scalar_wightman_propagator}. Expanding it at $s=0$  to the leading order in $s$ we get
\begin{equation}
\label{eq:finiteness_wightman_propagator}
\begin{aligned}
d=2\quad\text{and}\quad x^2>0:\qquad
&\Delta_W(x;s) = -\frac{\log(x^2)}{4\pi} - \frac{2\gamma+\log(s/4)}{4\pi} + O(s),\\
d\geq 3\quad\text{and}\quad x^2>0:\qquad
&\Delta_W(x;s) = \frac{\Gamma(\frac{d-2}{2})}{4\pi^{d/2}(x^2)^\frac{d-2}{2}} + O(s).
\end{aligned}
\end{equation} 
In the first expression $\gamma$ is the Euler constant. We see that for $d\geq 3$ the scalar Wightman propagator is completely finite at $s=0$. In $d=2$ it has a divergent part. For the space-like separation of points $x^2>0$ due to the relation \eqref{eq:space-like_relation} the identical expression holds for the Euclidean and Feynman propagators.
For the time-like separation of points in $d\geq 3$ we get
\begin{equation}
\label{eq:finiteness_wightman_propagator}
\begin{aligned}
x^2<0:\qquad
&\;\;\;\;\;\Delta_W(x;s) = \frac{(-i)^{d-2}\Gamma(\frac{d-2}{2})}{4\pi^{d/2}(-x^2)^\frac{d-2}{2}}
\left(\theta(x^0)+(-1)^{d-2}\theta(-x^0)\right) + O(s),\\
x^2<0:\qquad
&-i\Delta_F(x;s) = \frac{(-i)^{d-2}\Gamma(\frac{d-2}{2})}{4\pi^{d/2}(-x^2)^\frac{d-2}{2}}+ O(s).
\end{aligned}
\end{equation}

Spinning Wightman propagators are obtained by taking derivatives with respect to coordinates. Thus, spinning propagators are finite at $s=0$ for $d\geq 2$. The identical conclusion holds for the time-like separation.

\section{Spectral densities and central charges: technical details}
\label{app:technical_dentails}

In section \ref{sec:two-point_functions} we have derived the sum rules for the central charges $C_J$ and $C_T$ in terms of the Euclidean two-point functions of the conserved current and the stress-tensor respectively. They are given in \eqref{eq:integral_J} and \eqref{eq:integral_expression_T_form2}. In this appendix we will derive their implications for the spectral densities.

We can plug the expressions \eqref{eq:euclidean_J} and \eqref{eq:euclidean_T} into the sum rules \eqref{eq:integral_J} and \eqref{eq:integral_expression_T_form2}, and obtain the desired relation between the central charges and spectral densities. The details of this manipulations are subtle. In what follows we carefully derive the result for conserved currents and then simply state the answer for the stress-tensor.

\subsection*{Conserved currents}
Using \eqref{eq:euclidean_J} and \eqref{eq:feynman_propagator_J_Euclidean} we can rewire \eqref{eq:integral_J} as
\begin{align}
\label{eq:intermediate_J}
C_J^{UV}-C_J^{IR} = \lim_{r_{min}\rightarrow 0}
\lim_{r_{max}\rightarrow \infty}\int_{r_{min}}^{r_{max}}
dr\,\int_0^\infty ds \rho_J^1(s)\,f_d(r;s), 
\end{align}
where $r\equiv \sqrt{x^2}$ and we have defined
\begin{align}
f_d(r;s)=r^{2d-3}\,(\delta^{ab}+(d-2)r^{-2}x^a x^b)
 (s\,\delta^{ab}-\partial^a\partial^b)
\Delta_E(x;s).
\end{align}
We can explicitly evaluate this function  by using \eqref{eq:final_scalar_feynman} and taking its derivatives. Performing the straightforward algebra we get
\begin{equation}
\label{eq:function_G}
f_d(r;s) = \frac{d-1}{(2\pi)^{d/2}}\times s^\frac{3-d}{2}(r\sqrt s)^{\frac{3(d-2)}{2}}\times\left(
(r\sqrt{s}) K_{\frac{d+2}{2}}(r\sqrt{s})
	-2(d-1)K_{\frac{d}{2}}(r\sqrt{s})\right).
\end{equation}
We can now plug the function \eqref{eq:function_G} into \eqref{eq:intermediate_J}, exchange the order of integrals and make a change of variables from $r$ to the dimensionless quantity $u\equiv r\sqrt{s}$. We get
\begin{equation}
\label{eq:sum_rule}
C_J^{UV}-C_J^{IR}= \frac{d-1}{(2\pi)^{d/2}}\lim_{r_{min}\rightarrow 0}
\lim_{r_{max}\rightarrow \infty}\int_0^\infty \frac{ds}{s^{d/2-1}}\rho_J^1(s)\,
\Big(F_d(r_{max}\sqrt{s})-F_d(r_{min}\sqrt{s})\Big),
\end{equation}
where we have defined the function $F_d$ as an indefinite (primitive) integral
\begin{equation}
F_d(u)\equiv\int
du\, u^{\frac{3(d-2)}{2}}
\left(u K_{\frac{d+2}{2}}(u)
-2(d-1)K_{\frac{d}{2}}(u)\right).
\end{equation}

In order to obtain the final version of the sum rule \eqref{eq:sum_rule} first, we need to know the limits of the $F_d(u)$ functions. One can find that (up to an irrelevant additive constant)
\begin{equation}
\label{eq:limits}
\lim_{u\rightarrow 0} F_2(u) = -1,\quad
\lim_{u\rightarrow 0} F_{d\geq3}(u) = 0,\quad
\lim_{u\rightarrow \infty} F_{d\geq2}(u) = 0.
\end{equation}
It is important to stress that we cannot simply permute the limits with the integration in \eqref{eq:sum_rule}. To see it consider for example the limit $r_{min}\rightarrow\infty$. Due to \eqref{eq:limits} the integrand $F_d(M r_{max})$ in \eqref{eq:sum_rule} vanishes. However, the integral contains the region of small values of $s$, where 
\begin{equation}
\lim_{r_{max}\rightarrow \infty} \lim_{s\rightarrow 0} F_d(r_{max}\sqrt{s})
\end{equation}
can give a finite contribution. We are thus required to split the integral over $s$ in three pieces
\begin{equation}
\label{eq:sum_rule_J_expanded}
\begin{aligned}
C_J^{UV}-C_J^{IR} =& \frac{d-1}{(2\pi)^{d/2}}\lim_{\Lambda_{min}\rightarrow 0}\lim_{\Lambda_{max}\rightarrow \infty}
\lim_{r_{min}\rightarrow 0}
\lim_{r_{max}\rightarrow \infty}\Bigg(\\
&\int_0^{\Lambda_{min}} \frac{ds}{s^{d/2-1}}\rho_J^1(s)\,
\Big(F_d(r_{max}\sqrt{s})-F_d(r_{min}\sqrt{s})\Big)\\
+&\int_{\Lambda_{min}}^{\Lambda_{max}} \frac{ds}{s^{d/2-1}}\rho_J^1(s)\,
\Big(F_d(r_{max}\sqrt{s})-F_d(r_{min}\sqrt{s})\Big)\\
+&\int_{\Lambda_{max}}^{\infty} \frac{ds}{s^{d/2-1}}\rho_J^1(s)\,
\Big(F_d(r_{max}\sqrt{s})-F_d(r_{min}\sqrt{s})\Big)\Bigg).
\end{aligned} 
\end{equation}
Here the cut-off parameters $\Lambda_{min}$ and $\Lambda_{max}$ can be chosen arbitrarily since nothing depends on them explicitly. In the end of the analysis we will take their values to be very small and very big respectively. This explains the order of limits in \eqref{eq:sum_rule_J_expanded}.

We take the limits in $r_{min}$ and $r_{max}$ under the integral in \eqref{eq:sum_rule_J_expanded} where it is allowed and use \eqref{eq:limits}. We also replace the spectral density $\rho_J^1$ by the CFT ones for small and large energies. We arrive at the following expression then 
\begin{equation}
\label{eq:sum_rule_J_spectral}
\begin{aligned}
C_J^{UV}-C_J^{IR} &= \frac{d-1}{(2\pi)^{d/2}}
\Bigg(\\
&+\lim_{\Lambda_{min}\rightarrow 0}
\lim_{r_{max}\rightarrow \infty}
\int_0^{\Lambda_{min}} \frac{ds}{s^{d/2-1}}\rho_J^{1,\;\text{IR CFT}}(s)\,
\Big(F_d(r_{max}\sqrt{s})+\delta_{d,2})\Big)\\
&+\lim_{\Lambda_{min}\rightarrow 0}\lim_{\Lambda_{max}\rightarrow \infty}
\int_{\Lambda_{min}}^{\Lambda_{max}} ds\,\rho_J^1(s)\,
\delta_{d,2}\\
&-\lim_{\Lambda_{min}\rightarrow 0}\lim_{r_{max}\rightarrow \infty}
\int_{\Lambda_{max}}^{\infty} \frac{ds}{s^{d/2-1}}\rho_J^{1,\;\text{UV CFT}}(s)\,
F_d(r_{min}\sqrt{s})\Bigg).
\end{aligned} 
\end{equation}
We can now plug the explicit expression of the spectral density \eqref{eq:J_spin1_cft_cons} and obtain the final result.
In $d=2$ both UV and IR CFT spectral densities vanish. As a result we are left only with the second term in \eqref{eq:sum_rule_J_spectral}. We thus get the final answer
\begin{equation}
\label{eq:sum_rule_J_2d}
d=2:\quad C_J^{UV}-C_J^{IR} = \frac{1}{2\pi}\int_{0}^\infty ds\,\rho_J^1(s).
\end{equation}
In $d\geq 3$ the second term in \eqref{eq:sum_rule_J_spectral} vanishes instead and we are left only with the first and the third ones. The relation \eqref{eq:sum_rule_J_spectral} then gives us the constraint on the IR and UV CFT spectral densities $\rho_J^{1,\;\text{IR CFT}}(s)$ and $\rho_J^{1,\;\text{UV CFT}}(s)$. The latter have already been computed in \eqref{eq:J_spin1_cft_cons}. We can check the consistency of our manipulations here by plugging  \eqref{eq:J_spin1_cft_cons} into \eqref{eq:sum_rule_J_spectral} and switching to the $u$ variable. As a result we get
\begin{equation}
\label{eq:dgeq4}
\begin{aligned}
C_J^{UV}-C_J^{IR}= \frac{2(d-2)\,\kappa(d,d-1)}{(2\pi)^{d/2+1}} &\times
\Big(
C_J^{IR}\lim_{\Lambda_{min}\rightarrow 0}\lim_{r_{max}\rightarrow \infty}
\int_0^{\Lambda_{min}r_{max}} \frac{du}{u}F_d(u)\\
&-C_J^{UV}\lim_{\Lambda_{max}\rightarrow \infty}\lim_{r_{min}\rightarrow 0}
\int_{\Lambda_{max}r_{min}}^\infty \frac{du}{u}F_d(u)
\Big).
\end{aligned}
\end{equation}
Taking the limits\footnote{Notice, that after taking the limits $r_{min}$ and $r_{max}$ nothing depends on $\Lambda_{min}$ and $\Lambda_{max}$.} we arrive at the following non-obvious equality
\begin{equation}
\label{eq:sum_rule_J}
C_J^{UV}-C_J^{IR}=  (C_J^{IR}-C_J^{UV})\times
\frac{2(d-2)\kappa(d,d-1)}{(2\pi)^{d/2+1}}
\int_0^\infty \frac{du}{u} F_d(u).
\end{equation}
We have checked numerically that the following relation indeed holds true
\begin{equation}
\frac{2(d-2)\kappa(d,d-1)}{(2\pi)^{d/2+1}}
\int_0^\infty \frac{du}{u} F_d(u) = -1.
\end{equation}

\subsection*{Stress-tensor}
Analogously for the stress-tensor let us plug the spectral decomposition \eqref{eq:euclidean_T} into the sum rule \eqref{eq:integral_expression_T_form2}. One gets
\begin{equation}
C_T^{UV} - C_T^{IR} = \lim_{r_{min}\rightarrow 0}
\lim_{r_{max}\rightarrow \infty}
\int_{r_{min}}^{r_{max}}
dr\, r^{2d-1}
\int_0^\infty ds
\left(\rho_\Theta(s) h_d(r;s)+\rho_{\hat T}^2(s)l_d(r;s)\right),
\end{equation}
where we have defined
\begin{equation}
\begin{aligned}
h_d(r;s) &= \frac{1}{2\,(d-1)^3}\,R^{abcd}(x)
(\delta^{ab}-s^{-1}\partial^a\partial^b)
(\delta^{cd}-s^{-1}\partial^c\partial^d)
\Delta_E(x;s),\\
l_d(r;s) &=\frac{1}{2\,(d-1)}\,
R^{abcd}(x)
\Pi^{ab;\,cd}_2(\partial;s)
\Delta_{E}(x;s).
\end{aligned}
\end{equation}
We also remind that the object $R^{abcd}(x)$ was defined in \eqref{eq:definition_R}.

Applying the logic identical to the case of conserved currents in $d=2$ we simply get
\begin{equation}
C_T^{UV} - C_T^{IR} = \frac{6}{\pi}\int_0^\infty \frac{ds}{s^2}\rho_\Theta(s).
\end{equation}
In $d\geq 3$ instead we have a condition on the asymptotic behavior of the $\rho_T^{2}(s)$ component of the spectral density  at large and small values of $s$. This condition is compatible with the result \eqref{eq:asymptotics_T2hat}.

\section{Form factor normalization}
\label{app:FF_normalization}
The stress-tensor defines the generators of translations as
\begin{equation}
\label{eq:poincare_generators}
P^\mu \equiv \int d^{d-1} x T^{0\mu}(x).
\end{equation}
Let us now evaluate the matrix element of $P^\mu$ with one-particle states. By convention the one-particle states obey the following normalization
\begin{equation}
\<m_1,\vec p_1 |m_2,\vec p_2\> =
2p_1^0\delta_{m_1 m_2}\times
(2\pi)^{d-1}
\delta^{(d-1)}(\vec p_1-\vec p_2).
\end{equation}
Since the one-particle states are the eigenstates of translations, for identical particles one gets
\begin{equation}
\label{eq:rel2}
\<m,\vec p_1| P^\mu |m,\vec p_2\> = 2p_1^0p_1^\mu \times(2\pi)^{d-1}\delta^{(d-1)}(\vec p_1-\vec p_2),
\end{equation}
where from the definition of one-particle states one has
\begin{equation}
\label{eq:mass_shell}
p_1^2=p_2^2=-m^2.
\end{equation}
On the other hand using \eqref{eq:poincare_generators} and \eqref{eq:trans_invariance} one can write
\begin{align}
\nn
\<m,\vec p_1| P^\mu |m,\vec p_2\>
&= \<m,\vec p_1|T^{0\mu}(0)|m,\vec p_2\> \times \int_{-\infty}^{+\infty} d^{d-1}x\, e^{i(p_2-p_1)\cdot x} \\
&= \<m,\vec p_1|T^{0\mu}(0)|m,\vec p_2\>\times(2\pi)^{d-1}\delta^{(d-1)}(\vec p_1-\vec p_2).
\label{eq:rel3}
\end{align}
Combining together \eqref{eq:rel2} and \eqref{eq:rel3} we get
\begin{equation}
\label{eq:normalization_1}
\Big(\<m,\vec p_1|T^{0\mu}(0)|m,\vec p_2\>-2p_1^0p_1^\mu\Big)
\times(2\pi)^{d-1}\delta^{(d-1)}(\vec p_1-\vec p_2) = 0.
\end{equation}
Let us now recall the definition of the stress-tensor form factor \eqref{eq:ff_T}. Using crossing symmetry we conclude that\footnote{The crossing equations for the form factors in 2d are discussed for example in \cite{Karowski:1978vz} and \cite{doi:10.1142/1115}. In general dimensions they can be derived in the QFT framework using the LSZ procedure.
For the derivation of crossing equations in the case of scalar form factors in 4d see chapter 7.2 in \cite{barton1965introduction}.}
\begin{equation}
\label{eq:temp_1}
\<m,\vec p_1|T^{\mu\nu}(0)|m,\vec p_2\> = \mathcal{F}^{\mu\nu}_T(p_1,-p_2).
\end{equation}
Plugging \eqref{eq:temp_1} into \eqref{eq:normalization_1} we obtain 
\begin{equation}
\label{eq:normalization_2}
\Big(\mathcal{F}^{0\mu}_T(p_1,-p_2)-2p_1^0p_1^\mu\Big)
\times(2\pi)^{d-1}\delta^{(d-1)}(\vec p_1-\vec p_2) = 0.
\end{equation}
Using \eqref{eq:mass_shell} we can rewrite this as the following normalization condition of
the stress-tensor form factor
\begin{equation}
\label{eq:normalization_3}
\lim_{p_2 \rightarrow -p_1}\mathcal{F}^{0\mu}_T(p_1,p_2)=2p_1^0p_1^\mu.
\end{equation}

The remaining task is to find the consequence of the condition \eqref{eq:normalization_3}
on the components of the stress-tensor form factors. In order to do that we recall the decomposition of the stress-tensor form factor into tensor structure given by \eqref{eq:stress-tensor_form_factor}. It reads
\begin{equation}
\label{eq:stress-tensor_form_factor_app}
\begin{aligned}
\mathcal{F}^{\mu\nu}_T(p_1,p_2) =
&-\mathcal{F}'_{(0)}(s)
\times\left((p_1+p_2)^2\eta^{\mu\nu}-(p_1+p_2)^\mu(p_1+p_2)^\nu\right)\\
&+\mathcal{F}'_{(2)}(s)
\times(p_1-p_2)^\mu(p_1-p_2)^\nu.
\end{aligned}
\end{equation}
Here compared to \eqref{eq:stress-tensor_form_factor} we have slightly redefined the tensor structures in order to remove the kinematic singularities. The relation between the components of the form factor in \eqref{eq:stress-tensor_form_factor} and \eqref{eq:stress-tensor_form_factor_app} is given by
\begin{equation}
\label{eq:FF_relation_components}
\mathcal{F}'_{(0)}(s)\equiv \frac{\mathcal{F}_{(0)}(s)}{(p_1+p_2)^2},\qquad
\mathcal{F}'_{(2)}(s)\equiv \frac{\mathcal{F}_{(2)}(s)}{(p_1-p_2)^2}.
\end{equation}
Plugging \eqref{eq:stress-tensor_form_factor_app} into \eqref{eq:normalization_3} we obtain
\begin{equation}
\label{eq:result_1}
\lim_{s\rightarrow 0}\mathcal{F}'_{(0)}(s) = -\text{const},\qquad
\lim_{s\rightarrow 0}\mathcal{F}'_{(2)}(s) = \frac{1}{2},
\end{equation}
where const is an undetermined constant not fixed by the normalization condition \eqref{eq:normalization_3}. The minus is introduced for convenience.  

We can now translate the result \eqref{eq:result_1} to the original components of the form factor  \eqref{eq:stress-tensor_form_factor}. We simply have
\begin{equation}
\label{eq:result_2}
\lim_{s\rightarrow 0}s^{-1}\mathcal{F}_{(0)}(s) = \text{const},\qquad
\lim_{s\rightarrow 0}\mathcal{F}_{(2)}(s) = -2m^2,
\end{equation}
Furthermore using the expression of the trace of the stress-tensor form factor in terms of the $\mathcal{F}_{(0)}(s)$ and $\mathcal{F}_{(2)}(s)$ components given by \eqref{eq:trace_FF} and using the normalization conditions \eqref{eq:result_2} we obtain the following normalization of the trace of the stress-tensor form factor
\begin{equation}
\label{eq:result_3}
\lim_{s\rightarrow 0}\mathcal{F}_{\Theta}(s) = -2m^2.
\end{equation}

For works on the stress-tensor form factor normalization in the presence of particles of non-zero spin see for example \cite{Teryaev:2016edw,Lowdon:2020gsj} and references therein.

\bibliographystyle{JHEP}
\bibliography{refs}

\end{document}